\documentclass[12pt]{article} 

\usepackage[utf8]{inputenc}
\usepackage[T1]{fontenc} 
\usepackage{lmodern}
\usepackage[english]{babel}           

\usepackage{cancel}
\usepackage[pdftex]{graphicx}
\usepackage{epsfig}
\usepackage{graphicx}
\usepackage{comment}
\usepackage{latexsym}
\usepackage{hyperref}
\usepackage{amsmath}
\allowdisplaybreaks 
\usepackage{mathrsfs}
\usepackage[usenames, dvipsnames]{color}
\usepackage{amsbsy}
\usepackage{amssymb}
\usepackage{amsthm}
\usepackage{amsfonts}
\usepackage{cite}
\usepackage{enumitem}
\usepackage{xcolor}
\usepackage{color}
\usepackage{mathtools}
\usepackage{caption} 
\usepackage{subcaption} 

\usepackage{float}

\usepackage{diagbox}

\usepackage[title]{appendix}

\usepackage{tabularx}

\newcommand{\be}[0]{\begin{equation}}
\newcommand{\ee}[0]{\end{equation}}

\renewcommand{\thefootnote}{\fnsymbol{footnote}}

\newcommand{\Z}{\mathbb{Z}}
\renewcommand{\natural}{\mathbb{N}}

\renewcommand{\O}{{\cal O}}

\newcommand{\Str}{\textrm{Str}\,}

\newcommand{\diag}{{\rm diag}\,}

\newcommand{\ie}{{\em i.e.} }

\newcommand{\via}{{\it via} }
\newcommand{\apriori}{{\it a priori} }
\newcommand{\where}{\mbox{where}}

\renewcommand{\and}{\mbox{and}}
\newcommand{\for}{\mbox{for}}

\newcommand{\esps}{\phantom{\!\!\!\overset{|}{a}}}

\newcommand{\espD}{\phantom{\!\!\underset{\displaystyle |}{\cdot}}}
\newcommand{\espDD}{\phantom{\!\!\underset{\displaystyle |}{|}}}

\newcommand{\tabitem}{~~\llap{\tiny$\bullet$}~~}

\newcommand{\bm}{\boldmath} 

\newcommand{\V}{{\cal V}}
\newcommand{\W}{{\cal W}}
\newcommand{\F}{{\cal F}}
\newcommand{\N}{{\cal N}}
\newcommand{\J}{{\cal J}}
\newcommand{\I}{{\cal I}}
\newcommand{\K}{{\cal K}}
\renewcommand{\P}{{\cal P}}
\newcommand{\G}{{\cal G}}
\newcommand{\A}{{\cal A}}

\newcommand{\T}{{\cal T}}

\newcommand{\M}{{\cal M}}

\newcommand{\cR}{{\cal R}}
\newcommand{\cH}{{\cal H}}

\newcommand{\Ms}{M_{\rm s}}

\newcommand{\nF}{n_{\rm F}}
\newcommand{\nB}{n_{\rm B}}

\newcommand{\fundabar}{$\overline{\text{fundamental}}\text{ }$}

\newcommand{\half}{\frac{1}{2}}

\newcommand\dd{\text{d}}

\def\jac(#1,#2){%
\begin{bsmallmatrix}
#1\cr 
#2\cr
\end{bsmallmatrix}}

\catcode`\@=11
\def\marginnote#1{}
\newcount\hour
\newcount\minute
\newtoks\amorpm
\hour=\time\divide\hour by60 \minute=\time{\multiply\hour by60
\global\advance\minute by-\hour}
\edef\standardtime{{\ifnum\hour<12 \global\amorpm={am}%
        \else\global\amorpm={pm}\advance\hour by-12 \fi
        \ifnum\hour=0 \hour=12 \fi
        \number\hour:\ifnum\minute<10 0\fi\number\minute\the\amorpm}}
\edef\militarytime{\number\hour:\ifnum\minute<10 0\fi\number\minute}
\def\draftlabel#1{{\@bsphack\if@filesw {\let\thepage\relax
   \xdef\@gtempa{\write\@auxout{\string
      \newlabel{#1}{{\@currentlabel}{\thepage}}}}}\@gtempa
   \if@nobreak \ifvmode\nobreak\fi\fi\fi\@esphack}
        \gdef\@eqnlabel{#1}}
\def\@eqnlabel{}
\def\@vacuum{}
\def\draftmarginnote#1{\marginpar{\raggedright\scriptsize\tt#1}}
\def\draft{\oddsidemargin -.2truein
        \def\@oddfoot{\sl preliminary draft \hfil
        \rm\thepage\hfil\sl\today\quad\militarytime}
        \let\@evenfoot\@oddfoot \overfullrule 3pt
        \let\label=\draftlabel
        \let\marginnote=\draftmarginnote
   \def\@eqnnum{(\theequation)\rlap{\kern\marginparsep\tt\@eqnlabel}%
\global\let\@eqnlabel\@vacuum}  }
\def\thebibliography#1{
\vskip 0.5cm \centerline{\bf \Large References}
\list{
[\arabic{enumi}]}{\settowidth\labelwidth{[#1]}
\leftmargin\labelwidth
\advance\leftmargin\labelsep
\usecounter{enumi}}
\def\newblock{\hskip .11em plus .33em minus .07em}
\sloppy\clubpenalty4000\widowpenalty4000
\sfcode`\.=1000\relax}

\renewcommand{\theequation}{\arabic{section}.\arabic{equation}}
\renewcommand{\section}{\setcounter{equation}{0}\@startsection
{section}{1}{0mm}{-.\baselineskip}{0.5\baselineskip} {\normalfont\Large\bfseries}}
\renewcommand{\subsection}{\@startsection
{subsection}{2}{0mm}{-\baselineskip}{0.5\baselineskip} {\normalfont\large\bfseries}}
\renewcommand{\subsubsection}{\@startsection
{subsubsection}{3}{0mm}{-\baselineskip}{0.5\baselineskip}
{\normalfont\normalsize\slshape}}

\newcommand{\lattice}{\sum_{\vec{m},\vec{n}}\frac{\Lambda^{(4,4)}_{\vec{m},\vec{n}}}{\left|\eta^{4}\right|^{2}}}
\newcommand{\tetad}{\left|\frac{2\eta}{\vartheta_{2}}\right|^{4}}
\newcommand{\tetaq}{\left|\frac{\eta}{\vartheta_{4}}\right|^{4}}
\newcommand{\tetat}{\left|\frac{\eta}{\vartheta_{3}}\right|^{4}}

\newcommand{\latticeo}{\bigg(N^{2}\sum_{\vec{m}}\frac{P^{(4)}_{\vec{m}}}{\eta^{4}}+D^{2}\sum_{\vec{n}}\frac{W^{(4)}_{\vec{n}}}{\eta^{4}}\bigg)}
\newcommand{\tetado}{\left(\frac{2\eta}{\vartheta_{2}}\right)^{2}}
\newcommand{\tetaqo}{\left(\frac{\eta}{\vartheta_{4}}\right)^{2}}
\newcommand{\tetato}{\left(\frac{\eta}{\vartheta_{3}}\right)^{2}}

\newcommand{\latticem}{\bigg(N\sum_{\vec{m}}\frac{P^{(4)}_{\vec{m}}}{\hat{\eta}^{4}}+D\sum_{\vec{n}}\frac{W^{(4)}_{\vec{n}}}{\hat{\eta}^{4}}\bigg)}
\newcommand{\tetadm}{\left(\frac{2\hat{\eta}}{\hat{\vartheta}_{2}}\right)^{2}}

\newcommand{\latticek}{\bigg(\sum_{\vec{m}}\frac{P^{(4)}_{\vec{m}}}{\eta^{4}}+\sum_{\vec{n}}\frac{W^{(4)}_{\vec{n}}}{\eta^{4}}\bigg)}

\newcommand{\latticeowl}{\bigg(N_{ii'}N_{jj'}\sum_{\vec{m}}\frac{P^{(4)}_{\vec{m}+\vec{a}_{i}-\vec{a}_{j}}}{\eta^{4}}+D_{ii'}D_{jj'}\sum_{\vec{n}}\frac{W^{(4)}_{\vec{n}+\vec{a}_{i}-\vec{a}_{j}}}{\eta^{4}}\bigg)}

\newcommand{\latticeowlab}{\left(\sum_{\vec{m}}P^{(4)}_{\vec{m}+\vec{a}_{\alpha}-\vec{a}_{\beta}}P^{(2)}_{\vec{m}'+\vec{a}'_{\alpha}-\vec{a}'_{\beta}}+\sum_{\vec{n}}W^{(4)}_{\vec{n}+\vec{b}_{\alpha}-\vec{b}_{\beta}}P^{(2)}_{\vec{m}'+\vec{b}'_{\alpha}-\vec{b}'_{\beta}}\right)}

\newcommand{\latticeowlabshift}{\left(\sum_{\vec{m}}P^{(4)}_{\vec{m}+\vec{a}_{\alpha}-\vec{a}_{\beta}}P^{(2)}_{\vec{m}'+\vec{a}'_{\text{S}}+\vec{a}'_{\alpha}-\vec{a}'_{\beta}}+\sum_{\vec{n}}W^{(4)}_{\vec{n}+\vec{b}_{\alpha}-\vec{b}_{\beta}}P^{(2)}_{\vec{m}'+\vec{a}'_{\text{S}}+\vec{b}'_{\alpha}-\vec{b}'_{\beta}}\right)}

\newcommand{\latticemwl}{\bigg(N_{ii'}\sum_{\vec{m}}\frac{P^{(4)}_{\vec{m}}}{\hat{\eta}^{4}}+D_{ii'}\sum_{\vec{n}}\frac{W^{(4)}_{\vec{n}}}{\hat{\eta}^{4}}\bigg)}

\newcommand{\latticemwlab}{\left(\sum_{\vec{m}}P^{(4)}_{\vec{m}+2\vec{a}_{\alpha}}P^{(2)}_{\vec{m}'+2\vec{a}'_{\alpha}}+\sum_{\vec{n}}W^{(4)}_{\vec{n}+2\vec{b}_{\alpha}}P^{(2)}_{\vec{m}'+2\vec{b}'_{\alpha}}\right)}

\newcommand{\latticemwlabshift}{\left(\sum_{\vec{m}}P^{(4)}_{\vec{m}+2\vec{a}_{\alpha}}P^{(2)}_{\vec{m}'+\vec{a}'_{\text{S}}+2\vec{a}'_{\alpha}}+\sum_{\vec{n}}W^{(4)}_{\vec{n}+2\vec{b}_{\alpha}}P^{(2)}_{\vec{m}'+\vec{a}'_{\text{S}}+2\vec{b}'_{\alpha}}\right)}

\newcommand{\ndwl}{N_{ii'}D_{jj'}}
\newcommand{\rnwl}{\delta_{ij}\left(R_{ii'}^{\text{N}}R_{jj'}^{\text{N}}+R_{ii'}^{\text{D}}R_{jj'}^{\text{D}}\right)}

\newcommand{\rnrdwl}{e^{4i\pi\vec{a}_{i}\cdot\vec{a}_{j}}R_{ii'}^{\text{N}}R_{jj'}^{\text{D}}}

\newcommand{\vq}{V_{4}}
\newcommand{\oq}{O_{4}}
\newcommand{\sq}{S_{4}}
\newcommand{\cq}{C_{4}}

\newcommand{\opv}{\left|Q_{\text{O}}+Q_{\text{V}}\right|^{2}}
\newcommand{\omv}{\left|Q_{\text{O}}-Q_{\text{V}}\right|^{2}}
\newcommand{\spc}{\left|Q_{\text{S}}+Q_{\text{C}}\right|^{2}}
\newcommand{\smc}{\left|Q_{\text{S}}-Q_{\text{C}}\right|^{2}}

\newcommand{\opvo}{\left(Q_{\text{O}}+Q_{\text{V}}\right)}
\newcommand{\omvo}{\left(Q_{\text{O}}-Q_{\text{V}}\right)}
\newcommand{\spco}{\left(Q_{\text{S}}+Q_{\text{C}}\right)}
\newcommand{\smco}{\left(Q_{\text{S}}-Q_{\text{C}}\right)}

\newcommand{\opvm}{\left(\hat{Q}_{\text{O}}+\hat{Q}_{\text{V}}\right)}
\newcommand{\omvm}{\left(\hat{Q}_{\text{O}}-\hat{Q}_{\text{V}}\right)}

\newcommand{\RN}{R_{\text{N}}}
\newcommand{\RD}{R_{\text{D}}}

\def\car(#1,#2,#3,#4,#5,#6,#7,#8){
\def\ArgI{{#1}}
\def\ArgII{{#2}}
\def\ArgIII{{#3}}
\def\ArgIV{{#4}}
\def\ArgV{{#5}}
\def\ArgVI{{#6}}
\def\ArgVII{{#7}}
\def\ArgVIII{{#8}}
}

\def\carRelay(#1,#2,#3){
\left| \ArgI_{4}\ArgII_{4}#1\ArgIII_{4}\ArgIV_{4}#2\ArgV_{4}\ArgVI_{4}#3\ArgVII_{4}\ArgVIII_{4}\right|^{2}
}

\def\caro(#1,#2,#3,#4,#5,#6,#7,#8){
\def\ArgI{{#1}}
\def\ArgII{{#2}}
\def\ArgIII{{#3}}
\def\ArgIV{{#4}}
\def\ArgV{{#5}}
\def\ArgVI{{#6}}
\def\ArgVII{{#7}}
\def\ArgVIII{{#8}}
}

\def\carRelayo(#1,#2,#3){
\left( \ArgI_{4}\ArgII_{4}#1\ArgIII_{4}\ArgIV_{4}#2\ArgV_{4}\ArgVI_{4}#3\ArgVII_{4}\ArgVIII_{4}\right)
}

\def\carm(#1,#2,#3,#4,#5,#6,#7,#8){
\def\ArgI{{#1}}
\def\ArgII{{#2}}
\def\ArgIII{{#3}}
\def\ArgIV{{#4}}
\def\ArgV{{#5}}
\def\ArgVI{{#6}}
\def\ArgVII{{#7}}
\def\ArgVIII{{#8}}
}

\def\carRelaym(#1,#2,#3){
\left( \hat{\ArgI}_{4}\hat{\ArgII}_{4}#1\hat{\ArgIII}_{4}\hat{\ArgIV}_{4}#2\hat{\ArgV}_{4}\hat{\ArgVI}_{4}#3\hat{\ArgVII}_{4}\hat{\ArgVIII}_{4}\right)
}

\def\carq(#1,#2,#3,#4,#5){
\left|#1_{4}#2_{4}#5#3_{4}#4_{4}\right|^{2}
}

\def\cars(#1,#2,#3,#4,#5){
(#1_{4}#2_{4}#5#3_{4}#4_{4})
}

\def\carsm(#1,#2,#3,#4,#5){
(\hat{#1}_{4}\hat{#2}_{4}#5\hat{#3}_{4}\hat{#4}_{4})
}

\def\carsbar(#1,#2,#3,#4,#5){
(\bar{#1}_{4}\bar{#2}_{4}#5\bar{#3}_{4}\bar{#4}_{4})
}

\def\LAMBDA(#1,#2){
\ifnum #2=0
\ifnum #1=0
\Lambda^{(2,2)}_{\vec{m},\vec{n}}
\else
\Lambda^{(2,2)}_{\vec{m}+\half,\vec{n}}
\fi
\else
\ifnum #1=0
\ifnum #2=0
\Lambda_{\vec{m},\vec{n}}
\else
\Lambda^{(2,2)}_{\vec{m},\vec{n}+\vec{a_{\text{S}}}}
\fi
\fi
\fi
\ifnum #1=1
\ifnum #2=1
\Lambda^{(2,2)}_{\vec{m}+\half,\vec{n}+\vec{a_{\text{S}}}}
\fi
\fi
}

\def\LAMBDA(#1,#2){
\frac{\Lambda^{(2,2)}_{#1,#2}}{\left|\eta^{4}\right|^{2}}
}

\def\LAMBDAO(#1,#2){
\frac{\Lambda^{(2,2)}_{#1,#2}}{\eta^{4}}
}

\def\LAMBDAM(#1,#2){
\frac{\Lambda^{(2,2)}_{#1,#2}}{\hat{\eta}^{4}}
}

\makeatletter
\newcommand{\vast}{\bBigg@{3.5}}
\makeatother

\begin{document}


\begin{titlepage}
\begin{flushright}
IPPP/20/6, CPHT-RR012.032020, March 2020
\vspace{1.5cm}
\end{flushright}
\begin{centering}
{\bm\bf \Large On the stability of open-string orbifold models  \\ 
\vspace{0.2cm}with broken supersymmetry}

\vspace{7mm}

 {\bf Steven Abel$^1$, Thibaut Coudarchet$^2$ and Herv\'e Partouche$^2$}

 \vspace{4mm}

{\textsuperscript{1}Institute for Particle Physics Phenomenology, Durham University, and\\ Department of Mathematical Sciences,\\ South Road, Durham, U.K. \\ \textit{s.a.abel@durham.ac.uk}}

\vspace{4mm}
{\textsuperscript{2}CPHT, CNRS, Ecole Polytechnique, IP Paris, \\F-91128 Palaiseau, France \\ 
\textit{thibaut.coudarchet@polytechnique.edu}\\\textit{herve.partouche@polytechnique.edu}}

\end{centering}
\vspace{0.1cm}
$~$\\
\centerline{\bf\Large Abstract}\\
\vspace{-1cm}

\begin{quote}

\hspace{.6cm} 
We consider an open-string realisation of  $\N=2\to \N=0$ spontaneous breaking of supersymmetry in four-dimensional Minkowski spacetime. It is based on type~IIB orientifold theory compactified on $T^2\times T^4/\Z_2$, with Scherk--Schwarz supersymmetry breaking implemented along $T^2$. We show that in the regions of moduli space where the supersymmetry breaking scale is lower than the other scales, there exist configurations with minima that have massless Bose-Fermi degeneracy and hence vanishing one-loop effective potential, up to exponentially suppressed corrections. These backgrounds describe non-Abelian gauge theories, with all open-string moduli and blowing up modes of $T^4/\Z_2$ stabilized, while all untwisted closed-string moduli remain flat directions. Other backgrounds with strictly positive effective potentials exist, where the only instabilities arising at one loop are associated with the supersymmetry breaking scale, which runs away. All of these backgrounds are consistent non-perturbatively. 

\end{quote}

\end{titlepage}
\newpage
\setcounter{footnote}{0}
\renewcommand{\thefootnote}{\arabic{footnote}}
 \setlength{\baselineskip}{.7cm} \setlength{\parskip}{.2cm}

\setcounter{section}{0}


\tableofcontents

\section{Introduction}

\noindent The question of how moduli come to acquire masses in the true vacuum is central in the context of string phenomenology. Indeed the working hypothesis in much of string phenomenology is that the system is initially supersymmetric, with supersymmetry being a powerful guarantor of vacuum stability. Non-perturbative effects then induce a spontaneous breaking of supersymmetry at a scale much below the string scale $\Ms$~\cite{gc1,gc2,gc3,gc4,gc5,gc6} , introducing mild instabilities in only a very limited number of moduli that lead to phenomenologically desirable effects such as the Brout--Englert--Higgs mechanism. An alternative and arguably more honest approach is to implement  spontaneous supersymmetry breaking from the outset, at the classical level in flat space, and rely on perturbative calculations to derive interesting quantum physics.  In this approach,  loop corrections generate an effective potential for the entire system, in which one must seek local minima for the moduli. Moreover, very few of these minima would be expected to yield a cosmological constant that is close to zero.

This general route was advocated in Refs~\cite{Itoyama:1986ei,Abel:2015oxa,SNS1,SNS2,FR,Abel:2017rch,Abel:2017vos,CatelinJullien:2007hw,Bourliot:2009na,CFP,Borunda:2002ra}, and the question of stability 
was addressed in the heterotic string in~\cite{sta1,sta2,sta3,SNS1,SNS2,CoudarchetPartouche,Itoyama:2020ifw}, and more recently in the type~I framework in~\cite{PreviousPaper,Partouche:2019pgv}. In all these works, supersymmetry breaking was implemented by the string versions~\cite{SSstring1,SSstring11,SSstring2,SSstring3,SSstring4,openSS1,openSS2,openSS3,openSS4,openSS5,openSS6,openSS7, openSS8} of the Scherk--Schwarz mechanism~\cite{SS}, with the effective potential being studied  directly using string perturbation theory at one loop. The type I framework has the advantage of providing \via  T-dualities geometric descriptions of open-string moduli as positions of D-branes in the internal space~\cite{review-3}. 
The purpose of this paper is to demonstrate how the discussion can be  extended to more phenomenologically interesting cases that also contain  orbifolds.  

Let us begin by making some general remarks and observations about the setup. In practice, the scale $M$ of spontaneous supersymmetry breaking will be assumed to be lower than the other scales present, namely the string scale $\Ms=1/\sqrt{\alpha'}$, and the other scales arising from compactification. In other words the directions involved in the Scherk--Schwarz supersymmetry breaking are large compared to $\sqrt{\alpha'}$ and the other directions (or their T-duals). This restriction implies that the one-loop potential is dominated by the massless states and their Kaluza-Klein (KK) modes along the large ``Scherk--Schwarz directions'', and its dependence on the moduli fields becomes tractable. Moreover, any potential tree-level instabilities occurring when $M=\O(\Ms)$~\cite{PV1,PV2}, which are related to the Hagedorn transition, are avoided. Under this assumption,  in the string frame  the effective potential  will inevitably take the following form at an extremal point~\cite{Itoyama:1986ei,Abel:2015oxa,SNS1,SNS2,FR,Abel:2017rch,Abel:2017vos,CatelinJullien:2007hw,Bourliot:2009na,CFP,sta1,sta2,sta3,CoudarchetPartouche,Itoyama:2020ifw,PreviousPaper,Partouche:2019pgv}:
\begin{equation}
\V=\xi (\nF-\nB)M^d +\O\left((\Ms M)^{d\over 2}\, e^{-2\pi c\frac{\Ms }{M}}\right)\,,
\label{vd}
\end{equation}
where $d$ is the spacetime dimension.  In this expression, $\nF$ and $\nB$ are the numbers of precisely massless fermionic and bosonic degrees of freedom, while $\xi>0$ is a constant that accounts for the KK towers. 
Moreover, the exponentially suppressed terms arise from all other string states, where $c$ is an $\O(1)$ moduli-dependent quantity, with the exponential factor corresponding to their Yukawa potential across the compact Scherk--Schwarz volume.\footnote{\label{foo1}{Note that throughout our work, our use of the words ``extremal point of the potential''} is somewhat abusive, since $\V$ is in fact extremal with respect to all moduli except $M$ itself, which has a tadpole unless $\nF=\nB$.  In addition when we  assert properties such as ``tachyon free'', ``flat direction'', and so forth, these properties are all to be understood at one loop, and when all exponentially suppressed corrections are neglected.} 

Now let us  summarise the specific results for toroidal compactification in type~I found in Ref.~\cite{PreviousPaper}, and then anticipate and review those that we will find here.
Ref.~\cite{PreviousPaper} presented the rules for perturbatively consistent models to be tachyon free, which were based upon the  fact that, when an odd number of D$p$-branes is stacked on an orientifold plane (O$p$-plane), the position of one of the branes is rigid~\cite{Schwarz-rigid}, thus enhancing the stability of the setup. Most of these configurations yield $\nF-\nB<0$, while some others satisfy  $\nF-\nB=0$, which is an interesting choice for generating a small cosmological constant. The idea being that, if the one-loop effective potential is exponentially suppressed, then it may conspire with higher loops effects to stabilise $M$ and the dilaton, and eventually yield a cosmological term smaller than in  generic models. However, after imposing all known non-perturbative consistency conditions~\cite{triple, np-2,np-3,np-4,np-5} on configurations satisfying $\nF-\nB\ge 0$ for $d\ge 5$, it was found that there is only one survivor which has dimension $d=5$, and  $\nF-\nB=8\times 8$~\cite{APP}. T-dualizing the internal $T^5$, it corresponds to rendering all of the 32 D5-branes\footnote{We make the choice to call ``branes'' objects that live in the parent type~IIB theory, \ie before any orientifold (or orbifold) action is implemented. In other words, there are as many ``branes'' as Chan--Paton indices. In the descendant theories, these ``branes'' are  non-dynamically independent objects. } rigid, by distributing  them one by one on 32 distant O5-planes. The open-string ``gauge group'' denoted $SO(1)^{32}$ is trivial, where $SO(1)=\{e\}$, with $e$ being the neutral element. 

In the present work, we extend the above analysis to $d=4$ dimensions, when \mbox{$\N=2$} super\-symmetry is spontaneously broken to $\N=0$. {We show that there exist non-pertur\-ba\-ti\-vely consistent models that are tachyon free at one loop, with exponentially suppressed ($\nF-\nB=0$) or positive ($\nF-\nB>0$) potentials $\V$. We will construct them}  in the framework of the Bianchi--Sagnotti--Gimon--Polchinski (BSGP) model~\cite{BianchiSagnotti,GimonPolchinski,GimonPolchinski2}, with the type~I theory being compactified on  the partially orbifolded space $T^2\times T^4/\Z_2$. We choose the Scherk--Schwarz mechanism 
to act along the $T^2$~\cite{openSS1,openSS2,openSS3,openSS4,openSS5,openSS6,openSS7, openSS8, review-1, review-2}, which implies that the entire spectrum (including the ``twisted states'') is sensitive to the supersymmetry breaking. As well as the usual closed strings, the model contains open strings that have Neumann~(N) (or \mbox{Dirichlet~(D)}) boundary conditions when they are attached to one of the 32 D9-branes (or 32 D5-branes)~\cite{review-3}. There are corresponding moduli fields of various kinds, which will be the focus of our attention. Their masses arise at the quantum level once supersymmetry is broken, and can be  studied from various perspectives. Indeed one of the more general aspects of this paper is the array of tools that can be brought to bear on these questions. These will allow us to make the following conclusions about the behaviour of the zoo of moduli:

$\bullet$ Applying suitable T-dualities, all Wilson lines (WL's) on the worldvolumes of the D9- and D5-branes can be mapped into positions of 32+32 D3-branes. The one-loop effective potential is extremal with respect to these moduli when all D3-branes sit on O3-planes. We will derive the signs and magnitudes of the quadratic mass terms at one loop using two different (but related) methods. The first, which is purely algebraic, is based on the knowledge of the massless spectrum that is charged under the Cartan $U(1)$'s associated with the WL's. The second method is to evaluate the one-loop Coleman--Weinberg effective potential with  WL's switched on, and take the double-derivative at the origin of the WL moduli space. The mass matrices of these states is derived also taking into account the effect of six-dimensional anomaly-induced masses.

$\bullet$ In general the open-string sector also contains moduli in the ND sector, whose condensation if they are tachyonic would correspond to  ``recombinations of  branes''~\cite{recomb1,recomb2,recomb3,recomb4}.  One way to determine the masses of these states when the D3-branes sit on O3-planes is to compute the two points functions of ``boundary changing vertex operators''. The computation of such amplitudes in non-supersymmetric backgrounds is an interesting and delicate question, that will be presented in a companion paper~\cite{wip}.

$\bullet$ The closed strings also yield moduli, namely the internal metric and the dilaton in the Neveu--Schwarz-Neveu--Schwarz (NS-NS) sector, as well as the internal components of the Ramond-Ramond (RR) two-form. The expression of the one-loop potential $\V$ as a function of the metric can be derived explicitly. However, because this dependence becomes trivial when the potential is extremal with respect to the open-string WL's (see Eq.~(\ref{vd})), all degrees of freedom of the internal metric are flat directions (up to exponentially suppressed terms), except the supersymmetry breaking scale $M$ itself when $\nF\neq \nB$. Of course, the dilaton remains a flat direction at one loop. To study the dependence of $\V$ on the RR moduli, we use type~I/heterotic duality~\cite{HIdual1,HIdual2,HIdual3,HIdual4,dual0,dual1,dual2,dual3}, which maps the RR two-form to the  antisymmetric tensor. At one loop, the heterotic effective potential receives contributions from winding modes running in the virtual loop, whose masses depend on the antisymmetric tensor. Up to exponentially suppressed terms, there is no additional dependence of the potential on this tensor. Hence, because winding modes on the heterotic side are dual to non-perturbative D1-branes in type~I, we will conclude that $\V$ does not depend on the RR moduli (up to the exponentially suppressed terms). 

$\bullet$ Finally the  moduli arising in the twisted closed-string  sector belong to the quaternionic scalars of the 16 twisted hypermultiplets localized at the 16 fixed points of $T^2\times T^4/\Z_2$ in the BSGP model. Thanks to the generalized Green--Schwarz mechanism taking place in six dimensions~\cite{GimonPolchinski2}, between two and sixteen of these  moduli  acquire a large supersymmetric mass. We do not analyze the masses, which are generated at one loop by the supersymmetry breaking, of the remaining (up to fourteen) twisted quaternions. 

The plan of this work is as follows. In Sect.~\ref{2->0}, we describe the BSGP model on $T^2\times T^4/\Z_2$, with the Scherk--Schwarz mechanism implemented along $T^2$ to break \mbox{$\N=2\to \N=0$}. In particular, we derive the massless spectrum and the one-loop effective potential when all D3-branes (in suitable T-dual descriptions) sit on O3-planes. In Sect.~\ref{sc}, we determine the mass terms of the open-string WL's, the effects of the Green--Schwarz mechanism, and derive the flatness of the untwisted closed-string sector moduli.  In Sect.~\ref{stabmod}, we first discuss the stability/instability of representative examples of brane configurations, which belong to distinct non-perturbatively consistent components of the open-string moduli space~\cite{GimonPolchinski2}. 

We then perform a full scan of the hundreds of billions of possible distributions of the D3-branes on the O3-planes, which correspond to extremal points of the one-loop effective potential.$^{\ref{foo1}}$ We find that at the one-loop level, there are only two  non-perturbatively consistent marginally stable setups with exponentially suppressed effective potential ($\nF-\nB=0$). All open-string moduli are stabilised, together with the blowing up modes of the orbifold, while all untwisted closed-string moduli are flat directions. The anomaly free gauge symmetries are $U(1)\times SU(2)\times SU(5)^2\times SU(7)$ and $U(1)\times SU(3)\times SU(5)^2\times SU(6)$. There also exist four configurations that are tachyon free and have positive potential at one loop \mbox{($\nF-\nB>0$)}, implying that $M$ runs away. There are two further brane distributions that are tachyon free, but modulo possible instabilities associated with moduli existing in the ND sector: the relevant one-loop masses will be studied elsewhere~\cite{wip}. One of these models has $\nF-\nB=0$, while the other has \mbox{$\nF-\nB>0$}.

Our conclusions can be found  in Sect.~\ref{conclusion}. The core of the paper is accompanied by Appendices~\ref{A0} and~\ref{Apot}, which collect those technical details required for Sects.~\ref{2->0} and~\ref{sc}, respectively.


\vspace{0.6cm}
\section{\bm $\N=2\to \N=0$ open-string model}

\label{2->0}

\noindent In this section, we will describe the broad features of toroidal orbifold models of type~I that realize $\N=2\to \N=0$ spontaneous breaking of supersymmetry in four dimensions.  We will  consider the partition function that takes into account arbitrary  marginal deformations arising from the NN and DD sectors of the open strings, as well as from the NS-NS closed-string sector \ie the internal metric. We  also discuss the associated spectrum of the states that are massless at tree level. This will prepare us for the following sections, where we consider the response of the system to the breaking of supersymmetry, in particular its one-loop stability.


\subsection{The supersymmetric setup}
\label{21}

\paragraph{\em Original BSGP model: } Before implementation of the spontaneous breaking of supersymmetry, our framework is the Bianchi--Sagnotti--Gimon--Polchinski  model\cite{BianchiSagnotti,GimonPolchinski,GimonPolchinski2} compactified down to four dimensions.  It is obtained by applying an orientifold projection to the type~IIB theory, with background  
\begin{equation}
\mathbb{R}^{1,3}\times T^{2}\times {T^4\over \Z_2} \, , 
\end{equation}
where we will take Minkowski spacetime to span the directions $X^0,X^1,X^2,X^3$, while the $T^{2}$ torus directions are $X^4,X^5$. The remaining coordinates, corresponding to  the $T^4$ torus, are twisted by the $\Z_2$ orbifold generator,
\be
g:\quad (X^{6},X^{7},X^{8},X^{9})\longrightarrow(-X^{6},-X^{7},-X^{8},-X^{9})\, ,
\ee
implying that the model has  $\N=2$ supersymmetry. The background contains orientifold planes, which are the fixed loci of the orientifold generator $\Omega$ and of the combination $\Omega g$. Hence, an O9-plane lies along the nine spatial directions (the ``fixed locus'' of $\Omega$), while an O5-plane is located at each of the 16 fixed points of $T^4/\Z_2$. In order to cancel their RR charges, the open-string sector comprises $32$ D9-branes, as well as $32$ D5-branes transverse to the $T^{4}/\Z_2$ factor. Consistency conditions require the algebra of Chan--Paton factors to correspond to unitary or symplectic gauge groups rather than orthogonal ones\cite{GimonPolchinski}. The simplest configuration, which has a $U(16)\times U(16)$ open-string gauge group, is obtained when no WL deformations are introduced on the worldvolumes of the D9-branes and D5-branes, and when all D5-branes are coincident on a single O5-plane. The only marginal deformations in this system would be those associated with the NS-NS internal metric $G_{\I\J}$, $\I,\J=4,\dots,9$, which we can split into its $T^2$ components $G_{I'J'}$,  $I',J'=4,5$, and $T^4$ components $G_{IJ}$,  $I,J=6,\dots,9$. 

At one loop, the partition function includes contributions arising from worldsheets of closed strings and open strings, with the  topologies of a torus and Klein bottle, and an annulus and M\"obius strip respectively. Accordingly, the one-loop effective potential (which of course vanishes at this stage) involves  four vacuum-to-vacuum amplitudes $\T$, $\K$, $\A$, $\M$, as shown in Eq.~(\ref{Vdef}). Using the conventions for lattices and characters given in Appendix~\ref{conventions}, these contributions in the ``undeformed''  BSGP model are displayed in Appendix~\ref{GPPS}. 

\paragraph{\em Marginal deformations: } The original model with $U(16)\times U(16)$ open-string gauge group can be deformed by turning on (\ie giving a vev to) any of the available marginal deformations arising from the open-string or closed-string sectors.  In the effective supersymmetric theory these correspond to exactly $F$- and $D$-flat  directions. Let us first enumerate them and then describe them in detail: 
\begin{itemize}
\item[($i$)] Generic  positions of the D5-branes in  $T^{4}/\Z_2$. 
\item[($ii$)] Wilson lines along $T^{2}$ for the gauge group associated with the D5-branes (in the DD sector). 
\item[($iii$)] WL's along all of the six internal directions for the gauge group generated by the D9-branes (in the NN sector). In fact ``Wilson line'' is a misnomer along $T^4/\Z_2$ since we will see that non-trivial  vev's of these moduli reduce the rank of the gauge group. It is only in the $\N=4$ parent theory, without the orbifold generated by $g$, that these moduli are truly {WL's}. 

\item[($iv$)] Non-trivial vev's of the moduli in the ND sector. When the latter condense,  the background can be described in terms of brane recombinations or magnetized branes~\cite{recomb1,recomb2,recomb3,recomb4}. 
\item[($v$)] Non-trivial vev's of the RR moduli, namely  the 2-form components $C_{I'J'}$, $I',J'=4,5$,  and $C_{IJ}$, $I,J=6,\dots,9$. 
\item[($vi$)] Non-trivial vev's of the quaternionic scalars of the 16 twisted hypermultiplets in the closed-string sector. These are the blowing up modes of the orbifold, which are localized at the 16 fixed points of $T^4/\Z_2$. When they are turned on, the $T^4/\Z_2$ is deformed into a smooth $K3$ manifold.
 
\end{itemize}
In the present work, we will not consider deformations of the ND sector moduli ($iv$).\footnote{A subsequent work~\cite{wip}  will be entirely devoted to the delicate computation of their masses generated at one loop when supersymmetry is spontaneously broken.} On the contrary, we will justify that the RR moduli~($v$) do not yield relevant effects.  We will also discuss how the twisted quaternionic moduli in ($vi$) acquire supersymmetric masses thanks to a generalized Green--Schwarz  mechanism. 

Let us start the detailed discussion of actual deformations, with the moduli ($i$) corresponding to the positions along directions $X^6$, $X^7$, $X^8$, $X^9$ of the 32 D5-branes of the type~IIB theory. These must  be symmetric with respect to the generators $\Omega$ and $g$, hence the orientifold projection requires that if a  brane is located at $X^I$, $I=6,\dots,9$,  then a distinct brane  sits at $-X^I$~\cite{review-3}.\footnote{Before implementation of the $\Z_2$ orbifold action, this can be understood by T-dualizing $T^4$  in order to translate the D5-brane positions into D9-brane Wilson lines along the T-dual torus. These WL's are  associated with orthogonal gauge groups~\cite{review-3}.}   
Similarly, the $\Z_2$ twist projection correlates the position of a brane at $X^I$, with that of a brane (distinct or otherwise) at $-X^I$. Broadly speaking, in the type~I string theory, D5-brane positions  in $T^4/\Z_2$ vary in 4's. For instance, if $2n$ D5-branes are sitting at a fixed point, they support a gauge symmetry $U(n)$ that can be broken to $U(n-2k)\times USp(2k)$, with rank reduced to $n-k$,  if $2k$ branes move away from the fixed point together with their $2k$ ``mirror branes'' at the opposite coordinates. Hence the moduli space splits into disconnected components characterized by the value of $2n$ modulo 4, which can be either 0 or 2. In other words, the parity of $n$ matters.\footnote{Even though configurations with an odd number of D5-branes sitting on an O5-plane are symmetric under $X^I\to -X^I$, they are not allowed due to the unitary structure of the gauge group factors. }

\begin{figure}
\captionsetup[subfigure]{position=t}
\begin{center}
\begin{subfigure}[t]{0.48\textwidth}
\begin{center}
\includegraphics [scale=0.55]{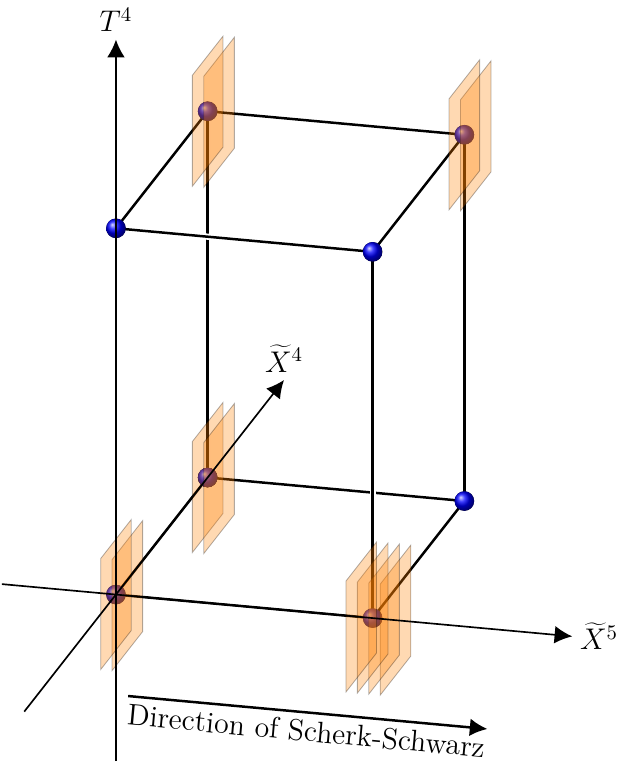}
\end{center}
\caption{\footnotesize A configuration of D3-branes associated with the D5-branes of the initial type~I theory, once $T^{2}$ is T-dualized. In this example, the D3-branes sit on O3-planes.}
\label{D5}
\end{subfigure}
\quad
\begin{subfigure}[t]{0.48\textwidth}
\begin{center}
\includegraphics [scale=0.55]{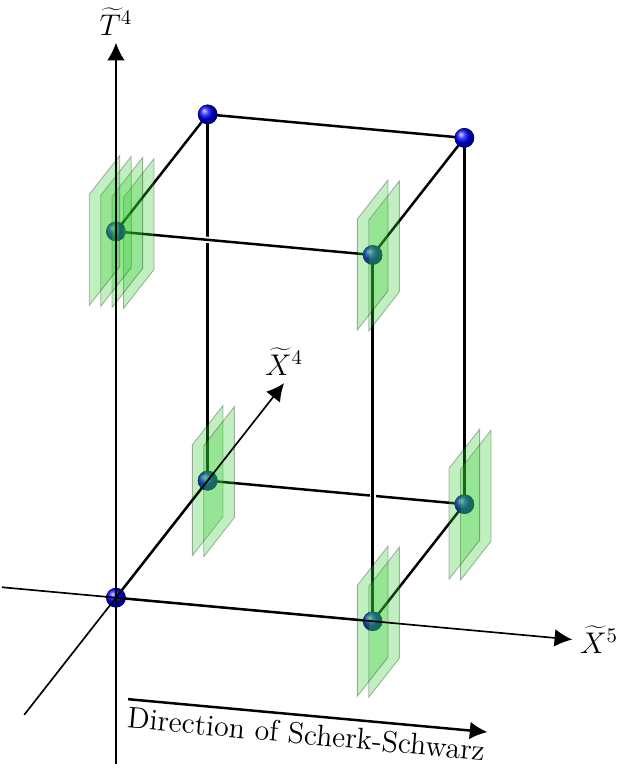}
\end{center}
\caption{\footnotesize A configuration of D3-branes associated with the D9-branes of the initial type~I theory, once both $T^{2}$ and $T^4/\Z_2$ are T-dualized. In this example, the D3-branes sit on O3-planes.}
\label{D9}
\end{subfigure}
\begin{subfigure}[t]{0.48\textwidth}
\begin{center}
\includegraphics [scale=0.55]{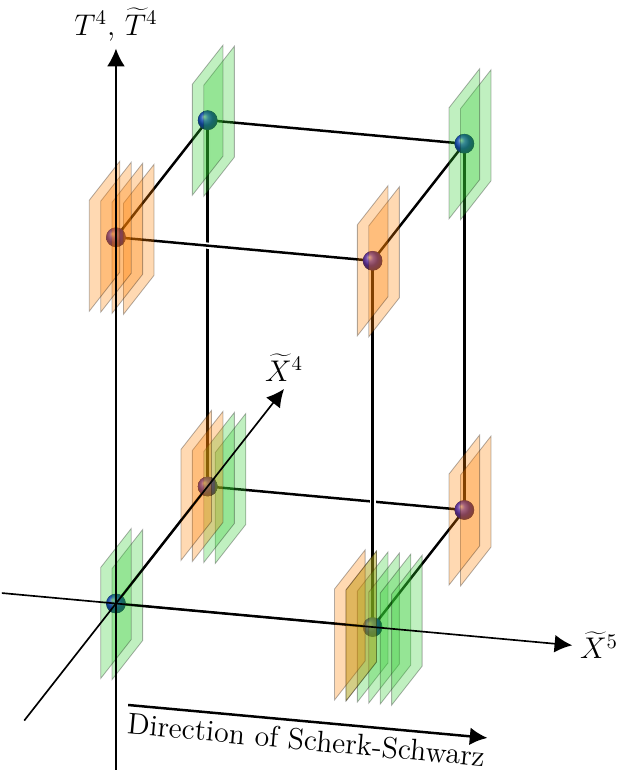}
\end{center}
\caption{\footnotesize Superposition of pictures (a) and (b). D3-branes associated with the  D5-branes (D9-branes) of the initial type~I theory are shown in orange (green).}
\label{D5D9}
\end{subfigure}
\quad
\begin{subfigure}[t]{0.48\textwidth}
\begin{center}
\includegraphics [scale=0.5]{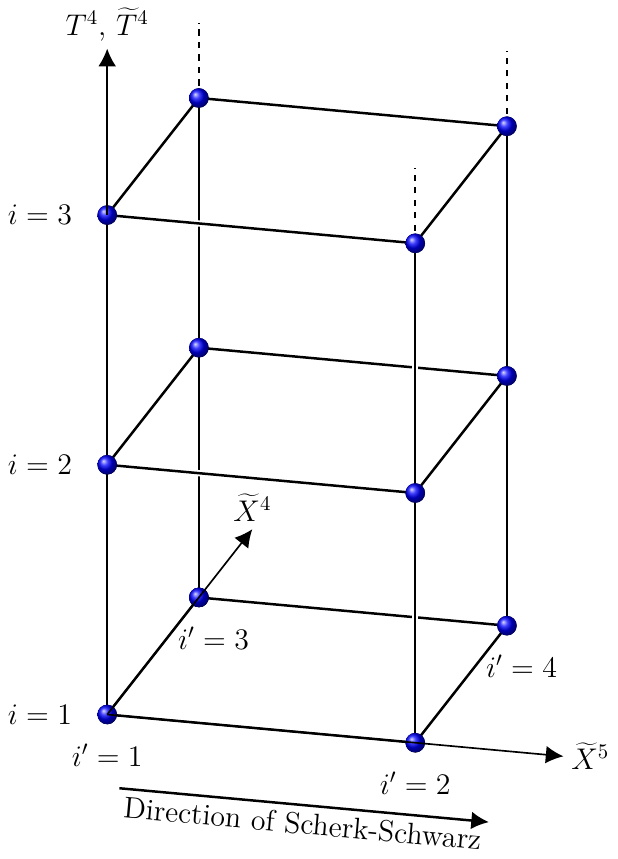}
\end{center}
\caption{\footnotesize Labelling of the $\tilde T^2/I_{45}$ fixed points $i'=1,2,3,4$, and schematic labelling of the $T^4/\Z_2$ or $\tilde T^4/\Z_2$ fixed points $i=1,\dots, 16$. Odd $i'$ correspond to  points located at $\tilde X^5=0$, while even $i'$ are associated with points at $\tilde X^5=\pi$, where $\tilde X^5$ is the coordinate T-dual to the direction along which the Scherk--Schwarz mechanism is implemented.}
\label{corners}
\end{subfigure}
\caption{\footnotesize Geometric T-dual description of the moduli arising from the NN and DD sectors of the orientifold theory. 
}
\label{D5D9D5D9}
\end{center}
\end{figure}

The Wilson lines ($ii$) along the $T^2$ of the D5 gauge groups parameterise the Coulomb branch of the gauge symmetry, and therefore preserve the rank. These also have a geometric interpretation. Upon T-dualizing $T^2$, the D5-branes become  D3-branes transverse to the six-dimensional internal space, and the WL's can then be thought of as the positions of the D3-branes along the T-dual torus $\tilde T^2$ of coordinates $\tilde X^4$,~$\tilde X^5$. Moreover, the 16 O5-planes become 64 \mbox{O3-planes} sitting at the fixed loci of $\Omega I_{45}g$, where $I_{45}$ is the inversion $(\tilde X^4,\tilde X^5)\to (-\tilde X^4,-\tilde X^5)$. Similarly to the deformations ($i$), the position of a D3-brane in $\tilde X^{I'}$, $I'=4,5$, is correlated with that of a distinct partner D3-brane at $-\tilde X^{I'}$. Hence, brane positions along $\tilde T^2/I_{45}$ vary in 2's. In this T-dual geometric picture, the six-dimensional internal space can be thought of as a  ``box'',  a generalization of a one-dimensional segment, with an O3-plane sitting at each of its 64 corners.  This box along with the D3-branes sitting on O3-planes is depicted in Fig.~\ref{D5}.

In the original type~I picture, D5-branes and D9-branes are on an equal footing, in the sense that a T-duality on $T^4/\Z_2$ turns the former into the latter and {\em vice versa}. Hence, the  moduli ($iii$) associated with the gauge group induced by the D9-branes can also be given a geometric interpretation in terms of positions of D3-branes, upon T-dualizing {\em all} the directions of $T^2\times T^4/\Z_2$. An example of a configuration in which the resulting D3-branes sit on O3-planes is shown in Fig.~\ref{D9}, where $\tilde T^4$ denotes the T-dual four-dimensional torus. 

Despite the fact that Figs~\ref{D5} and~\ref{D9} refer to T-dual theories, it is convenient to represent all the D-branes on a single picture, as shown in Fig.~\ref{D5D9}. Although this depiction is certainly abusive, it turns out to be very useful to understand and manipulate various moduli configurations. In practice,  we will refer interchangeably to ``positions'' and ``Wilson lines'' bearing in mind that they refer to the appropriate \mbox{T-dual} pictures. 

Let us now define the Wilson lines in detail. 
We should repeat that the denomination ``Wilson line'' is only fully justified along the $T^2$, or in the parent type~I model, when no orbifold action is implemented. In such an $\N=4$ theory, a Wilson line matrix living in the Cartan subgroup of the D9-brane $SO(32)$ gauge group can be associated with every direction in $T^2\times T^4$. For $\I=4,\dots,9$, it can be parameterised as 
\begin{align}
\begin{split}
\W_{\I}^{\text{D9}}&=\diag\left(e^{2i\pi a_{\alpha}^{\I}},\alpha=1,\dots,32\right)\\
&=\diag\left(e^{2i\pi a_{1}^{\I}},e^{-2i\pi a_{1}^{\I}},e^{2i\pi a_{2}^{\I}},e^{-2i\pi a_{2}^{\I}},\dots,e^{2i\pi a_{16}^{\I}},e^{-2i\pi a_{16}^{\I}}\right),
\end{split}
\end{align}
where $\alpha$ labels the 32 D9-branes, and the corresponding D3-brane positions in $\tilde T^2\times \tilde T^4$ are $\tilde X^\I= 2\pi a_\alpha^\I$. 
In the orbifold model, the number of degrees of freedom of the matrices associated with the $T^4/\Z_2$ directions is reduced, and there are nine disconnected components in the moduli space corresponding to different numbers of fixed points supporting 2 modulo~4 branes:

$\bullet$   The first component of moduli space contains a Higgs branch parameterised by 
\be
\W_{I}^{\text{D9}}=\diag\big(e^{2i\pi a_{1}^{I}},e^{-2i\pi a_{1}^{I}},\dots,e^{2i\pi a_{8}^{I}},e^{-2i\pi a_{8}^{I}},e^{-2i\pi a_{1}^{I}},e^{2i\pi a_{1}^{I}},\dots,e^{-2i\pi a_{8}^{I}},e^{2i\pi a_{8}^{I}}\big)~,
\ee
where $I=6,\dots,9$. Generically this yields a gauge symmetry $USp(2)^8$ of rank 8, whose  Coulomb branch is parameterised by the WL matrices $I'=4,5$,
\be
\W_{I'}^{\text{D9}}=\diag\big(e^{2i\pi a_{1}^{I'}},e^{-2i\pi a_{1}^{I'}},\dots, e^{2i\pi a_{8}^{I'}},e^{-2i\pi a_{8}^{I'}},e^{2i\pi a_{1}^{I'}},e^{-2i\pi a_{1}^{I'}},\dots,e^{2i\pi a_{8}^{I'}},e^{-2i\pi a_{8}^{I'}}\big)~,
\label{wlpart}
\ee
and along which the gauge symmetry is reduced at generic points to $U(1)^8$. 
However, $USp(2)^8$ can be initially enhanced up to $U(16)$ of rank 16  at the points $a^I_1=\cdots=a^I_8\in\{0,\half\}$, $I=6,\dots,9$, and the Coulomb branch is then parameterised by
\be
\W_{I'}^{\text{D9}}=\diag\big(e^{2i\pi a_{1}^{I'}},e^{-2i\pi a_{1}^{I'}},e^{2i\pi a_{2}^{I'}},e^{-2i\pi a_{2}^{I'}},\dots,e^{2i\pi a_{16}^{I'}},e^{-2i\pi a_{16}^{I'}}\big)
\ee
for $I'=4,5$. This  leads generically to an Abelian symmetry $U(1)^{16}$,  with the 8 positions in $\tilde T^4/\Z_2$ stabilised.\footnote{From the gauge theory perspective, they acquire tree level Higgs masses. From the geometric point of view, two pairs of D3-branes at a fixed point of $\tilde T^4/\Z_2$ can only move away from it if the coordinates of the pairs along $\tilde T^2/I_{45}$ match, in order to respect the $\Z_2$ symmetry in $\tilde T^4$. When this is the case for all 8 pairs of pairs, the Coulomb branch takes consistently the form given in Eq.~(\ref{wlpart}).  }

$\bullet$ A second component of the moduli space contains a Higgs branch that may be parameterised as
\be
\begin{aligned}
&\W_{I}^{\text{D9}}=\diag\big(e^{2i\pi a_{1}^{I}},e^{-2i\pi a_{1}^{I}},\dots, e^{2i\pi a_{7}^{I}},e^{-2i\pi a_{7}^{I}}, \eta^I_{8}, \eta^I_{8},\\
&\hspace{5.73cm}e^{-2i\pi a_{1}^{I}},e^{2i\pi a_{1}^{I}},\dots, e^{-2i\pi a_{7}^{I}},e^{2i\pi a_{7}^{I}}, \eta^I_{16}, \eta^I_{16}\big)~,\espD \\
\where \quad &\eta^I_{8},\eta^I_{16}\in \{1,-1\}~,\quad (\eta^6_{8},\eta^7_{8},\eta^8_{8},\eta^9_{8})\neq (\eta^6_{16},\eta^7_{16},\eta^8_{16},\eta^9_{16})~.
\end{aligned}
\ee
Generically, the gauge symmetry is $USp(2)^7\times U(1)^2$, which can again be enhanced up to $U(15)\times U(1)$. In the former case, the gauge group in the Coulomb branch is $U(1)^{9}$ for generic matrices $\W^{\text{D9}}_{I'}$, while in the second case it is $U(1)^{16}$ with all positions in $\tilde T^4/\Z_2$ stabilised.

$\bullet$ There are seven more disconnected components of moduli space. In the ultimate one, the Higgs branch is zero-dimensional, the positions of all 32 branes in $\tilde T^4/\Z_2$ being rigid. To be specific, we have   
\be
\begin{aligned}
&\W_{I}^{\text{D9}}=\diag\big(\eta^I_{1}, \eta^I_{1},\dots,\eta^I_{16},\eta^I_{16}\big)~,\espD \\
\where \quad &\eta^I_{\alpha}\in \{1,-1\}~ , ~\alpha=1,\dots, 16~,\quad (\eta^6_{\alpha},\eta^7_{\alpha},\eta^8_{\alpha},\eta^9_{\alpha})\neq (\eta^6_{\beta},\eta^7_{\beta},\eta^8_{\beta},\eta^9_{\beta})~,~ \alpha\neq \beta~.
\end{aligned}
\ee
There is only a Coulomb branch with the gauge symmetry always being $U(1)^{16}$, regardless of the WL's along $T^2$, 
\be
\W_{I'}^{\text{D9}}=\diag\big(e^{2i\pi a_{1}^{I'}},e^{-2i\pi a_{1}^{I'}},e^{2i\pi a_{2}^{I'}},e^{-2i\pi a_{2}^{I'}},\dots,e^{2i\pi a_{16}^{I'}},e^{-2i\pi a_{16}^{I'}}\big)~.
\ee

Similarly, the positions in $\tilde T^2\times T^{4}/\Z_2$ of the D3-branes T-dual to D5-branes \mbox{$\alpha=1,\dots, 32$}  can be written as $\tilde X^{I'}=2\pi b_{\alpha}^{I'}$, $I'=4,5$, $X^I=2\pi b_\alpha^I$, $I=6,\dots, 9$. They span 9 disconnected components that admit various Higgs, Coulomb or mixed Higgs/Coulomb branches. The latter  can be  parameterised with matrices $\W^{\rm D5}_{\I}$ exactly analogous to those of the D9-branes, up to the exchange $a_\alpha^\I\to b_\alpha^{\I}$. 


\paragraph{\em Discrete deformations: } In what follows we will be mostly interested in configurations where all branes are located at the corners of the appropriate six-dimensional ``boxes.''\footnote{We will see in Sect.~\ref{sc}  that in the presence of spontaneous supersymmetry breaking, such configurations yield  extrema of the effective potential.} 
In order to write the corresponding one-loop amplitudes, we label the 64 corners by a pair of indices $ii'$, where $i\in\{1,\dots,16\}$ refers to the $T^{4}/\Z_2$ (or its T-dual counterpart) fixed points, and  $i'\in\{1,\dots, 4\}$ specifies the $\tilde T^{2}/I_{45}$ fixed points. 
Figure \ref{corners} shows schematically how the labelling works. At a given corner $ii'$, we denote $N_{ii'}$ the number of D3-branes T-dual to D9-branes, and $D_{ii'}$ the number of D3-branes T-dual to D5-branes.
In this setup, the Wilson lines/D3-brane positions $2\pi a_\alpha^\I$ and $2\pi b_\alpha^\I$, $\alpha=1,\dots, 32$, associated with the D9-branes and  D5-branes take values equivalent to the coordinates of some {corner} $ii'$, which we denote by  the six-vectors $2\pi \vec{a}_{ii'}$. It is also convenient to write $\vec a_{ii'}\equiv (\vec a_{i'},\vec a_i)$, where $\vec a_{i'}$, $\vec a_i$ are two- and four-vectors, whose components take values 0 or $\half$. With these definitions, the amplitudes $\A$ and $\M$ arising from the open-string sector are as shown in Appendix~\ref{GPPSWL}. In the closed-string sector, the amplitudes $\T$ and $\K$ are independent of the WL's/brane positions, and their expressions are simply those of the ``undeformed'' $U(16)\times U(16)$ BSGP model (see Appendix~\ref{GPPS}). On the contrary, $\A$ and $\M$ involve the numbers of branes 
$N_{ii'}$, $D_{ii'}$, as well as their counterparts $R^{\text{N}}_{ii'}$ and $R^{\text{D}}_{ii'}$ under the orbifold action. These coefficients can be parameterised as  
\begin{equation}
\label{unitaryparam}
N_{ii'}=n_{ii'}+\bar{n}_{ii'}~,\quad D_{ii'}=d_{ii'}+\bar{d}_{ii'}~,\quad R^{\text{D}}_{ii'}=i(n_{ii'}-\bar{n}_{ii'})~,\quad R^{\text{D}}_{ii'}=i(d_{ii'}-\bar{d}_{ii'})~,
\end{equation}
where $n_{ii'}=\bar{n}_{ii'}$ and $d_{ii'}=\bar{d}_{ii'}$  are positive integers. The tadpole cancellation condition then implies
\begin{equation}
\sum_{i,i'}n_{ii'}=16~,\qquad \sum_{i,i'}d_{ii'}=16~,
\end{equation}
which leads to the open-string gauge group 
\begin{equation}
\G_{\rm open}=\prod_{ii' / n_{ii'}   \neq 0} U(n_{ii'})  \times \prod_{ jj' / d_{jj'}   \neq 0} U(d_{jj'})~.
\end{equation}


\paragraph{\em Non-perturbative consistency: } Although consistent at the perturbative level, the models constructed so far must satisfy additional requirements to remain valid at the non-perturbative level~\cite{GimonPolchinski2}. To state these additional constraints, let us first consider the BSGP model in six dimensions. We have seen that the moduli space of the positions of the D5-branes in $T^4/\Z_2$ splits into 9 disconnected pieces. These are characterized by the even number $\cR=0,2,\dots, 16$ of pairs of D5-branes mirror to each other with respect to $\Omega$ that have rigid positions at distinct fixed points of $T^4/\Z_2$. To be consistent non-perturbatively, a model must have $\cR=0$, 8 or 16. 
When $\cR=8$, the mirror pairs must sit on the 8 corners of one of the hyperplanes $X^I=0$ or $\pi$, $I=6,\dots,9$. Similarly, the number of mirror pairs of D5-branes T-dual to the D9-branes with rigid positions in $\tilde T^4/\Z_2$ must be $\tilde \cR=0$, 8 or 16.  
Hence, there are only $3\times 3$ fully consistent components in the moduli space, which can be further reduced to 6 by T-duality:\footnote{They can  be connected to each other by deforming $T^4/\Z_2$ into smooth $K3$ manifolds~\cite{GimonPolchinski2}.}
\be
(\cR,\tilde \cR)=(0,0)~,~~ (0,8)~,\quad (0,16)~,~~ (8,8)~,~~ (8,16)~,~~ (16,16)~. 
\ee
Compactifying down to four dimensions and T-dualizing $T^2$, there are no additional constraints on the distribution of D3-branes. The latter, including the $2\cR+2\tilde \cR$ ones with rigid positions in $T^4/\Z_2$ or $\tilde T^4/\Z_2$, can move in pairs along the directions of $\tilde T^2/I_{45}$. 


\subsection{Spontaneous breaking of supersymmetry} 
\label{sec22}
What remains to be implemented is the spontaneous breaking of 
$\N=2$ supersymmetry. This can be done  \via a stringy version~\cite{openSS1,openSS2,openSS3,openSS4,openSS5,openSS6,openSS7, openSS8} of the Scherk--Schwarz mechanism~\cite{SS}. To this end, we consider an additional $\Z_2$ orbifold shift on the fifth direction, $X^{5}\rightarrow X^{5}+\pi$,  coupled to $(-1)^{F}$, where $F$ is the spacetime fermion number. 
Denoting  the integer momenta along $T^2$ in the ``undeformed'' supersymmetric  BSGP model by $\vec m'\equiv (m_4,m_5)$ , the combined effects of  the continuous deformations considered so far plus the extra freely acting orbifold action amounts to the following shifts:
\be
\label{shifts}
\begin{aligned}
& \vec m'~\longrightarrow ~\vec m'+F\,\vec a'_{\rm S}&& \quad \mbox{in the closed-string sector}~, \\
&\vec m'~\longrightarrow~ \vec m'+F\,\vec a'_{\rm S}+\vec a'_\alpha-\vec a'_\beta&&\quad \mbox{in the NN sector}~,\\
&\vec m'~\longrightarrow ~\vec m'+F\,\vec a'_{\rm S}+\vec b'_\alpha-\vec b'_\beta&&\quad \mbox{in the DD sector}~,\\
&\vec m'~\longrightarrow ~\vec m'+F\,\vec a'_{\rm S}+\vec a'_\alpha-\vec b'_\beta&&\quad \mbox{in the ND sector}~.\\
\end{aligned}
\ee
In the above, we have defined 
\be
\vec a'_{\rm S}=\Big(0,\half\Big)~,
\ee
while $\vec a'_\alpha\equiv(a^4_\alpha,a_\alpha^5)$ and $\vec b'_\alpha\equiv(b^4_\alpha,b_\alpha^5)$, $\alpha=1,\dots,32$, denote the WL's along $T^2$. Equivalently,  in the D3-brane picture where $2\pi \vec a'_\alpha$ (or $2\pi \vec b'_\alpha$) and $2\pi \vec a'_\beta$ (or $2\pi \vec b'_\beta$) are the positions of the two ends of the open strings in $\tilde T^2$, the components of $\vec m'$ are winding numbers. The key point is of course that the gravitini have acquired masses  
 \begin{equation}
\label{breakingscale}
M=\frac{\sqrt{G^{55}}}{2}\,\Ms~,
\end{equation}
showing that the breaking of $\N=2\to \N=0$ supersymmetry is spontaneous.   Moreover, $M$ itself is one of the marginal deformations, provided it is less than the critical value of order of the string scale $\Ms$, at which a tree-level tachyonic instability arises~\cite{PV1,PV2}. In the language of supergravity, the background is then a ``no-scale model''~\cite{noscale}, which means that  the tree-level potential is positive, semi-definite, and admits a flat direction parameterised by~$M$.  

As described above, when the WL deformations are discrete (the D3-branes sit on the O3-planes of the  six-dimensional boxes), the vectors $\vec a_\alpha'$ and $\vec b_\alpha'$ take values equal to the appropriate $\vec a_{i'}$, $i'=1,\dots,4$. This has an important consequence for the light spectrum, because KK modes in the open-string sector are massless if
\be
\vec m'+F\,\vec a'_S+\vec a_{i'}-\vec a_{j'} = \vec 0~.
\label{T2la}
\ee
This equation admits solutions for both bosons $(F=0)$ and fermions ($F=1$) depending on the relative displacements. This will be detailed in  the next paragraph.


\paragraph{\em  The potential and tree-level massless spectrum:}
\label{massless_spectrum_section}

The one-loop effective potential in  the non-supersymmetric case no longer vanishes. For discrete WL deformations, the amplitudes $\T$, $\K$, $\A$ and $\M$ take the form displayed in Appendix~\ref{GPPSWLSS}. They are expressed in terms of partition functions, from which we can derive the massless bosonic and fermionic spectra. To this end, it is  useful to specify the labelling of the $\tilde T^2/I_{45}$ fixed points as follows: we will denote by $i'=1,3$ those located at the origin of the T-dual Scherk--Schwarz direction, $\tilde X^5=0$, and by $i'=2,4$ those at $\tilde X^5=\pi$ (see Fig.~\ref{corners}). 
From Eqs~(\ref{torus})--(\ref{mobius}), we can then read off  the massless spectrum of the \mbox{$\N=2\to \N=0$} model when the WL's take discrete values as described above. Knowledge of the massless-state representations will be important to derive conditions for the stability of the one-loop potential using a simple algebraic method in Sect.~\ref{algebraic}. 

\begin{figure}[t]
\captionsetup[subfigure]{position=t}
\begin{center}
\begin{subfigure}[t]{0.48\textwidth}
\begin{center}
\includegraphics [trim=0cm 16cm 10cm 0cm,clip,scale=0.55]{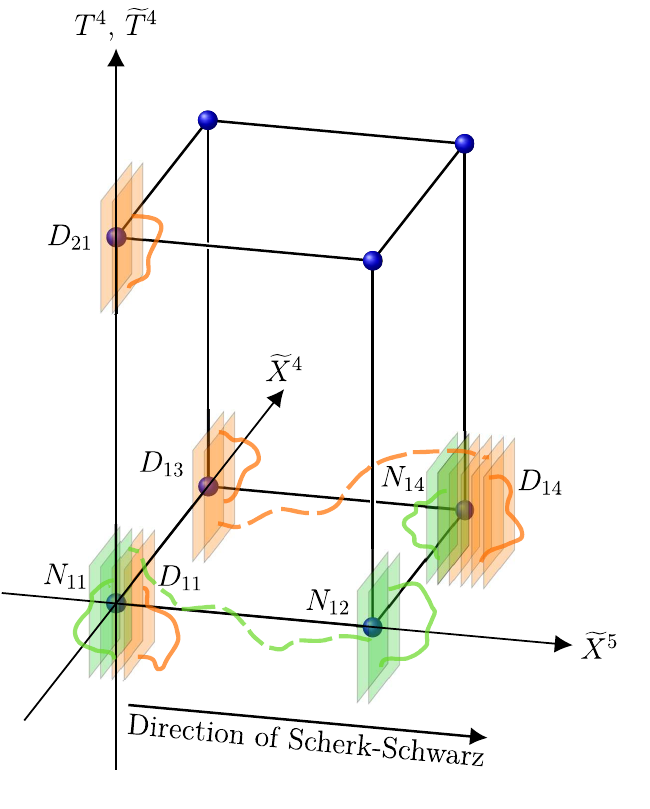}
\end{center}
\caption{Bosonic NN and DD states  (solid strings) are massless when they correspond in the D3-brane picture to strings with both ends attached to the same stack of branes. 
By contrast fermionic NN and DD states  (dashed strings) are massless when they correspond to strings stretched between corners of the six-dimensional box that are adjacent along the \mbox{T-dual} Scherk--Schwarz direction.}
\label{massless_NNDD}
\end{subfigure}
\quad
\begin{subfigure}[t]{0.48\textwidth}
\begin{center}
\includegraphics [trim=0cm 16cm 10cm 0cm,clip,scale=0.55]{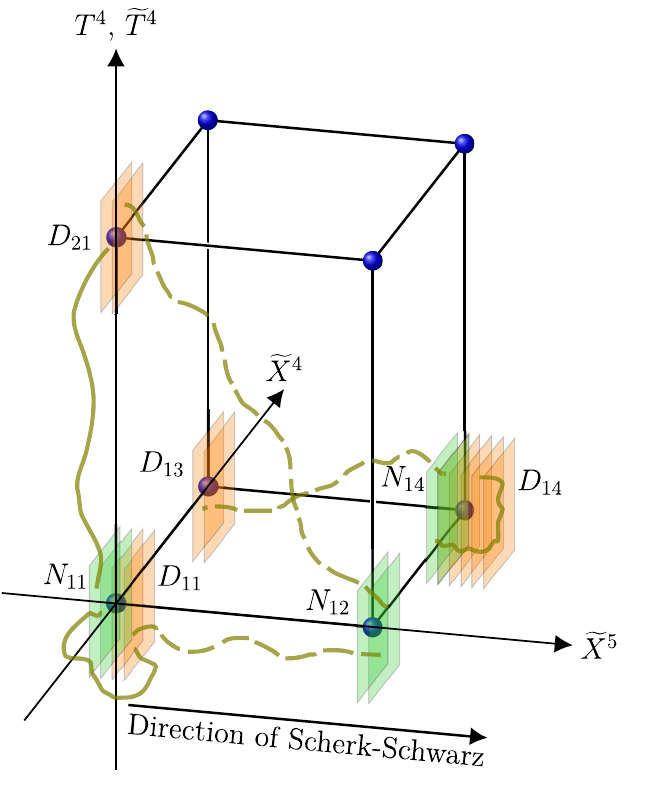}
\end{center}
\caption{ND states correspond to strings stretched between a stack of D3-branes T-dual to D9-branes and a stack of D3-branes T-dual to D5-branes. Bosonic ND states  (solid strings) are massless when the stacks are located on corners with common coordinates in $\tilde T^2/I_{45}$. Fermionic ND states  (dashed strings)  are massless when the corners have common coordinate $\tilde X^4$ and distinct coordinate~$\tilde X^5$. }
\label{massless_ND}
\end{subfigure}
\caption{Open-string massless modes.}
\label{massless_picture}
\end{center}
\end{figure}

In the open-string sector, the massless states arise from characters appearing in $\A$ and $\M$ at the origin of the $T^2$ and $T^4$ lattices. Eq.~(\ref{T2la}), which defines the origin of the $T^2$ lattice, implies that massless bosons require the ends of the strings (in the D3-brane picture) to be located on fixed points of coordinates $\vec a_{ii'}\equiv (\vec a_{i'},\vec a_i)$ and $\vec a_{jj'}\equiv (\vec a_{j'},\vec a_j)$ satisfying
\be
\mbox{\em massless bosons:}\quad \vec a_{i'}-\vec a_{j'}=\vec 0 ~~ \Longleftrightarrow~~ i'=j'~ . 
\ee
On the contrary,  massless fermions require
\be
\mbox{\em massless fermions:}\quad \vec{a}_{i'}-\vec{a}_{j'}=\mp \vec{a}'_{\text{S}} ~~ \Longleftrightarrow~~ \left\{\begin{array}{l}i'=2i^{\prime\prime}-1~,~j'=2i^{\prime\prime}\\\mbox{or}\\
i'=2i^{\prime\prime}~,~j'=2i^{\prime\prime}-1
\end{array}\right.\\,\quad i^{\prime\prime}=1,2~,
\ee
implying that in the $\tilde T^2/I_{45}$, the string is stretched along the T-dual Scherk--Schwarz direction $\tilde X^5$. For such states the contributions to the mass induced by the spontaneous breaking of supersymmetry and by the WL's cancel exactly, \ie the superHiggs and the Higgs mechanisms offset each other.  
In the NN  and DD sectors, whose contributions to the partition functions  involve respectively $T^{4}$ momentum and $T^4$ winding number lattices (in the D9- and D5-brane picture), massless states must also satisfy
\be
\mbox{ \em massless NN or DD states:}\quad \vec a_{i}-\vec a_{j}=\vec 0 ~~ \Longleftrightarrow~~ i=j~.
\ee
Finally, because the ND sector does not involve $T^4$ lattices, $i$ and $j$ need not be correlated to yield massless states, hence
\be
\mbox{\em massless ND states:}\quad  i ,j~~\mbox{arbitrary}~.
\ee
To illustrate the above considerations, Fig.~\ref{massless_NNDD} displays massless states arising in the NN sector (green) and DD sector (orange) that are bosonic (solid strings) or fermionic (dashed strings). Similarly, Fig.~\ref{massless_ND} shows massless strings in the ND sector (khaki) which are bosonic (solid strings) or fermionic (dashed strings).  

At the origin of the lattices appearing in the amplitude $\A+\M$, the massless states arise from the constant terms in the combinations of characters $O_4/\eta^4$, $V_4/\eta^4$, $S_4/\eta^4$, $C_4/\eta^4$ (see Eqs~(\ref{annulus}),~(\ref{mobius})) (\ie the terms $q^0$ in the notations of Appendix~\ref{A0}, where $q=e^{-\pi\tau_2}$ and $\tau_2$ is the Schwinger parameter).\footnote{$O_4,V_4,S_4,C_4$ are $SO(4)$ affine characters arising from the breaking of the ten-dimensional little group $SO(8)\to SO(4)\times SO(4)$ imposed by the $\Z_2$-orbifold action.} These combinations are dressed with coefficients which can be expressed using the unitary parameterisation (\ref{unitaryparam}). For the bosons and fermions, the relevant characters are  respectively
\begin{align}
\label{massless_spectrum}
&\text{\textbf{Bosons: }}{1\over \eta^8}\sum_{i,i'}\bigg\{\vq\oq\left[n_{ii'}\bar{n}_{ii'}+d_{ii'}\bar{d}_{ii'}\right]\nonumber\\
&+\oq\vq\left[\frac{n_{ii'}(n_{ii'}-1)}{2}+\frac{\bar{n}_{ii'}(\bar{n}_{ii'}-1)}{2}+\frac{d_{ii'}(d_{ii'}-1)}{2}+\frac{\bar{d}_{ii'}(\bar{d}_{ii'}-1)}{2}\right]\nonumber\\
&+\oq\cq\sum_{j}\bigg[\frac{1-e^{4i\pi\vec{a}_{i}\cdot\vec{a}_{j}}}{2}\left(n_{ii'}d_{ji'}+\bar{n}_{ii'}\bar{d}_{ji'}\right)+\frac{1+e^{4i\pi\vec{a}_{i}\cdot\vec{a}_{j}}}{2}\left(n_{ii'}\bar{d}_{ji'}+\bar{n}_{ii'}d_{ji'}\right)\bigg]\bigg\}~,\nonumber\\
&\text{\textbf{Fermions: }}{1\over \eta^8}\sum_{i,i''}\bigg\{\cq\cq\left[n_{i,2i''-1}\bar{n}_{i,2i''}+\bar{n}_{i,2i''-1}n_{i,2i''}+d_{i,2i''-1}\bar{d}_{i,2i''}+\bar{d}_{i,2i''-1}d_{i,2i''}\right]\nonumber\\
&+\sq\sq\left[n_{i,2i''-1}n_{i,2i''}+\bar{n}_{i,2i''-1}\bar{n}_{i,2i''}+d_{i,2i''-1}d_{i,2i''}+\bar{d}_{i,2i''-1}\bar{d}_{i,2i''}\right]\\
&+\sq\oq\sum_{j}\bigg[\frac{1-e^{4i\pi\vec{a}_{i}\cdot\vec{a}_{j}}}{2}\left(n_{i,2i''-1}d_{j,2i''}+\bar{n}_{i,2i''-1}\bar{d}_{j,2i''}+n_{i,2i''}d_{j,2i''-1}+\bar{n}_{i,2i''}\bar{d}_{j,2i''-1}\right)\nonumber\\
&\hspace{1cm}+\frac{1+e^{4i\pi\vec{a}_{i}\cdot\vec{a}_{j}}}{2}\left(n_{i,2i''-1}\bar{d}_{j,2i''}+\bar{n}_{i,2i''-1}d_{j,2i''}+n_{i,2i''}\bar{d}_{j,2i''-1}+\bar{n}_{i,2i''}d_{j,2i''-1}\right)\bigg]\bigg\}~.\nonumber
\end{align}
We can immediately read off from these formulae the numbers of massless bosonic and fermionic open-string degrees of freedom:
\be
\begin{aligned}
\nB^{\text{open}}&=4\bigg[2\sum_{ii'}\left(n_{ii'}^{2}+d_{ii'}^{2}\right)+\sum_{i,i',j}n_{ii'}d_{ji'}  -32\bigg]~,\\
\nF^{\text{open}}&=4\bigg[4\sum_{i,i''}\left(n_{i,2i''-1}n_{i,2i''}+d_{i,2i''-1}d_{i,2i''}\right)+\sum_{i,i'',j}\left(n_{i,2i''-1}d_{j,2i''}+n_{i,2i''}d_{j,2i''-1}\right)\bigg]~.
\end{aligned}
\ee
We can also deduce the representations in which these massless modes are organized. For the bosons, the first line in Eq.~(\ref{massless_spectrum}) corresponds to the bosonic content of $\N=2$ vector multiplets in the adjoint representations of the $U(n_{ii'})$ and $U(d_{ii'})$ gauge groups. The second line is associated with the scalars of $\N=2$ hypermultiplets in the antisymmetric $\oplus$ 
$\overline{\text{antisymmetric}}$ representations of $U(n_{ii'})$ and $U(d_{ii'})$. Finally, the last line corresponds to the scalars of hypermultiplets in the ND sector, which are in  bifundamental representations of $U(n_{ii'})\times U(d_{ji'})$. To be more precise, they are in tensor products of fundamental $\otimes$ fundamental or \fundabar representations, depending on the parity of $4\vec a_i\cdot \vec a_j\in \Z$. The massless fermions in the NN , DD and ND sectors are those of hypermultiplets, all in various bifundamental representations of unitary gauge groups supported on stacks of D3-branes separated along the T-dual Scherk--Schwarz direction (and possibly  for the ND states also along $T^4$ or $\tilde T^4$).  

For later use in Sect.~\ref{algebraic}, it is relevant to perform a precise counting of the representations of each individual unitary gauge group factor. In Table~\ref{spectrum} we gather    the massless states charged under  $U(n_{i,2i''-1})$ and $U(n_{i,2i''})$ for given $i=1,\dots, 16$ and $i''=1,2$, which are found  from Eq.~(\ref{massless_spectrum}).  The counting for the gauge groups $U(d_{i,2i''-1})$ and  $U(d_{i,2i''})$, which are generated by the D5-branes, is of course identical, up to the exchange of all coefficients $n_{kk'}\leftrightarrow d_{kk'}$. 
\begin{table}[h!]
\!\!\!\!\!\begin{tabular}{|l|l|}
\hline
\multicolumn{2}{|c|}{Massless representations of $U(n_{i,2i''-1})$}\\
\hline\hline
\underline{Bosonic degrees of freedom:$\esps$} & \underline{Fermionic degrees of freedom:}$\espD$\\ 
\tabitem $4$ adjoint & \tabitem $8 n_{i,2i''}$  (fundamental$~\oplus~\overline{\text{fundamental}}$)\\
\tabitem $4$ (antisymmetric$~\oplus~\overline{\text{antisymmetric}})$ & \tabitem $\displaystyle 2\sum_{j}d_{j,2i''}$  (fundamental$~\oplus~\overline{\text{fundamental}}$) \\
\tabitem $\displaystyle 2\sum_{j}d_{j,2i''-1}$  (fundamental$~\oplus~\overline{\text{fundamental}}$) & \\
\hline
\multicolumn{2}{c}{}\\
\hline
\multicolumn{2}{|c|}{Massless representations of  $U(n_{i,2i''})$}\\
\hline\hline
\underline{Bosonic degrees of freedom:$\esps$} & \underline{Fermionic degrees of freedom:}$\espD$\\
\tabitem $4$ adjoint & \tabitem $8 n_{i,2i''-1}$  (fundamental$~\oplus~\overline{\text{fundamental}}$)\\
\tabitem $4$ (antisymmetric$~\oplus~\overline{\text{antisymmetric}}$) & \tabitem $\displaystyle 2\sum_{j}d_{j,2i''-1}$  (fundamental$~\oplus~\overline{\text{fundamental}}$) \\
\tabitem $\displaystyle 2\sum_{j}d_{j,2i''}$  (fundamental$~\oplus~\overline{\text{fundamental}}$) & \\
\hline
\end{tabular}
\caption{Representations of $U(n_{i,2i''-1})$ and $U(n_{i,2i''}) $ into which the massless degrees of  {freedom} are organized. }
\label{spectrum}
\end{table}

In the closed-string sector, all the initially massless  fermions in the BSGP model acquire a mass $M$ after implementation of the Scherk--Schwarz mechanism.  The massless spectrum thus reduces to the bosonic one encountered in the BSGP model, and is more easily described from a six-dimensional point of view. In the untwisted sector, we have the components of $(G+C)_{\hat \mu\hat\nu}$, $\hat\mu,\hat \nu=2,\dots, 5$, and the internal components $(G+C)_{IJ}$, $I,J=6,\dots,9$, which yield in light-cone gauge $(6-2)\times (6-2)+4\times 4$ degrees of freedom. Moreover, there are also the scalars of the 16 twisted hypermultiplets. Hence, we  obtain a total of \be
\nB^{\text{closed}}=4\times (4+4+16)~,\qquad \nF^{\text{closed}}=0
\ee
bosonic and fermionic degrees of freedom. In terms of six dimensional $\N=1$ supermultiplets, the $\nB^{\text{closed}}$ states comprise the bosonic components of the gravity multiplet $(g_{\hat \mu\hat\nu},C^+_{\hat \mu\hat\nu})$, where $g_{\hat \mu\hat\nu}$ is the traceless graviton and $C^+_{\hat \mu\hat\nu}$ is a self-dual 2-form, a tensor multiplet $(C^-_{\hat \mu\hat\nu},\phi)$, where $C^-_{\hat \mu\hat\nu}$ is an anti self-dual 2-form and $\phi$ is the dilaton, and $4+16$ hypermultiplets.

Taking into account both the closed-string and open-string sectors, the numbers $\nF$ and $\nB$ of massless fermionic and bosonic degrees of freedom in the $\N=2\to \N=0$ model that includes discrete WL deformations satisfy
\begin{align}
\begin{split}
\label{nfnb}
\nF-\nB=4\Big[8&-2\sum_{i,i''}\left(n_{i,2i''-1}-n_{i,2i''}\right)^{2}-2\sum_{i,i''}\left(d_{i,2i''-1}-d_{i,2i''}\right)^{2}\\
&-\sum_{i,i'',j}\left(n_{i,2i''-1}-n_{i,2i''}\right)\left(d_{j,2i''-1}-d_{j,2i''}\right)\Big]~.
\end{split}
\end{align}


\vspace{0.6cm}
\section{Stability conditions}
\label{sc}

\noindent Let us now consider the model described in the previous section at those points in moduli space where the WL's take discrete values. In this section we will show that, at such points, the one-loop effective potential is extremal with respect to the WL's\footnote{It is also extremal with respect to the scalars in the ND sector~\cite{wip}.}, and we will derive the masses of these moduli at the quantum level. We will also determine the masses of (some of) the 16 twisted quaternionic moduli acquired  by a generalized Green--Schwarz mechanism in six dimensions.
For the WL's, we use an algebraic method based on our knowledge of the representations of the massless spectrum, as well as a direct derivation from the one-loop effective potential. We will see that the final answer for the WL masses is obtained by combining these results with a detailed analysis of the one-loop anomaly cancellation mechanism that involves couplings of anomalous $U(1)$ gauge bosons to  twisted Stueckelberg fields.  


\subsection{Signs of the Wilson line masses}
\label{algebraic}

In this and the following subsection, we consider the WL mass terms arising from the one-loop Coleman--Weinberg effective potential. However, we will see in Sect.~\ref{GSm} that additional large contributions (still proportional to the open-string coupling) arise from a generalized Green--Schwarz mechanism that takes place in six dimensions. This effect implies that tachyonic instabilities at the one-loop level can only arise in submanifolds of the WL moduli space described in Sect.~\ref{21}. Therefore, negative signs of the WL mass terms derived in the present subsection do not necessarily imply tachyonic instabilities, as will be seen in Sect.~\ref{stabmod}.  

In Refs~\cite{SNS1,SNS2,CoudarchetPartouche}, an expression for the one-loop effective potential $\V$ was derived for heterotic string compactified on a torus, when supersymmetry is broken by  the Scherk--Schwarz mechanism acting along one compact coordinate, say $X^5$. It applies in the local neighborhood of points in moduli space where extra massless states arise, and is valid provided the size of $X^5$ is greater than the string length as well as all the other compactification length scales (or their T-dual counterparts). In four dimensions, denoting the WL of the $r$-th Cartan $U(1)$ of the gauge group $\G$ along the internal direction $X^\I$ by $y_{r}^{\I}$, we can develop the potential to second order around a point of enhanced massless spectrum as follows: 
\be
\V=M^4(\nF-\nB)\xi~+~ M^4\Big(\sum_{{\rm weights}\, Q \in {\cal R}_{\rm B}}\hspace{-0.1cm} -\hspace{-0.1cm} \sum_{{\rm weights}\, Q\in {{\cal R}_{\rm F}}}\Big)\xi'Q_rQ_s\Big(\sum_{\I=4\atop \phantom{\I}\neq 5}^9{y_{r}^\I y_{s}^\I\over 3G^{55}}+y_{r}^5 y_{s}^5\Big)~+~\cdots~,
\label{VexpD}
\ee
where $\xi,\xi'>0$, the supersymmetry breaking scale is $M$, and where $\nF$, $\nB$ denote the numbers of massless fermionic and bosonic degrees of freedom at $y_r^\I=0$, living respectively in reducible representations ${\cal R}_{\rm F}$, ${\cal R}_{\rm B}$ of $\G$. Note that there is no WL tadpole. This follows from the fact that linear terms in WL's are also linear in Cartan charges $Q_r$ and that the latter can be paired for particles and antiparticles. Writing the gauge group as  $\G\equiv\prod_\kappa\G_\kappa$, the sums over the weights of ${\cal R}_{\rm F}$, ${\cal R}_{\rm B}$ can be expressed in terms of Dynkin indices $T_{{\cal R}_u^{(\kappa)}}$ of irreducible representations ${\cal R}^{(\kappa)}_u$ of the gauge group factors $\G_\kappa$, using the relation  
\begin{equation}
T_{{\cal R}_u^{(\kappa)}}\delta_{rs}=\half\sum_{{\rm weights}\, Q \in {\cal R}_u^{(\kappa)}}Q_{r}Q_{s}~, \qquad  r,s=1,\dots,\text{rank}\, \G_{\kappa}~.
\label{tdyn}
\end{equation}
Indeed, we may write (with no sum over $r$ and $\I$)
\begin{equation}
\frac{\partial^{2}\V}{(\partial y_{r}^{\I})^{2}}\bigg|_{y=0}~\propto~ \sum_u T_{{\cal R}_{{\rm B} u}^{(\kappa)}}-\sum_u T_{{\cal R}_{{\rm F}u}^{(\kappa)}}~,\quad r=1,\dots, \text{rank}\, \G_{\kappa}~,\quad \I=4,\dots,9~,
\label{m2}
\end{equation}
where ${\cal R}_{{\rm B} u}^{(\kappa)}$ and ${\cal R}_{{\rm F} u}^{(\kappa)}$ are the bosonic and fermionic massless representations of $\G_\kappa$. 

Note that in Eq.~(\ref{VexpD}) the coefficients $\xi,\xi'$ capture the contributions of the KK modes propagating along the large extra dimension $X^5$, while all corrections arising from the other massive states (level-matched or not) are exponentially suppressed. Therefore, the resulting expression holds in more general contexts, such as the type~I string theory compactified on tori studied in Ref.~\cite{PreviousPaper}, or in the orbifold model considered in the present work, for the WL's along $T^2$. In particular, the signs of the one-loop contributions to their squared masses can be found by subtracting the Dynkin indices of the fermionic representations from those of the bosonic ones. From Table~\ref{spectrum}, which lists the relevant representations of $SU(q)$, and Table~\ref{Dynkin_SU}
 which gives the associated Dynkin indices, {we find that the one-loop contributions to the squared masses of the WL's along $T^2$, of the special unitary groups supported by the stacks of D9-branes  and D5-branes are proportional (up to positive dressing factors) to\footnote{The effect of a generalised Green--Schwarz mechanism must be taken into account to determine if the WL's along $T^2$ are stable or not (see Sect.~\ref{GSm}).}
\begin{align}
\begin{split}
\label{stabT2bis}
& 4(n_{i,2i''-1}-n_{i,2i''}-1)+\sum_{j=1}^{16}(d_{j,2i''-1}-d_{j,2i''}) \qquad \for\qquad U(n_{i,2i''-1})~,\\
&4(n_{i,2i''}-n_{i,2i''-1}-1)+\sum_{j=1}^{16}(d_{j,2i''}-d_{j,2i''-1})\qquad \for\qquad U(n_{i,2i''})~, \\
&4(d_{i,2i''-1}-d_{i,2i''}-1)+\sum_{j=1}^{16}(n_{j,2i''-1}-n_{j,2i''})  \qquad \for\qquad U(d_{i,2i''-1})~,\\
&4(d_{i,2i''}-d_{i,2i''-1}-1)+\sum_{j=1}^{16}(n_{j,2i''}-n_{j,2i''-1}) \qquad \for\qquad U(d_{i,2i''})~.
\end{split}
\end{align}
\begin{table}[h]
\begin{center}
\begin{tabular}{|c||c|c|c|}\hline
 Gauge factor $\G_{\kappa}$ & Representation $\cR_u^{(\kappa)}$ & $\dim\cR_u^{(\kappa)}$ & $T_{\cR_u}^{(\kappa)}$\\
\hline\hline
$SO(p)$, $p\geq 2$ & fundamental & $p$ & $1$\\
 & adjoint & $\frac{p(p-1)}{2}$ & $p-2$\\
\hline\hline
$SU(q)$, $q\geq 2$ & fundamental & $q$ & $1$\\
 & $\overline{\text{fundamental}}$ & $q$ & $1$\\
 & adjoint & $q^{2}-1$ & $2q$\\
 & antisymmetric & $\frac{q(q-1)}{2}$ & $q-2$\\
 &  $\overline{\text{antisymmetric}}$ & $\frac{q(q-1)}{2}$ & $q-2$\\
 \hline
\end{tabular}
\end{center}
\caption{Dimensions and Dynkin indices of representations of special orthogonal and unitary groups. The Dynkin indices of the fundamental representations are normalized to $1$ by convention.}
\label{Dynkin_SU}
\end{table}
Note that at this stage, these mass-term coefficients}  have been derived assuming $n_{ii'}\ge 2$ and $d_{ii'}\ge 2$. To extend them to the case where $n_{ii'}=1$ or $d_{ii'}=1$, one may consider Eq.~(\ref{m2}) where the adjoint representations have vanishing charges and the antisymmetric representations are zero-dimensional, so that only ``fundamental'' or ``$\overline{\text{fundamental}}$'' representations contribute. Then Eq.~(\ref{m2}) is still applicable but the corresponding coefficients $T_{{\cal R}_{{\rm B} u}^{(\kappa)}}$ and $T_{{\cal R}_{{\rm F} u}^{(\kappa)}}$  are no longer strictly speaking Dynkin indices. As the associated $U(1)$ charges are universal Chan--Paton factors, one finds that the conditions~(\ref{stabT2bis}) remain valid. 

On the contrary, because WL is a misnomer for the moduli describing the positions of the D3-branes along $\tilde T^4/\Z_2$ (or $T^4/\Z_2$), the signs of their squared masses  cannot be determined by applying Eq.~(\ref{m2}) for unitary groups. However, inspecting the amplitude $\A+\M$ in Eqs~(\ref{annulus}),~(\ref{mobius}), we see that small (continuous) deformations of these positions appear only in the NN sector (or DD sector), when  the $\Z_2$-orbifold generator $g$ does not act.\footnote{Explicit expressions are actually given in Eqs.~(\ref{annulus_stab}) and~(\ref{mobius_stab}).} Consequently, up to an overall factor of $\half$, the NN sector contribution is simply that of the open-string sector in the parent $\N=4\rightarrow\N=0$ model studied in \cite{PreviousPaper}, which has orthogonal gauge groups.  The signs of the moduli masses arising at one loop can therefore be found using Dynkin indices of representations of special orthogonal groups, which are shown in Table~\ref{Dynkin_SU}. In the parent $\N=4\rightarrow\N=0$ model,  a pair of stacks of $N_{i,2i''-1}$ and $N_{i,2i''}$ D3-branes T-dual to D9-branes produces an $SO(N_{i,2i''-1})\times SO(N_{i,2i''})$ gauge factor. The states  charged under $SO(N_{i,2i''-1})$ are $8$ bosons in the adjoint representation,  and  $8N_{i,2i''}$ fermions in the fundamental arising from bifundamentals of $SO(N_{i,2i''-1})\times SO(N_{i,2i''})$. The  representations of the degrees of freedom  charged under $SO(N_{i,2i''})$ are identical, up to the exchange $N_{i,2i''-1}\leftrightarrow N_{i,2i''}$. 
 The end result is that the masses of the WL's along $T^4$ of the special orthogonal groups are non-negative when  
\begin{align}
\begin{split}
&N_{i,2i''-1}-N_{i,2i''}-2~\geq ~0\qquad \for\qquad SO(N_{i,2i''-1})~,\\
&N_{i,2i''}-N_{i,2i''-1}-2~\geq ~0\qquad \for\qquad SO(N_{i,2i''})~.
\end{split}
\end{align}
In the $\N=2\to \N=0$ orbifold model, this result implies that 
the masses of the position moduli of the D3-branes in $\tilde T^4/\Z_2$ (or $T^4/\Z_2$) are  non-negative when 
\begin{align}
\begin{split}
&n_{i,2i''-1}-n_{i,2i''}~\geq ~1\qquad \for\qquad U(n_{i,2i''-1})~,\quad n_{i,2i''-1}\ge 2~,\\
&n_{i,2i''}-n_{i,2i''-1}~\geq ~1\qquad \for\qquad U(n_{i,2i''})~,\qquad n_{i,2i''}\ge 2~,\\
&d_{i,2i''-1}-d_{i,2i''}~\,\geq ~1\qquad \for\qquad U(d_{i,2i''-1})~,\quad \,d_{i,2i''-1}\ge 2~,\\
&d_{i,2i''}-d_{i,2i''-1}~\,\geq ~1\qquad \for\qquad U(d_{i,2i''})~,\qquad \,d_{i,2i''}\ge 2~.
\end{split}
\label{sc2}
\end{align}
In the above, the conditions for the D5-brane locations are deduced by  
T-dualizing $T^4/\Z_2$, which amounts to changing all coefficients $n_{kk'}\rightarrow d_{kk'}$. Finally we recall the special cases: namely that {when $n_{i,2i'-1}$, $n_{i,2i'}$, $d_{i,2i'-1}$ or $d_{i,2i'}=1$, the antisymmetric and $\overline{\text{antisymmetric}}$ representations are zero-dimensional, so the positions of the D3-branes in $\tilde T^4/\Z_2$ or $T^4/\Z_2$ are no longer moduli fields.\footnote{As explained in Sect.~\ref{21}, the cause of the rigidity of the position in  $\tilde T^4/\Z_2$ or $T^4/\Z_2$ of a pair of coincident D3-branes can be six-dimensional (in all components of the moduli space with $(\cR,\tilde \cR)\neq (0,0)$). Or it can be four-dimensional, by splitting two pairs of D3-branes at fixed points $ii'$ and $ij'$, where $i'\neq j'$.} Notice that the conditions~(\ref{sc2}) are valid even when there are fewer than 8 dynamical positions in $\tilde T^4/\Z_2$ or $T^4/\Z_2$ (see Sect.~\ref{21}), \ie when there are $U(k)$ gauge group factors with odd $k$'s. This follows from the fact that the remaining dynamical positions of the branes generating the $U(k)$'s must not be tachyonic.

Notice that the two first (last) inequalities in~(\ref{sc2}) are incompatible. Hence, one of them must be absent, which means that either $n_{i,2i''-1}$ or $n_{i,2i''}$ ($d_{i,2i''-1}$ or $d_{i,2i''}$) must be 0 or 1. In other words, the WL positions along $\tilde T^4/\Z_2$ and $T^4/\Z_2$ are non-tachyonic if and only if the configuration satisfies
\be
\forall~i\,,i'':~ (n_{i,2i''-1},n_{i,2i''})~,~ (d_{i,2i''-1},d_{i,2i''})\in \big\{(0,p),\,(p,0),\,(1,p),\,(p,1)~\,\where\,~p\in\natural \big\}\,.
\label{condiwl}
\ee} 


\subsection{Wilson line masses and effective potential}
\label{mT1}

Prior to taking into account the effect of the Green--Schwarz mechanism in the next subsection, let us also discuss how the signs and absolute values of the open-string WL  masses  may be inferred from the one-loop Coleman-Weinberg effective potential $\V$. This is a check on the above stability conditions. To this end, the potential may be evaluated for arbitrary (continuous) D3-brane positions $2\pi a_\alpha^\I$, $2\pi b_\alpha^\I$, $\alpha=1,\dots, 32$, and Taylor expanded at quadratic order around the backgrounds of interest corresponding to branes localized on O3-planes. Hence, we define the WL fluctuations as
\begin{align}
\begin{split}
\label{WLbis}
a_{\alpha}^{\I}&=\langle a_{\alpha}^{\I}\rangle+\epsilon_{\alpha}^{\I}~,\qquad \langle a_{\alpha}^{\I}\rangle~\in~\left\{0,\frac{1}{2}\right\}\,,\\
b_{\alpha}^{\I}&=\langle b_{\alpha}^{\I}\rangle+\xi_{\alpha}^{\I}~,\qquad  \,\langle b_{\alpha}^{\I}\rangle~\in~\left\{0,\frac{1}{2}\right\}\,.
\end{split}
\end{align}
As in the previous subsection, we are interested in regions of moduli space in which the KK mass scale associated with the large Scherk--Schwarz direction $X^5$ is lower than the string scale as well as all other mass scales induced by the compactification moduli $G_{\I\J}$. In practice, this translates to the conditions 
\begin{equation}
G^{55}\,\ll\, G_{44},|G_{IJ}|\,\ll\, G_{55}~,~\,\quad |G_{45}|, |G_{5J}|\,\ll\,\sqrt{G_{55}}~,~\,\quad  I,J=6,\dots,9~,~\,\quad G_{55}\gg 1~.
\end{equation}

The detailed computation of the open-string contribution to the one-loop potential is performed in Appendix~\ref{Apot}. For the closed-string sector, the derivation proceeds as in the  $\N=4$ case in four dimensions which can be found in Ref.~\cite{PreviousPaper}.  The full result takes the  form
\begin{equation}
\mathcal{V}=\frac{\Gamma\big(\frac{5}{2}\big)}{\pi^{\frac{13}{2}}}M^{4}\sum_{l_{5}}\frac{\N_{2l_{5}+1}(\epsilon,\xi,G)}{|2l_{5}+1|^{5}}+\O\left((\Ms M)^{2}e^{-2\pi c\frac{\Ms }{M}}\right)\,,
\label{pofi}
\end{equation}
where $c$ is a positive constant of order 1. In this expression, we have defined 
\be
\N_{2l_{5}+1}(\epsilon,\xi,G)=\nF^{\rm closed}-\nB^{\rm closed}+ \N^{\rm open}_{2l_{5}+1}(\epsilon,\xi,G)~,
\ee
where $\N^{\rm open}_{2l_{5}+1}(\epsilon,\xi,G)$ is given in Eq.~(\ref{N}). The above quantity captures the dominant contributions to $\V$, which arise from the massless states as well as their towers of KK modes propagating along the direction $X^5$. As compared to $M$, all other string modes are super heavy, yielding (together with the non level-matched states in the closed-string sector) exponentially suppressed corrections, as indicated in Eq.~(\ref{pofi}).  Hence, $\N^{\rm open}_{2l_{5}+1}(\epsilon,\xi,G)$ is expressed as a sum over massless open strings stretched between pairs $(\alpha,\beta)$ of branes in the NN, DD or ND sectors. The dependencies on the WL fluctuations $\epsilon^\I_\alpha$, $\xi^\I_\alpha$ appear in the arguments taken by a  function $\cH_{5\over 2}$ given in Eq.~(\ref{cH}), which is dressed by oscillatory cosines. Finally, the definition of $\hat G^{44}$ can be found in Eq.~(\ref{ghat}).  

In order to find the effective potential contribution to the WL masses, we must expand $\N_{2l_{5}+1}(\epsilon,\xi,G)$  to quadratic order using the small argument behaviour of the function $\cH_{5\over 2}$ shown in Eq.~(\ref{expsup}). As seen in Sect.~\ref{21}, the  $\epsilon^\I_\alpha$, $\xi^\I_\alpha$ are however correlated or frozen to zero. To take this fact into account, we label the independent degrees of freedom with indices $r$ and $r'$ as follows, 
\begin{equation}
\label{epxi}
\begin{alignedat}{3}
&\epsilon_{r}^{I}~,\quad && I=6,\dots,9,\quad && r=1,\dots,\sum_{i,i'}\Big\lfloor\frac{N_{ii'}}{4}\Big\rfloor=\sum_{i,i'}\Big\lfloor\frac{n_{ii'}}{2}\Big\rfloor  \le8-{\tilde \cR\over 2}~,\\
&\xi_{r}^{I}~,\quad && I=6,\dots,9,\quad && r=1,\dots,\sum_{i,i'}\Big\lfloor\frac{D_{ii'}}{4}\Big\rfloor=\sum_{i,i'}\Big\lfloor\frac{d_{ii'}}{2}\Big\rfloor  \le8-{\cR\over 2}~,\\
&\epsilon_{r'}^{I'}~, ~\xi_{r'}^{I'}~,\quad && I'=4,5,\quad &&r'=1,\dots,16~,
\end{alignedat}
\end{equation}
where $\tilde \cR$ and $\cR$ were defined previously as the numbers of pairs of D3-branes with rigid positions either in $\tilde T^4/\Z_2$ or $T^4/\Z_2$. 
To write the expansion in compact notations, it is convenient to introduce the following notations:

$\bullet$  $i(r)i'(r)$ denotes the corner of $\tilde T^2/I_{45}\times\tilde T^4/\Z_2$  around which $2\pi \epsilon_r^I$ fluctuates, and $i(r)\hat \imath'(r)$ denotes the adjacent corner along the Scherk--Schwarz direction~$\tilde X^5$. Note that because $\epsilon_r^I$ is dynamical, the two pairs of D3-branes whose position it describes are at the same fixed point of $\tilde T/I_{45}$.

$\bullet$ Similarly,  $j(r)j'(r)$ denotes the corner of $\tilde T^2/I_{45}\times T^4/\Z_2$ around which $2\pi \xi_r^I$ fluctuates, and $j(r)\hat \jmath'(r)$ denotes the adjacent corner along the Scherk--Schwarz direction~$\tilde X^5$.

$\bullet$  $i(r')i'(r')$ denotes the corner of $\tilde T^2/I_{45}\times\tilde T^4/\Z_2$  around which $2\pi \epsilon_{r'}^{I'}$ fluctuates, and $i(r')\hat \imath'(r')$ the adjacent corner along the Scherk--Schwarz direction~$\tilde X^5$.

$\bullet$  Similarly, $j(r')j'(r')$ denotes the corner of  $\tilde T^2/I_{45}\times T^4/\Z_2$ around which $2\pi \xi_{r'}^{I'}$ fluctuates, and $j(r')\hat \jmath'(r')$ the adjacent corner along the Scherk--Schwarz direction~$\tilde X^5$. 

\noindent With these conventions, we obtain 
 \begin{align}
\N_{2l_{5}+1}(\epsilon,\xi,G)&\;= ~\nF-\nB+32\pi^{2}(2l_{5}+1)^{2}\Bigg\{ \nonumber \\
&\;~~\sum_{r}\left(n_{i(r)i'(r)}-n_{i(r)\hat \imath'(r)}-1\right)\epsilon_{r}^{I}\Delta^{IJ}\epsilon_{r}^{J}+\sum_{r}\left(d_{j(r)j'(r)}-d_{j(r)\hat \jmath'(r)}-1\right)\xi_{r}^{I}\Delta_{IJ}\xi_{r}^{J}\nonumber \\
&+\sum_{r'}\left(n_{i(r')i'(r')}-n_{i(r')\hat \imath'(r')}-1+\frac{1}{4}\sum_{i}\left(d_{ii'(r')}-d_{i\hat \imath'(r')}\right)\right)\epsilon_{r'}^{I'}\Delta^{I'J'}\epsilon_{r'}^{J'}\label{N_final} \\
&+\sum_{r'}\left(d_{j(r')j'(r')}-d_{j(r')\hat \jmath'(r')}-1+\frac{1}{4}\sum_{j}\left(n_{jj'(r')}-n_{j\hat \jmath'(r')}\right)\right)\xi_{r'}^{I'}\Delta^{I'J'}\xi_{r'}^{J'}\nonumber\\
&+\O\left(\epsilon^{4},\xi^{4}\right)\Bigg\}\;,\nonumber 
\end{align} 
where we have defined 
\begin{equation}
\Delta^{I'J'}=\frac{1}{3}\left(\frac{G^{I'J'}}{G^{55}}+2\, \frac{G^{5I'}}{G^{55}}\frac{G^{5J'}}{G^{55}}\right)\,,~~\quad\Delta^{IJ}={2\over 3}\frac{G^{IJ}}{G^{55}}~,~~\quad\Delta_{IJ}={2\over 3}\frac{G_{IJ}}{G^{55}}~.
\end{equation}
Because the above tensors have positive eigenvalues, the signs of the WL masses reproduce exactly the results displayed in Eqs~(\ref{sc2}) and~(\ref{stabT2bis}).


\subsection[Mass generation {\em via} generalized Green--Schwarz mechanism]{Mass generation {\bf \em via} generalized Green--Schwarz mechanism}
\label{GSm}

In this subsection, we discuss how Abelian vector bosons in six dimensions become massive thanks to a generalized Green--Schwarz mechanism~\cite{GimonPolchinski2}. As a result, their WL's along $T^2$ are automatically heavy, improving the overall stability of the models. 

Since all $\N=1$ supersymmetric theories are chiral, anomaly cancellations in the BSGP type~IIB orientifold model proceed in a non-trivial way. For any values of the WL's along  $T^4/\Z_2$  for the D9-brane gauge group, and arbitrary positions of  the D5-branes in $T^4/\Z_2$, the fermionic spectrum ensures the cancellation of the irreducible gauge and gravitational anomalies. However, there are residual reducible anomalies, which are described by an anomaly polynomial $I_8$ explicitly written down  in Ref.~\cite{GimonPolchinski2}. When the WL's and positions take discrete values $\vec a_i$, the gauge symmetry generated by the D9-branes and D5-branes is a product of unitary groups,
\be
\prod_{i/n_i\neq 0} U(n_i)\times \prod_{j/d_j\neq 0}U(d_j)~,\quad \where \quad \sum_i n_i=\sum_i d_i=16~,
\label{uni}
\ee 
and where the rank is 32.  As usual in six dimensions, the anomaly polynomial $I_8$ does not factorise, reflecting the fact that massless forms transform nonlinearly under  gauge transformations and diffeomorphisms. In the case at hand, these forms are RR fields belonging to the closed-string spectrum: there is the 2-form $C$ in the untwisted sector, as well as sixteen 4-forms $C^i_4$ in the twisted sector. By Hodge duality ($\dd C^i_4=*\dd C^i_0$), the magnetic 4-form degrees of freedom are equivalent to electric pseudoscalars $C^i_0$. Each of them combines with 3 NS-NS scalars of the twisted sector, thus realizing the bosonic part of the massless twisted hypermultiplet localized at the fixed point $i$ of $T^4/\Z_2$.

Anomaly cancellation requires the effective action to contain tree-level couplings proportional to 
\be
\int C \wedge X_4~~  \quad \mbox{or} ~~\quad \sum_{i,a}c_{ia}  \int C_0^i \wedge F^3_a + \sum_{i,a}c_{ia}  \int C_4^i \wedge F_a ~ , \label{mass2}
\ee
where $F_a$, $a=1,\dots,16$, are the field strengths of the Cartan $U(1)$ generators of $\prod_{i/d_i\neq 0} U(d_i)$, while $F_a$, $a=17,\dots,32$, are those of $\prod_{i/n_i\neq 0} U(n_i)$. Similar couplings involving ${\rm tr} \,R^2$ also exist.  In the above expressions, the coefficients are 
\be
\begin{aligned}
c_{ia}&=4\delta_{a\in i}~,&&\mbox{for $a=1,\dots,16$}~,\\
c_{ia}&= -e^{4i\pi \vec a_i\cdot \vec a_{j(a)}}~, &&\mbox{for $a=17,\dots,32$}~,
\end{aligned}
\ee
where $\delta_{a\in i}=1$ when the  $a$-th $U(1)$ belongs to the Cartan subalgebra of $U(d_i)$, and $\delta_{a\in i}=0$ otherwise. Moreover, we denote by $2\pi \vec a_{j(a)}$ the coordinate vector of the corner of $\tilde T^4/\Z_2$ which supports the Cartan $U(1)$  labelled by $a$ of  $\prod_{j/n_j\neq 0} U(n_j)$  (in a T-dual description).
The Lagrangian can be cast into a local form by dualizing the last term in Eq.~(\ref{mass2}),
which becomes
\be
\sum_{i}\int \big(C_0^i +\sum_{a} c_{ia} A_a\big) \wedge * \big(C_0^i + \sum_{b}c_{ib} A_b\big)~, 
\label{mass3}
\ee
where the $A_a$'s denote the Abelian vector potentials, $F_a=\dd A_a$. As a result, the latter admit a tree-level mass term 
\be
\half \sum_{a,b}A_a \M_{ab}^2 A_b~,\quad \where \quad \M_{ab}^2=\sum_i c_{ia}c_{ib}~.
\ee
The mass matrix $\M^2$ can be diagonalized by an orthogonal transformation, $A_a = \P_{ab} \hat A_b$. Denoting the eigenvalues by $\hat {\cal M}^2_a$, the nonzero ones (which are actually positive) are in one-to-one correspondence with  the Stueckelberg fields $C^i_0$ which are eaten by the $\hat A_a$'s that gain a mass. One can see that if there are 16 or fewer unitary factors in Eq.~(\ref{uni}), all of them are broken to $SU$ groups, while if there are more than 16 unitary factors, exactly 16 are broken to $SU$ groups~\cite{GimonPolchinski2}. By supersymmetry, all twisted hypermultiplets initially containing the $C^i_0$'s which are eaten also become massive. They combine with Abelian vector multiplets to become long massive vector multiplets. As a result, there are between 2 and 16 twisted quaternionic scalars for which stability is automatically guaranteed.   

Compactifying down to four dimensions, we may define the WL's along $T^2$ as $\hat A^{I'}_a=\hat \xi^{I'}_a$,  and write their total mass terms by adding the tree-level contributions to the one-loop effective potential corrections, 
\be
\hat \xi^{I'}_a\left[\hat{\M}^2_a\,\delta_{ab}\,\delta_{I'J'}+ \P_{ca}\,{\partial V\over \partial \xi^{I'}_c\partial \xi^{J'}_d}\,\P_{db}\right] \hat \xi^{J'}_b~,
\ee
where $(\xi^{I'}_1,\dots,  \xi^{I'}_{32})\equiv (\xi^{I'}_1,\dots,  \xi^{I'}_{16},\epsilon^{I'}_1,\dots,  \epsilon^{I'}_{16})$. In the above formula, both contributions are proportional to the open-string coupling. However, while the first one is a supersymmetric mass term proportional to $\Ms^2$, the second one scales like $(M^2/\Ms)^2$, which is always subdominant in the regime $M<\Ms$. Hence, all  WL's of massive $\hat A_a$'s are super heavy and can be safely set to zero in a study of moduli stability,
\be
\hat \xi^{I'}_a\equiv 0~,\quad \mbox{when $\hat{\M}^2_a>0$}~.
\ee
For the remaining WL's denoted $\hat \xi^{I'}_u$ to be non-tachyonic at one loop, one needs to find brane configurations such that the mass matrix 
 \be
 \P_{cu}\,{\partial V\over \partial \xi^{I'}_c\partial \xi^{J'}_d}\,\P_{dv}~,\quad \mbox{for $u,v$ such that $\hat{\M}^2_u,\hat{\M}^2_v= 0$}~, 
 \ee
has non-negative eigenvalues.


\subsection{Untwisted closed-string moduli}
\label{RRmod}

So far, we have mainly discussed the generation of masses for the open-string moduli, as well as for those arising in the closed-string twisted sector. We continue the discussion by considering the dependencies of the effective potential on the closed-string untwisted moduli.

We see from Eqs~(\ref{pofi}) and ~(\ref{N_final}) that when the vev's of the WL's vanish, the  one-loop effective potential reduces to 
\begin{equation}
\mathcal{V}=\xi (\nF-\nB)M^4 +\O\left((\Ms M)^{2}e^{-2\pi c\frac{\Ms }{M}}\right)\,,\quad \where \quad \xi=\frac{\Gamma\big(\frac{5}{2}\big)}{\pi^{\frac{13}{2}}}\sum_{l_{5}}\frac{1}{|2l_{5}+1|^{5}}~.
\end{equation}
Up to the exponentially suppressed corrections, the dependence on the NS-NS internal metric  $G_{\I\J}$ has disappeared, except \via the supersymmetry breaking scale $M$. Therefore, when the D3-branes sit on O3-planes, all components of the (inverse) metric except $G^{55}$ are flat directions. Moreover, unless the potential vanishes \ie $\nF=\nB$,  $G^{55}=4M^2$ has a tadpole and must  run away.  In the NS-NS sector, the remaining untwisted modulus is the dilaton.  However, since the one-loop potential is independent of it, that remains a flat direction at this order. 

The components $C_{I'J'}$ of the RR two-form along $T^2$ can be interpreted as  Wilson lines of Abelian vector bosons $C_{\hat \mu J'}$ in six dimensions. Therefore the algebraic method presented in Sect.~\ref{algebraic} can be applied to determine their masses at the quantum level. Using the fact that the perturbative type~I spectrum does not admit charged states under the RR gauge fields, we can conclude that the moduli $C_{I'J'}$ remain massless at one loop.  It is however possible to draw much  stronger statements using heterotic/type~I duality as follows. For the case at hand, we have been careful to consider type~I models that are expected to be well defined at the non-perturbative level, so that heterotic duals should exist. 
In four dimensions, the above equivalence of the two theories compactified on $T^2\times T^4/\Z_2$ turns out to be a weak coupling/weak coupling duality~\cite{dual0,dual1,dual2,dual3}.  Using the adiabatic argument~\cite{adiabatic}, the equivalence remains valid once the Scherk--Schwarz breaking of supersymmetry is implemented along the large periodic direction $X^5$. 

Let us consider first the case when the $\Z_2$ action generated by $g$ is not implemented yet. The duality maps the type~I variables $(G+C)_{\I\J}$ into  $(G+B)_{\I\J}$ on the heterotic side, where $B_{\I\J}$ is the internal antisymmetric tensor. The moduli deformations of the Narain lattice $\Gamma_{6,6+16}$ can be parameterised by $(G+B)_{\I\J}\equiv Y_{\I\J}$, $\I,\J=4,\dots,9$, as well as the WL's of $SO(32)$ along $T^6$ denoted as $Y_{\I\J}$, $\J=10,\dots,25$. Actually, all of these $6\times (6+16)$ moduli are the WL's of $SO(44)$ along $T^6$. At a generic point in moduli space (the Coulomb branch), the gauge symmetry is reduced to $U(1)^6\times U(1)^{16}$. Conversely, non-Abelian gauge symmetries are restored at enhanced gauge symmetry points. In particular, non-Cartan states charged under $U(1)^6$,  which are generically massive, become massless at special values of $(G+B)_{\I\J}\equiv Y_{\I\J}$. Their Cartan charges are the winding numbers $n_\I$, $\I=4,\dots,9$. Because the Coleman--Weinberg effective potential is expressed in terms of the tree-level mass spectrum, its dependence on $(G+B)_{\I\J}\equiv Y_{\I\J}$ can arise only from the aforementioned non-Cartan states running in the loop.\footnote{We always assume that $M<\Ms$, which implies the contributions of the non-level matched states to be suppressed.} 
Turning back to the type~I picture, these windings states are D1-branes, which belong to the non-perturbative spectrum. As a result, when $M<\Ms$, the one-loop effective potential {\em does not depend on} $C_{\I\J}$, $\I,\J=6,\dots,9$, up to exponentially suppressed corrections. 

Notice however that even though the masses of these D1-branes scale like the inverse string coupling, there is a  moduli-dependent dressing that can vanish, implying such states to be in principle observable in low energy experiments. In the spirit of the seminal works of Seiberg and Witten~\cite{SW} or Strominger~\cite{Strominger:1995cz}, their effects in virtual loops are also captured by the heterotic effective potential~\cite{Estes:2011iw,Liu:2011nw}. In that case, some of the scalars $(G+C)_{\I\J}$, or rather $(G+B)_{\I\J}$, can be stabilised at the enhanced gauge symmetry points described above~\cite{GVenhanced}. As shown in Ref.~\cite{CoudarchetPartouche}, all components  $(G+B)_{\I\J}$, $\I\neq 5$, $\J\neq 5$ can be stabilised. Moreover, the potential is periodic in all $(G+B)_{\I5}$ and the latter can also  be stabilised. Finally, the moduli $(G+B)_{5\J}$ remain flat directions.\footnote{We stress that this assumes $M$ to be lower than the string scale \ie the direction $X^5$ to be large.} 

Re-introducing the $\Z_2$-orbifold action generated by $g$, none of the states arising from the twisted sector in heterotic string can induce an enhancement of the gauge symmetry.\footnote{This follows from the fact that the zero-point energy of the twisted vacuum is higher than that of the untwisted sector. } They can however have non-trivial winding numbers along $T^2$ and thus introduce extra dependencies of the Coleman--Weinberg effective potential on the WL's $(G+B)_{I'J'}$, \mbox{$I',J'=4,5$}. However, due to their high masses, their contributions are exponentially suppressed. The type~I counterparts of these states are ``twisted D1-branes'', which would not be taken into account in perturbation theory. 

One might question the extensive use of heterotic/type~I duality, because the open-string side contains a D5-brane sector, which is  mapped to a non-perturbative NS5-brane sector on the heterotic side.  However, the states that are potentially responsible for  the non-perturbative stabilisation of type~I moduli $(G+C)_{I'J'}$, $I',J'=4,5$  and   $(G+C)_{IJ}$, $I,J=6,\dots,9$, are D1-branes. The latter are electrically charged under  the  two-form $C$, and magnetically neutral (they are not dyonic  D1-D5 bound states). As a result, the stabilisation mechanism is independent of the existence of a D5-brane sector.


\vspace{0.6cm}
\section{Stability analysis of the models}
\label{stabmod}

\noindent {Let us now turn to the analysis of the one-loop stability of the moduli (or at least a sub-set of them) encountered when all  D3-branes are located at corners of the six-dimensional box depicted schematically in Fig.~\ref{corners}. We will restrict the discussion to the configurations satisfying the non-perturbative constraints presented at the end of Sect.~\ref{21}. 
The mass terms of the WL's can be read from Eq.~(\ref{N_final}), but a projection on the submanifold of the moduli  not acquiring a six-dimensional supersymmetric mass from the Green--Schwarz mechanism must simultaneously be applied. In our  study, stability of the twisted quaternionic moduli is only guaranteed when they become massive due to this mechanism. We will not determine their stability at one loop when they remain massless in six dimension. Finally, a sufficient condition for instabilities not to arise from the ND sector of the theory is simply the absence of ND moduli, which is ensured if none of the D3-branes T-dual to the D9-branes and none of the D3-branes T-dual to the D5-branes share the same position in $\tilde T^2/I_{45}$,
\begin{equation}
\mbox{\em no ND-sector moduli:}\quad n_{ii'}d_{ji'}=0~\quad\mbox{for all } i,j,i'~\mbox{(no sum on $i'$)}~.
\label{no_ND}
\end{equation}
If this condition is not satisfied, then the radiatively induced masses-squared of the moduli in the ND sector must be computed. This can be done by considering the two-point functions of ``boundary changing vertex operators''. This is an interesting problem in its own right, which will be studied in a companion paper~\cite{wip}.

In what follows, we will first present simple examples lying in the $(\cR,\tilde \cR)=(0,0)$ and $(\cR,\tilde \cR)=(16,16)$ components of the moduli space to get familiar  with the implementation of the  generalized Green--Schwarz mechanism. Thanks to a numerical exploration of all brane configurations, we then list all setups that yield a vanishing or positive one-loop potential and that are tachyon free  (up to exponentially suppressed terms). 
} 


\subsection[Simple configurations in the component $(\cR,\tilde \cR)=(0,0)$]{\bm Simple configurations in the component $(\cR,\tilde \cR)=(0,0)$} 

At tree level in the branch $(\cR,\tilde \cR)=(0,0)$ of the WL moduli space,  all 32+32 \mbox{D3-branes} are free to move in 4's in $T^4/\Z_2$ or $\tilde T^4/\Z_2$. Let us consider the simplest configuration where all D3-branes T-dual to the D5-branes have the same positions $2\pi \vec a_{i_0}$ in $T^4/\Z_2$, while those T-dual to the D9-branes have common positions   $2\pi \vec a_{j_0}$ in $\tilde T^4/\Z_2$. In six dimensions, the open-string gauge group before taking into account the Green--Schwarz mechanism is thus $U(16)\times U(16)$. To determine the anomalous $U(1)$'s that become massive, we need to write the mass matrix squared $\M_{ab}^2$ of the 32 Abelian  vector potentials $A_a$ in six dimensions. To this end, it is convenient to refine our labelling as follows: 
\begin{equation}
\nonumber 
\begin{aligned}
&\mbox{$a\equiv r'=1,\dots ,16 \qquad ~~~:~$ Cartan generators of the $U(16)$ arising from the D5-branes,} \\
\quad &\mbox{$a\equiv \tilde r'+16=17,\dots,32 ~:~$ Cartan generators of the $U(16)$ arising from the D9-branes. }
\end{aligned}
\end{equation}
With this notation the mass matrix squared is
\begin{equation}
\label{param_M}
\M^2=\left(\begin{array}{@{}c|c@{}}
\M^2_{r's'} & \M^2_{r'\tilde{s}'} \\
\hline
\M^2_{\tilde{r}'s'} & \M^2_{\tilde{r}'\tilde{s}'}
\end{array}\right)\,,
\end{equation}
where the $16\times 16$ blocks are given by 
\begin{equation}
\begin{aligned}
&\M^2_{r's'}=16~,\quad&&\M^2_{r'\tilde{s}'}=-4\,e^{4i\pi\vec{a}_{i_0}\cdot\vec{a}_{j_0}}~,\\
&\M^2_{\tilde{r}'s'}=-4\,e^{4i\pi\vec{a}_{j_0}\cdot\vec{a}_{i_0}}~,\quad&&\M^2_{\tilde{r}'\tilde{s}'}=16~.
\end{aligned}
\end{equation}
Among the 32 eigenvalues, 2 are  positive while the others vanish. Setting to zero the vev's of the massive eigenvectors yields the conditions
\begin{equation}
-\sum_{r'}A_{r'}+\sum_{\tilde{r}'}A_{\tilde{r}'}=0~~\quad\text{and}~~\quad\sum_{r'}A_{r'}+\sum_{\tilde{r}'}A_{\tilde{r}'}=0~,
\label{vbm}
\end{equation}
implying that  $U(16)\times U(16)$ is actually reduced to $SU(16)\times SU(16)$, as expected. 

To proceed, let us consider the examples where all D3-branes T-dual to the D5-branes are coincident at $2\pi \vec a_{i_0'}$ in $\tilde T^2/I_{45}$, and similarly those T-dual to the D9-branes are stacked at $2\pi \vec a_{j_0'}$. The gauge symmetry in four dimensions is therefore still $SU(16)\times SU(16)$. The mass terms of the moduli/positions $\xi^I_r$ along $T^4/\Z_2$ and $\epsilon^I_r$ along $\tilde T^4/\Z_2$ (see Eq.~(\ref{epxi})), $I=6,\dots,9$, $r=1,\dots,8$, can be read from Eq.~(\ref{N_final}). Omitting all dressing factors, they are given by $n_{ii'}$- and  $d_{jj'}$-dependent coefficients equal to $(16-0-1)=15$, which is positive. Hence, the positions of the D3-branes along the internal directions $I=6,\dots,9$ are stabilised.

\begin{figure}[h!]
\captionsetup[subfigure]{position=t}
\begin{center}
\begin{subfigure}[t]{0.31\textwidth}
\begin{center}
\includegraphics [scale=0.55]{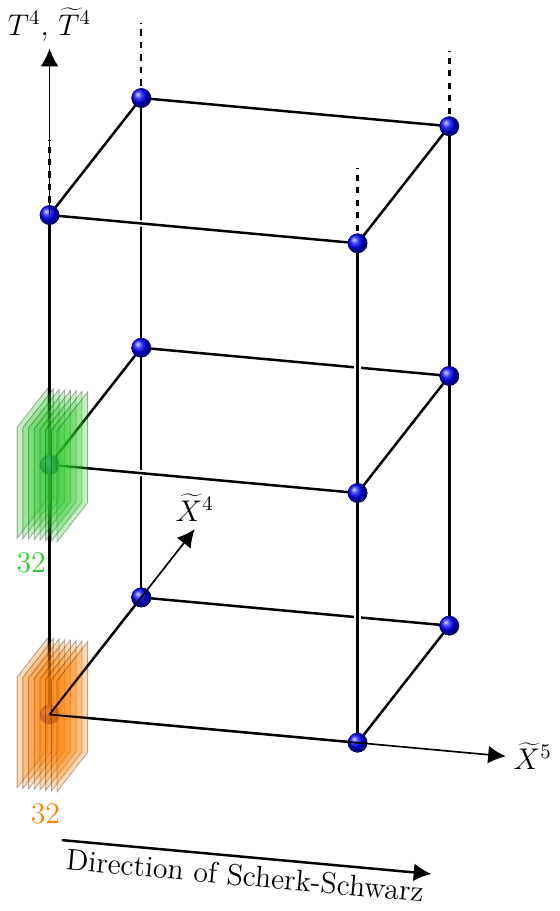}
\end{center}
\caption{\footnotesize The two stacks of 32 D3-branes T-dual to the D5- or D9-branes have common positions in $\tilde T^2/I_{45}$. }
\end{subfigure}
\quad
\begin{subfigure}[t]{0.31\textwidth}
\begin{center}
\includegraphics [scale=0.55]{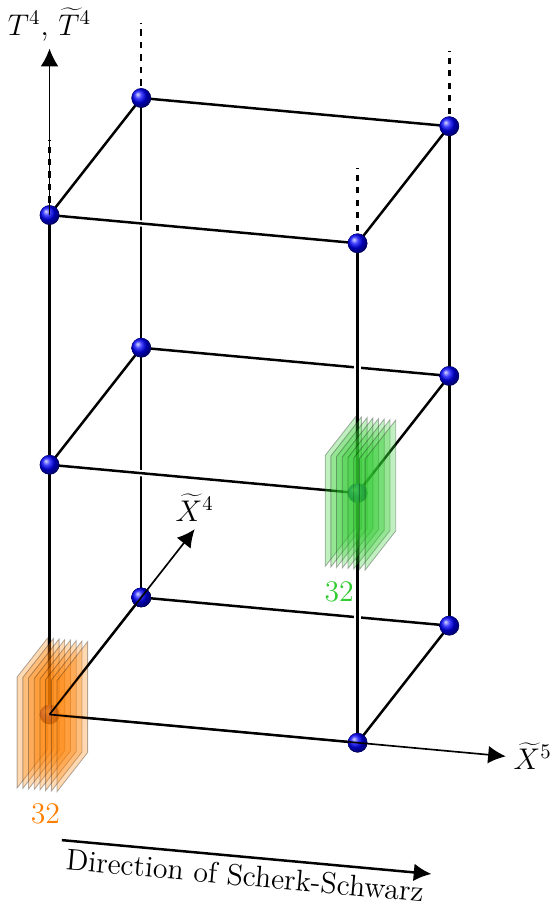}
\end{center}
\caption{\footnotesize The two stacks have the same coordinates along $\tilde X^4$ and distinct coordinates along $\tilde X^5$.}
\end{subfigure}
\quad
\begin{subfigure}[t]{0.31\textwidth}
\begin{center}
\includegraphics [scale=0.55]{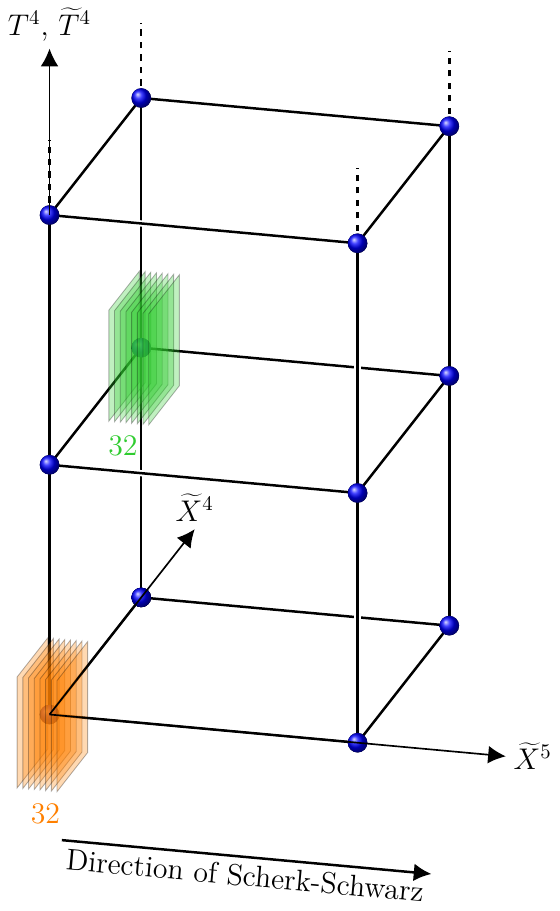}
\end{center}
\caption{\footnotesize The two stacks have distinct  coordinates along $\tilde X^4$.  }
\end{subfigure}
\caption{\footnotesize D3-brane configurations  in component $(\cR,\tilde \cR)=(0,0)$ of the WL moduli space. }
\label{16-abc}
\end{center}
\end{figure}

 As seen in Eq.~(\ref{N_final}), the mass terms of the $T^2$ WL's  $\xi^{I'}_{r'}$ and $\epsilon^{I'}_{\tilde r'}$ arising from the one-loop effective potential depend on the precise locations of the stacks in $\tilde T^2/I_{45}$.  Omitting irrelevant dressings as earlier, they are given by coefficients $(16-0-1+{\delta\over 4} \,16)=15+4\delta$, where
\begin{enumerate}[label=(\alph*)]
\item $\delta=+1$ if $i_0'=j_0'$,
\item $\delta=-1$ if the corners $i_0'$ and $j_0'$ of $\tilde T^2/I_{45}$ are facing each other along the Scherk--Schwarz direction~$\tilde X^5$,
\item $\delta=0$~~\,  if the  corners $i_0'$ and $j_0'$ of $\tilde T^2/I_{45}$ have distinct positions along $\tilde X^4$.
\end{enumerate}
The three possibilities are depicted in Fig.~\ref{16-abc}.
Note that $\delta=+1$ ($\delta=-1$) in Case~(a) (Case~(b))  thanks to the existence at tree level of massless scalars (fermions) in the ND sector. Because these mass terms are positive, we can immediately conclude that all positions in $\tilde T^2/I_{45}$ are stabilised. However, it is instructive to also take into account the effect of the  generalized Green--Schwarz mechanism, which makes  the components $I'=4,5$ of the linear combinations of six-dimensional vector bosons of Eq.~(\ref{vbm}) even more massive. Indeed, this can be used to eliminate say $\xi_{1}^{I'}$ and $\epsilon_{1}^{I'}$,
\begin{equation}
\xi_1^{I'}=-\sum_{r'\neq 1}\xi_{r'}^{I'}~,~~\quad\epsilon_1^{I'}=-\sum_{\tilde{r}'\neq 1}\epsilon_{\tilde{r}'}^{I'}~,
\end{equation}
in the mass terms of Eq.~(\ref{N_final}). This results in a new $30\times 30$ mass matrix squared for the remaining moduli $\xi_{r'}^{I'}, \epsilon_{\tilde r'}^{I'}$, which of course has only strictly positive eigenvalues.\footnote{14 are equal and the last one is 16 times larger.} 

To conclude on the above examples, the masses of the moduli we have not analyzed are those of the 14 remaining hypermultiplets in the twisted closed-string sector, as well as those of the hypermultiplet in the single bifundamental of $SU(16)\times SU(16)$ arising from the open-string ND sector in Case~(a). Using Eq.~(\ref{nfnb}), we have $\nF-\nB=-4064-1024\, \delta<0$, which implies that the supersymmetry breaking scale (\ie gravitino mass) $M$ runs away, while all other components of the NS-NS metric $G_{\I\J}$ and the dilaton as well as the RR two-form $C_{\I\J}$ are flat directions.


\subsection[Simple configurations in the component $(\cR,\tilde \cR)=(16,16)$]{\bm Simple configurations in the component $(\cR,\tilde \cR)=(16,16)$} In this case, all D3-branes positions in $T^4/\Z_2$ or $\tilde T^4/\Z_2$ are rigid. Indeed, there is a mirror pair (with respect to the orientifold projection) of \mbox{D3-branes} T-dual to the D5-branes at each of the 16 fixed point of $T^4/\Z_2$, and similarly  a mirror pair of D3-branes T-dual to the D9-branes at each fixed point of $\tilde T^4/\Z_2$. Before taking into account the effect of the Green--Schwarz mechanism, the gauge symmetry is $U(1)^{16}\times U(1)^{16}$. Hence, all antisymmetric representations are zero dimensional (see Eq.~(\ref{massless_spectrum}) or Table~\ref{spectrum}) and there is indeed no position modulus among them to consider. In this component of the moduli space, the only freedom is in the coordinates of the mirror pairs in $\tilde T^2/I_{45}$, which in our case of interest coincide with the positions of the four fixed points.  

To study the masses of the moduli/positions along $\tilde T^2/I_{45}$, as well as those of the twisted quaternionic scalars in the closed-string sector, our starting point is the mass matrix squared $\M^2_{ab}$ of the 32 Abelian vector potentials present in the six-dimensional theory. Its components are given by 
\begin{equation}
\begin{aligned}
&\M^2_{r's'}=16\,\delta_{r's'}~,\quad&&\M^2_{r'\tilde{s}'}=-4\,e^{4i\pi\vec{a}_{i(r')}\cdot\vec{a}_{i(\tilde{s}')}}~,\\
&\M^2_{\tilde{r}'s'}=-4\,e^{4i\pi\vec{a}_{i(\tilde{r}')}\cdot\vec{a}_{i(s')}}~,\quad&&\M_{\tilde{r}'\tilde{s}'}=16\,\delta_{\tilde{r}'\tilde{s}'}~.
\end{aligned}
\end{equation}
Because the gauge group contains more than 16 unitary factors, the matrix has 16 positive eigenvalues and 16 vanishing ones. This  implies that the gauge symmetry $U(1)^{32}$ is actually reduced to $U(1)^{16}$, and that all of the 16 twisted quaternionic scalars are massive, ensuring that $T^4/\Z_2$ will not undergo deformation into a smooth K3 manifold. Setting to zero all massive linear combinations of vector potentials, we obtain for their components along $\tilde T^2/I_{45}$ the relations
\begin{equation}
\label{rel_GS}
4\epsilon_{\tilde{r}'}^{I'}=-\sum_{s'}e^{4i\pi\vec{a}_{i(\tilde{r}')}\cdot\vec{a}_{i(s')}}\xi_{s'}^{I'}~,
\end{equation}
showing that all $\epsilon_{\tilde r'}^{I'}$ can be eliminated in terms of the $\xi^{I'}_{r'}$'s. Let us now consider various D3-brane configurations  and explore their stability along $\tilde T^2/I_{45}$.

\begin{figure}
\captionsetup[subfigure]{position=t}
\begin{center}
\begin{subfigure}[t]{0.31\textwidth}
\begin{center}
\includegraphics [scale=0.55]{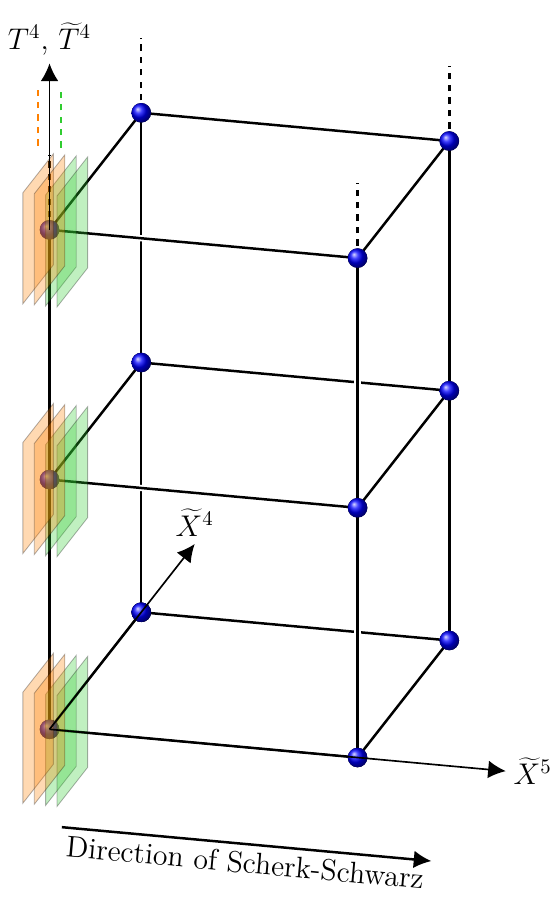}
\end{center}
\caption{\footnotesize The $16+16$ pairs of D3-branes T-dual to the D5- or D9-branes have common positions in $\tilde T^2/\I_{45}$.}
\end{subfigure}
\quad
\begin{subfigure}[t]{0.31\textwidth}
\begin{center}
\includegraphics [scale=0.55]{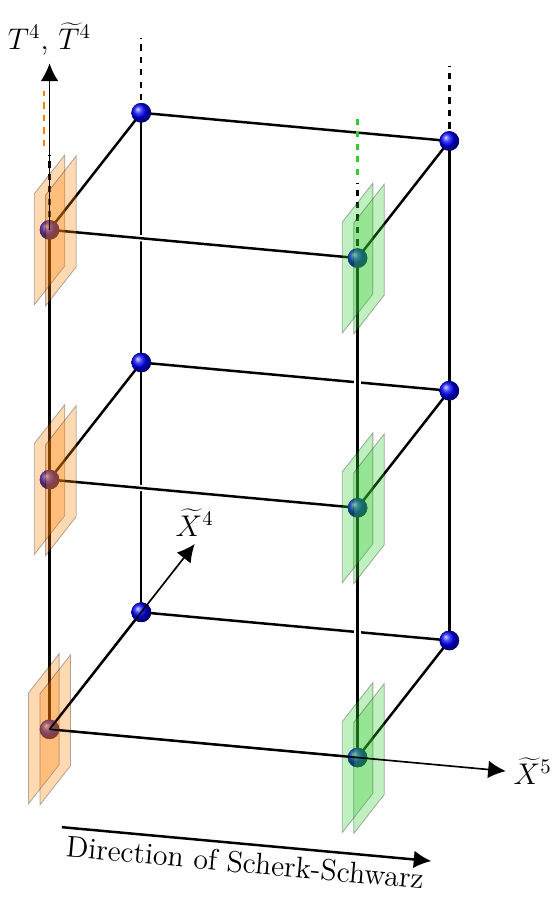}
\end{center}
\caption{\footnotesize Compared to Case~(a), all D3-branes T-dual to the D9-branes are displaced along  the Scherk--Schwarz direction~$\tilde X^5$.}
\end{subfigure}
\quad
\begin{subfigure}[t]{0.31\textwidth}
\begin{center}
\includegraphics [scale=0.55]{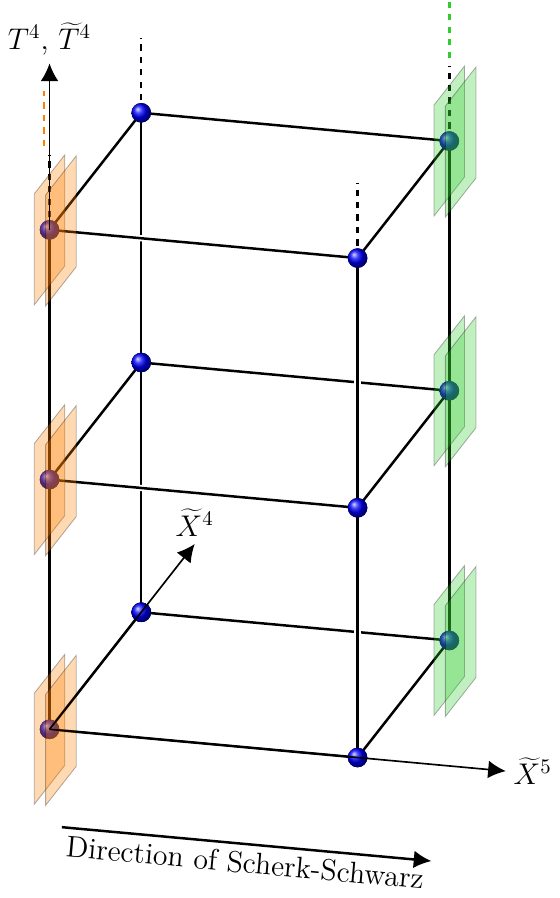}
\end{center}
\caption{\footnotesize Compared to either Case~(a) or~(b), all D3-branes T-dual to the D9-branes are displaced along~$\tilde X^4$.}
\end{subfigure}
\quad
\begin{subfigure}[t]{0.5\textwidth}
\begin{center}
\includegraphics [scale=0.55]{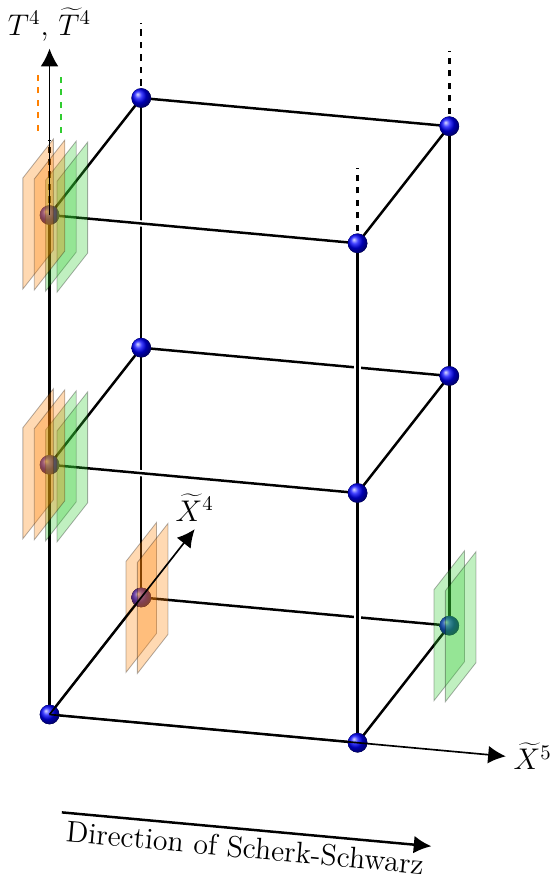}
\end{center}
\caption{\footnotesize Compared to Case~(a), one pair of D3-branes \mbox{T-dual} to D5-branes is displaced along $\tilde X^4$, while its initially coincident pair of D3-branes T-dual to D9-branes is moved along both $\tilde X^4$ and $\tilde X^5$. }
\label{0-d}
\end{subfigure}
\caption{\footnotesize D3-brane configurations in component $(\cR,\tilde \cR)=(16,16)$ of the WL moduli space.}
\label{0-abcd}
\end{center}
\end{figure}

\paragraph{\em Example 1:} The simplest setup amounts to having all D3-branes T-dual to the D5-branes at the same position $2\pi\vec a_{i_0'}$ of $\tilde T^2/I_{45}$, and similarly all D3-branes T-dual to the D9-branes at some common position $2\pi  \vec a_{j_0'}$.  Three cases (a), (b), (c) can be distinguished however, since all mass-term coefficients of the $\xi^{I'}_{r'}$ and $\epsilon^{I'}_{\tilde r'}$ read from Eq.~(\ref{N_final}) are $(1-0-1+{\delta\over 4} \,16)=4\delta$, where $\delta$ is defined as explained below Eq.~(\ref{vbm}). Fig.~\ref{0-abcd}
shows the three possibilities for distributing the pairs of branes. Therefore, we can conclude even before taking into account the Green--Schwarz mechanism that the  positions of all the D3-branes are stabilised in Case~(a), are unstable in Case~(b), and are massless in Case~(c). However, eliminating the $\epsilon_{\tilde r'}^{I'}$ thanks to the relations~(\ref{rel_GS}), it turns out that the mass terms of the remaining degrees of freedom $\xi_{r'}^{I'}$ are simply multiplied by a factor of~2. Moreover, $\nF-\nB=-224-1024\, \delta$, implying that $M$ has a tadpole and runs away.  In detail the behaviour of the configurations are as follows:

$\bullet$ In Case~(a), the potential is negative, and there are $16^2$ massless quaternionic scalars charged under $U(1)^{16}$   arising from the ND sector. Their masses must be determined to make a conclusion about the stability/instability  of the configuration, which we discuss in~\cite{wip}. Note however that in  component $(\cR,\tilde \cR)=(16,16)$ of the moduli space, Case~(a) yields the most negative value of $\nF-\nB$. Hence, we do not expect the moduli of the ND sector to be tachyonic at one loop, and expect the configuration to be stable, except for the supersymmetry breaking scale $M$, and for the remaining closed-string moduli $G_{\I\J}$, $C_{\I\J}$ and $\phi$ which  are flat directions. The possibility that the model leads to brane recombination \via condensation of the ND-sector moduli remains a possibility  that is discussed further in \cite{wip}. 

$\bullet$ In Case~(b), the potential is positive but the D3-brane positions are unstable, so the distribution will evolve  in $\tilde T^2/I_{45}$.    

$\bullet$ In Case~(c), the potential is negative  and the WL's are massless.  It turns out that (up to exponentially suppressed terms) the one-loop effective potential does not depend on these moduli, which  are therefore flat directions.\footnote{The one-loop potential dependencies on $U(1)$ WL's are identical to those of $SO(2)$ factors treated in Ref.~\cite{PreviousPaper}, which turn out to be  trivial.}  Hence,  the configuration is marginally stable. 


\paragraph{\em Example 2:} Thus far, conclusions about the stability/instability of the WL positions in $\tilde T^2/I_{45}$ could be drawn without taking into account the effect of the Green--Schwarz mechanism. In fact,  this is possible  only for particularly simple choices of brane setups, when all mass terms of the $\xi^{I'}_{r'}, \epsilon_{\tilde r'}^{I'}$ in Eq.~(\ref{N_final}) have the same sign.  To construct a more generic brane configuration, consider Case~(a) of Example~1, and move along $\tilde X^4$ one pair of D3-branes T-dual to D5-branes, and move along $\tilde X^4$ and $\tilde X^5$ its initially coincident pair of D3-branes T-dual to D9-branes. The new configuration, denoted~(d), is shown in Fig.~\ref{0-d}. The mass coefficients of fifteen $\xi^{I'}_{r'}$ and fifteen $\epsilon^{I'}_{\tilde r'}$ are ${15\over4}$, while those of the last two positions are $-{1\over 4}$. Hence, \apriori   the configuration seems unstable. However, eliminating in Eq.~(\ref{N_final}) all $\epsilon^{I'}_{\tilde r'}$'s by using Eq.~(\ref{rel_GS})  yields a new $16\times 16$ mass-squared matrix  for the $\xi_{r'}^{I'}$'s which has only positive eigenvalues. As a result, the brane configuration turns out to actually be stable, provided the $15^2$ quaternionic  moduli of the ND sector do not introduce instabilities, as already mentioned in Case~(a) of Example~1. In the present Case~(d), $\nF-\nB=-1120$ is higher than in Case~(a), but it remains negative.


\subsection{Full scan of the six components of the moduli space}

Among the configurations that have been presented so far, none of them is tachyon free with a positive or exponentially suppressed potential at one loop. In fact, setups with these properties are expected to be rare. For instance, in the case of a compactification on $T^6$ realising $\N=4\to \N=0$ breaking, this fact can be understood qualitatively by inspecting Eq.~(\ref{VexpD}), where the massless fermions contribute positively to the potential and negatively to the WL squared masses, and \emph{vice versa} for the massless bosons. Hence, the more positive the potential is, the more tachyonic instabilities are likely to arise.
For instance, for toroidal compactifications in dimension $d\ge 5$, it was shown in Refs~\cite{PreviousPaper,APP} that there exists only one  orientifold model\footnote{The assumptions are that ($i$) the Scherk--Schwarz mechanism is implemented along a single direction, ($ii$) there are no exotic orientifold planes, and ($iii$) there is no discrete background for the internal NS-NS antisymmetric tensor.} which is non-perturbatively consistent, tachyon-free at one loop and which has non-negative potential. It is defined  in five dimensions, has a trivial open-string gauge group\footnote{$SO(1)$ denotes the group containing only the neutral element.} $SO(1)^{32}$, and satisfies $\nF-\nB=8\times 8$.  

To determine if tachyon free brane configurations with zero or positive one-loop potentials exist in the  $\Z_2$-orbifold case, we have performed a computer scan of all brane configurations as follows:
\begin{itemize}
\item[($i$)] In each of the six non-perturbatively consistent components of the moduli space, we  loop over all distributions of mirror pairs (with respect to the orientifold action) of D3-branes in $T^4/\Z_2$ and $\tilde T^4/\Z_2$. 

\item[($ii$)] For each configuration, we derive the squared-mass matrix $\M^2$ of the 32 Cartan $U(1)$'s.

\item[($iii$)] We then loop over all possible distributions of the pairs along $\tilde T^2/I_{45}$. We restrict to the configurations that respect the condition~(\ref{condiwl}) for the positions in $ T^4/\Z_2$ and $\tilde T^4/\Z_2$ not to be tachyonic, and eliminate those for which $\nF-\nB<0$. 

\item[($iv$)] For each distribution satisfying the above constraints, we then compute the one-loop contributions to the mass terms of the brane positions in $\tilde T^2/I_{45}$ (see Eq.~(\ref{N_final})), and project out those combinations of moduli that become massive \via the Green--Schwarz mechanism. We obtain the true squared-mass matrix of the remaining dynamical positions in $\tilde T^2/I_{45}$ and eliminate all configurations for which this matrix admits at least one strictly negative eigenvalue. 

\end{itemize}
Among the hundreds of billions of initial possibilities,\footnote{When moving a stack of branes from one fixed point to another the massless spectrum is invariant, so we count only one of these configurations. However, since the spectra whose masses are of the order of the string scale will in general differ, our counting of the inequivalent configurations is actually greatly underestimated.} only eight emerge from the scan: six of them are  tachyon free,  and two are tachyon free up to possible instabilities that may arise from ND-sector moduli. Most interestingly,  two out of the six, and one out of the two configurations have vanishing one-loop potential $(\nF-\nB=0)$, up to exponentially suppressed terms. Let us summarise them: 


\paragraph{\em Exponentially suppressed one-loop potentials:}  \phantom{q}

\begin{figure}[t]
\captionsetup[subfigure]{position=t}
\begin{center}
\begin{subfigure}[t]{0.31\textwidth}
\begin{center}
\includegraphics [scale=0.52]{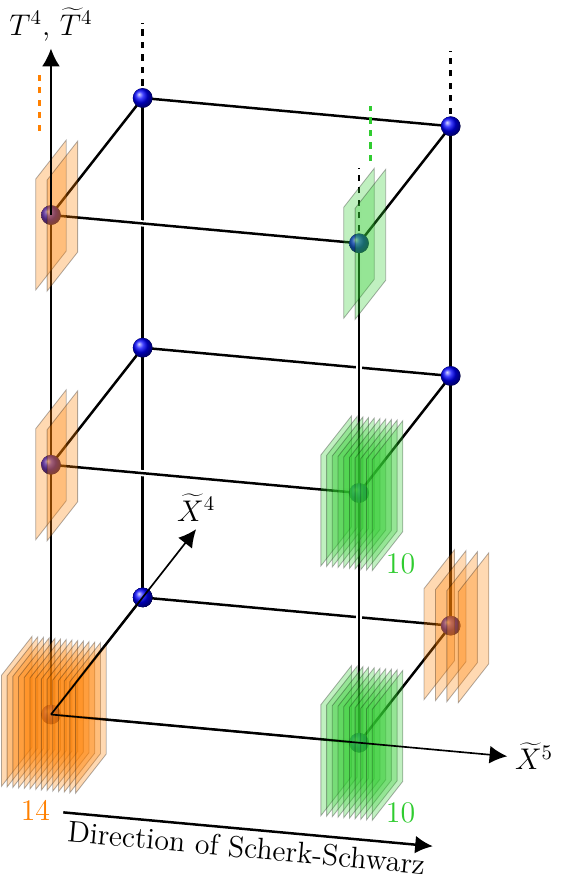}
\end{center}
\caption{\footnotesize Configuration tachyon free at one loop, with $\nF-\nB=0$.}
\label{nfnb0_a}
\end{subfigure}
\quad
\begin{subfigure}[t]{0.31\textwidth}
\begin{center}
\includegraphics [scale=0.52]{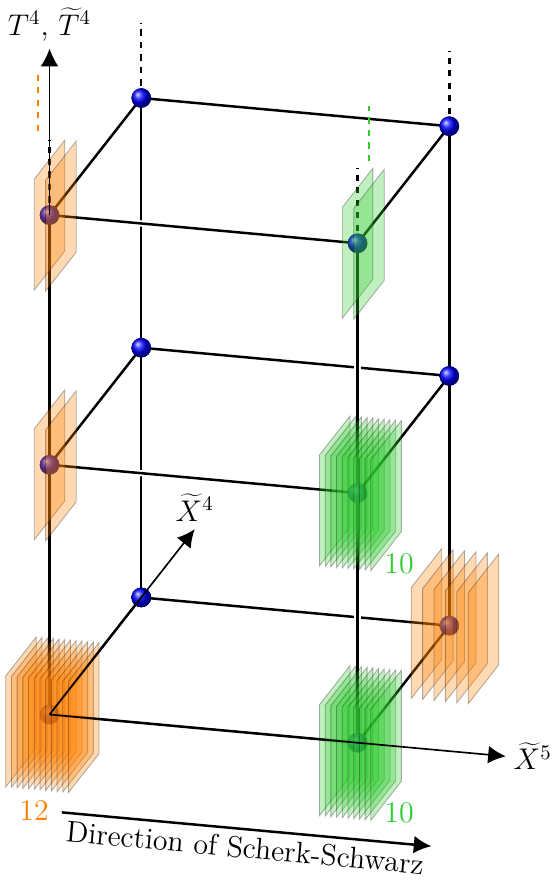}
\end{center}
\caption{\footnotesize Configuration tachyon free at one loop, with $\nF-\nB=0$.}
\label{nfnb0_b}
\end{subfigure}
\quad
\begin{subfigure}[t]{0.31\textwidth}
\begin{center}
\includegraphics [scale=0.52]{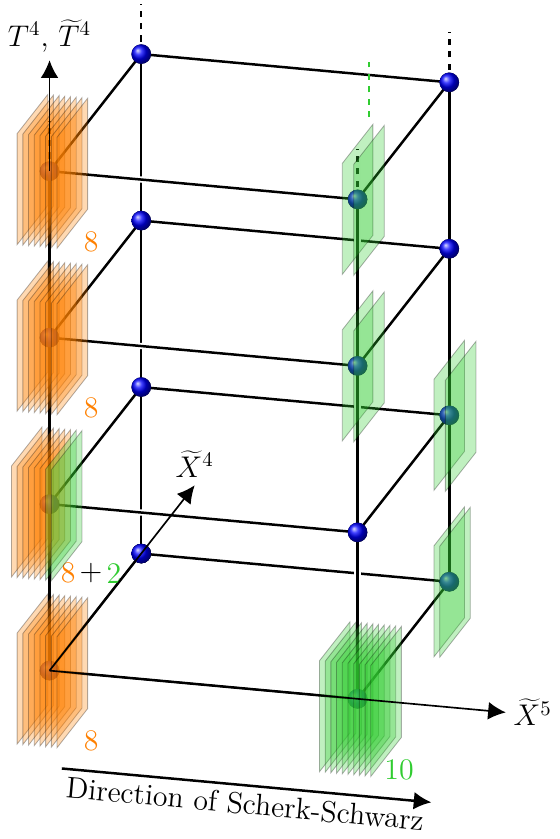}
\end{center}
\caption{\footnotesize Configuration  with $\nF-\nB=0$ but admitting moduli in the ND sector.}
\label{nfnb0_c}
\end{subfigure}
\caption{\footnotesize Two brane configurations (a) and (b) in component $(\cR,\tilde \cR)=(8,8)$  of the moduli space, and one configuration (c) in component $(\cR,\tilde \cR)=(0,8)$.}
\label{nfnb0}
\end{center}
\end{figure}

$\bullet$ In the component $(\cR,\tilde \cR)=(8,8)$  the scan finds  two configurations referred to as Case~(a) and~(b), which are free of tachyons and satisfy  $\nF-\nB=0$. The gauge groups generated by the D5-branes and D9-branes are, including anomalous $U(1)$'s,
\be
\begin{aligned}
&\mbox{Case (a)~:}\quad \left[U(1)^7\times U(2)\times U(7)\right]_{\text{DD}}\times \left[U(1)^6\times U(5)^2\right]_{\text{NN}}\, ,\\
&\mbox{Case (b)~:}\quad \left[U(1)^7\times U(3)\times U(6)\right]_{\text{DD}}\times \left[U(1)^6\times U(5)^2\right]_{\text{NN}}\, .
\end{aligned}
\ee
The D3-brane configurations are depicted in Figs~\ref{nfnb0_a} and \ref{nfnb0_b}, respectively. 
In the first case, the D3-branes T-dual to the D5-branes are distributed in $T^4/\Z_2$ as  7 pairs and one stack of 18 D3-branes, which is split in $\tilde T^2/I_{45}$ into $14+4$ branes.  The D3-branes T-dual to the D9-branes are distributed as 6 pairs and two stacks of 10. The  second configuration is identical to the previous one, up to the splitting of the 18 D3-branes now  into $12+6$.

In both cases, all dynamical brane positions in $T^4/\Z_2$ or $\tilde T^4/\Z_2$ are stabilised. They are associated with the stacks of $2n$ branes with $n\geq 2$, and their masses read from Eq.~(\ref{sc2}) are proportional to $n-1>0$. All other pairs of branes have rigid positions in $T^4/\Z_2$ or $\tilde T^4/\Z_2$ from the outset. Because there are initially 16 unitary gauge group factors in six dimensions,~
\be
\left[U(1)^7\times U(9)\right]_{\text{DD}}\times \left[U(1)^6\times U(5)^2\right]_{\text{NN}}\, ,
\ee
there are 16 anomalous $U(1)$'s becoming massive at the same time as the 16 blowing-up modes arising from the twisted closed-string sector, thanks to the Green--Schwarz mechanism. The true gauge symmetry in six dimensions is then $[U(9)/U(1)_{\rm diag9}]_{\rm DD}\times [SU(5)^2]_{\rm NN}$, where $U(1)_{\rm diag9}$ denotes the anomalous $U(1)$ associated with the $U(9)$ factor. However, in four dimensions, the WL's on $T^2$ break $U(9)\to U(2)\times U(7)$ in Case (a), which can be written as $U(1)_{\rm diag9}\times U(1)_{\bot 9}\times SU(2)\times SU(7)$, where $U(1)_{\bot 9}$ denotes the $U(1)$ orthogonal to $U(1)_{\rm diag9}$. Modding by $U(1)_{\rm diag9}$, we obtain
\be
\begin{aligned}
&\mbox{Case (a)~:}\quad \big[U(1)_{\bot 9} \times SU(2)\times SU(7)\big]_{\rm DD}\times \big[SU(5)^2\big]_{\rm NN}\, ,\\
&\mbox{Case (b)~:}\quad  \big[U(1)_{\bot 9}\times SU(3)\times SU(6)\big]_{\rm DD}\times \big[SU(5)^2\big]_{\rm NN}\,, 
\label{gsab}
\end{aligned}
\ee
where we use a similar reasoning for Case~(b).
Along $\tilde T^2/I_{45}$, all D3-brane positions are also stabilised, after freezing the super heavy combinations associated with the anomalous $U(1)$'s.   
The ND sector does not contain moduli fields since condition~(\ref{no_ND}) is satisfied. Thus, in Cases~(a) and~(b) and at the one-loop level,  no tachyons are present and the potential admits flat directions parameterised by the internal metric (including $G^{55}$ \ie $M$, as justified in the next paragraph), the dilaton,  and the RR two-form moduli. Notice that these configurations exist in four dimensions but not in five.

The massless spectrum of these two models contains the bosonic parts of $\N=2$ vector multiplets transforming under the adjoint representations of the groups given in Eq.~(\ref{gsab}), along with the scalars of $\N=2$ hypermultiplets in antisymmetric $\oplus$ $\overline{\text{antisymmetric}}$ representations of each non-Abelian factors.  In terms of degrees of freedom, this yields $\nB^{\text{open}}=800$ in Case~(a), and $\nB^{\text{open}}=736$ in Case~(b). To this, one must add the $96$ degrees of freedom coming from the closed-string sector yielding $\nB=896$ in Case~(a), and $\nB=832$ in Case~(b). Finally, the massless spectrum contains the fermionic degrees of freedom of hypermultiplets in the ND sector. They transform under bifundamental representations of all pairs of gauge group factors supported by stacks of D3-branes (T-dual to D5-branes) and stacks of D3-branes (T-dual to D9-branes) facing each other along the T-dual Scherk--Schwarz direction $\tilde X^5$. This leads to  $\nF=4\times 14\times 16=896$ in Case~(a), and $\nF=4\times 13\times 16=832$ in Case~(b), which exactly equals the numbers of bosonic degrees of freedom.

$\bullet$ The scan also selects a third configuration with $\nF-\nB=0$,  in component $(\cR,\tilde \cR)=(0,8)$ of the moduli space, which we will refer to as Case~(c). The gauge symmetry is, including anomalous $U(1)$'s,   
\begin{equation}
\mbox{Case (c)~:}\quad \left[U(4)^4\right]_{\text{DD}} \times \left[U(1)^{11}\times U(5)\right]_{\text{NN}}\,,
\label{u16c}
\end{equation}
and the configuration of branes is shown in Fig.~\ref{nfnb0_c}. The D3-branes T-dual to the D5-branes are distributed in $T^4/\Z_2$ as 4 stacks of 8. The D3-branes T-dual to the D9-branes are distributed as 8 pairs (with rigid positions in $\tilde T^4/\Z_2$),  one stack of 4 split in $\tilde T^2/I_{45}$ into $2+2$, and one stack of 12 split in $\tilde T^2/I_{45}$ into $10+2$.

All positions along $T^4/\Z_2$ and $\tilde T^4/\Z_2$ are rigid or massive. Because there are 14 unitary factors in six dimensions, 
\be
 \left[U(4)^4\right]_{\text{DD}} \times \left[U(1)^{8}\times U(2)\times U(6)\right]_{\text{NN}}\,,
\ee
there are 14 anomalous $U(1)$'s that are actually massive, along with 14 twisted moduli in the closed-string sector. Hence, the mass acquired at one loop by the remaining 2 blowing up modes of $T^4/\Z_2$ should be computed in order to conclude whether the internal space is stabilized at the orbifold  point or not. The true gauge symmetry in six dimensions is $[SU(4)^4]_{\text{DD}} \times [U(2)/U(1)_{\rm diag2}\times U(6)/U(1)_{\diag6}]_{\text{NN}}$. Using notations similar to those introduced in Cases~(a) and~(b), the WL's on $T^2$ break $U(2)\to U(1)_{\rm diag2}\times U(1)_{\rm\bot 2}$ and $U(6)\to U(1)_{\rm diag6}\times U(1)_{\rm\bot 6}\times SU(5)$. Modding by $U(1)_{\rm diag2}$ and $U(1)_{\rm diag6}$,  we obtain the true gauge group in four dimensions
\begin{equation}
\mbox{Case (c)~:}\quad  \left[SU(4)^4\right]_{\text{DD}} \times \big[U(1)_{\rm\bot 2}\times U(1)_{\rm\bot 6}\times SU(5)\big]_{\text{NN}}~.
\end{equation}

Taking into account the Green--Schwarz mechanism, the remaining positions along $\tilde T^2/I_{45}$ are all massless at one loop, except one which is massive. The internal metric and RR two-form, as well as the dilaton are flat directions of the one-loop potential (up to exponentially suppressed terms). However, we cannot determine if this configuration is fully marginally stable without also considering the masses of the ND sector moduli which are also present in this case: this is left for future work.

The massless bosonic degrees of freedom include those of an $\N=2$ vector multiplet in the adjoint representation of  the group~(\ref{u16c}), along with the scalars of $\N=2$ hypermultiplets in antisymmetric $\oplus$ $\overline{\text{antisymmetric}}$ representations for each non-Abelian factor. There are also bosonic degrees of freedom transforming under four  bifundamental representations of $U(4)_{\text{DD}}\times U(1)_{\text{NN}}$.  Taking into account the closed-string degrees of freedom, we obtain \mbox{$\nB=832$}. The massless fermionic degrees of freedom are in the bifundamental representations of all pairs of gauge group factors supported by stacks of D3-branes (T-dual to D5-branes) and stacks of D3- branes (T-dual to D9-branes) facing each other along the T-dual Scherk--Schwarz direction $\tilde X^5$. Their number is given by $\nF=4\times 16\times 13=832$, again equating to $\nB$.


\paragraph{\em Positive potentials:}  There also exist five configurations with strictly positive potentials. They all lie in component  $(\cR,\tilde \cR)=(8,8)$ and have an identical open-string (anomalous) gauge group 
\be
\left[U(1)^6\times U(5)^2\right]_{\text{DD}}\times \left[U(1)^6\times U(5)^2\right]_{\text{NN}}~.
\ee 
\begin{figure}[H]
\vspace{-0.3cm}
\captionsetup[subfigure]{position=t}
\begin{center}
\begin{subfigure}[h!]{0.31\textwidth}
\begin{center}
\includegraphics [scale=0.52]{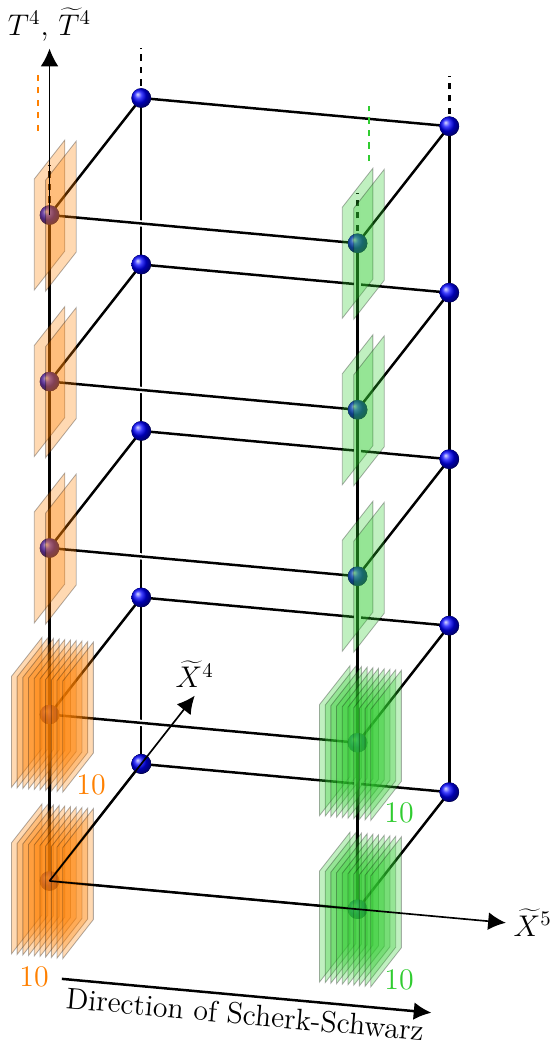}
\end{center}
\caption{\footnotesize Configuration tachyon free at one loop, with $\nF-\nB=40$.}
\label{nfnbpos_a}
\end{subfigure}
\quad
\begin{subfigure}[h!]{0.31\textwidth}
\begin{center}
\includegraphics [scale=0.52]{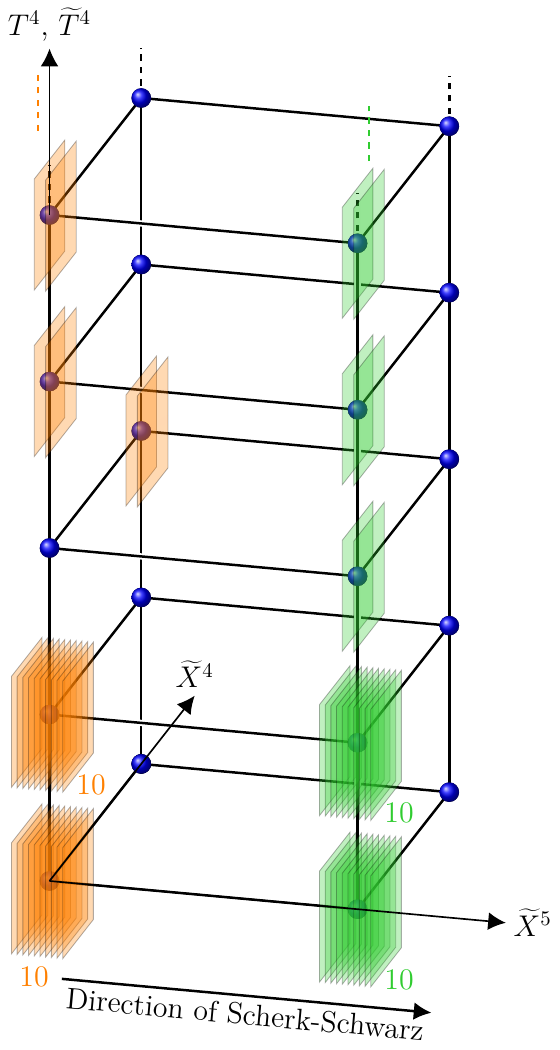}
\end{center}
\caption{\footnotesize Configuration tachyon free at one loop, with $\nF-\nB=24$.}
\label{nfnbpos_b}
\end{subfigure}
\quad
\begin{subfigure}[h!]{0.3\textwidth}
\begin{center}
\includegraphics [scale=0.52]{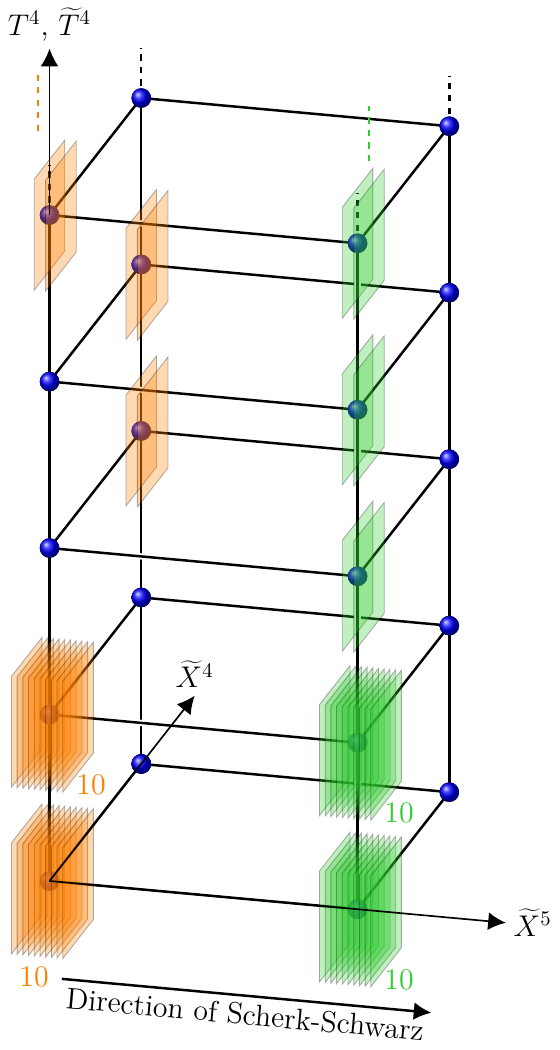}
\end{center}
\caption{\footnotesize Configuration tachyon free at one loop, with $\nF-\nB=8$.}
\label{nfnbpos_c}
\end{subfigure}
\quad
\begin{subfigure}[t]{0.31\textwidth}
\begin{center}
\includegraphics [scale=0.52]{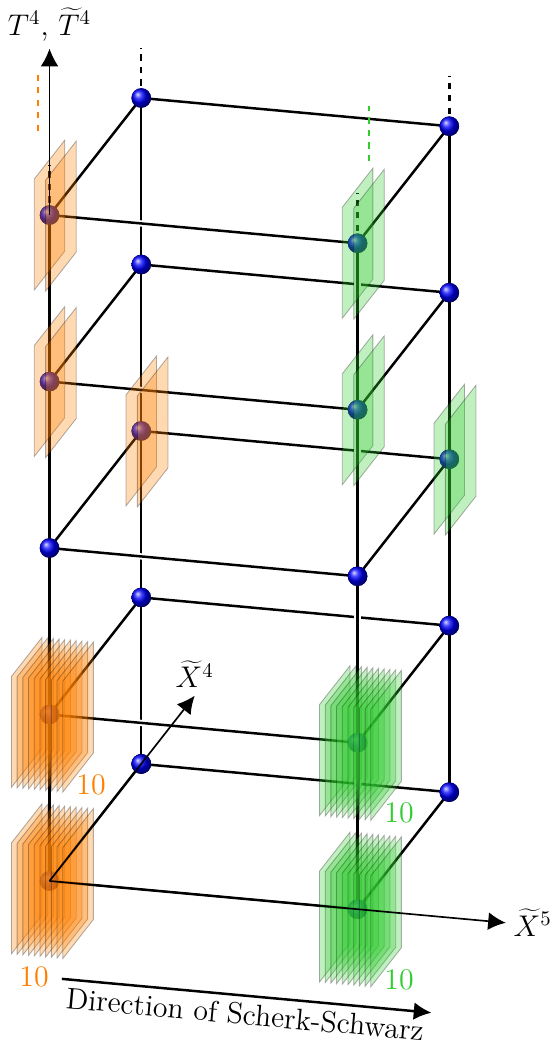}
\end{center}
\caption{\footnotesize Brane configuration tachyon free with \mbox{$\nF-\nB=10$}.}
\label{nfnbpos_d}
\end{subfigure}
\quad
\begin{subfigure}[t]{0.31\textwidth}
\begin{center}
\includegraphics [scale=0.52]{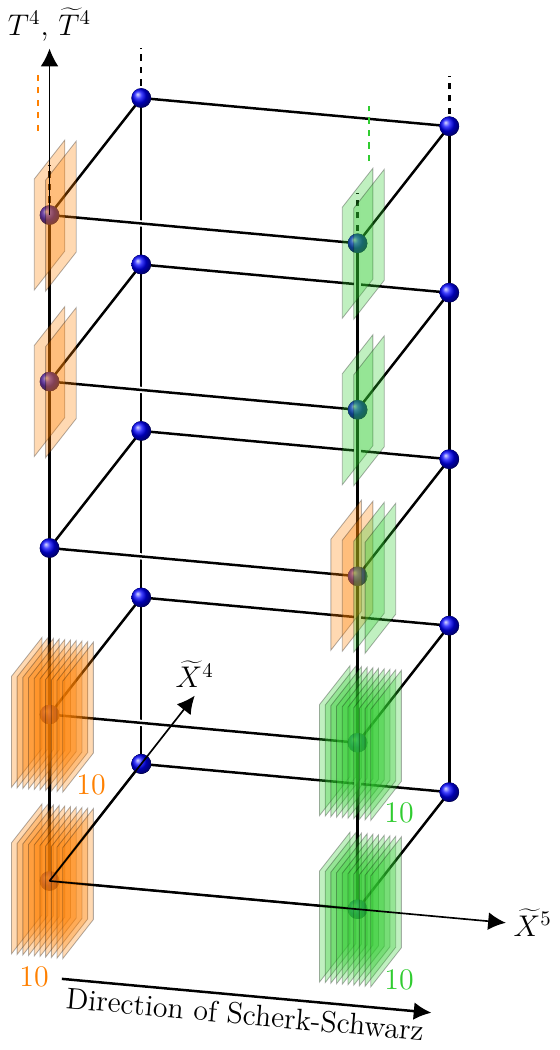}
\end{center}
\caption{\footnotesize Brane configuration with \mbox{$\nF-\nB=8$} and moduli in the ND sector.}
\label{nfnbpos_e}
\end{subfigure}
\caption{\footnotesize Brane configurations in component $(\cR,\tilde \cR)=(8,8)$  of the moduli space.}
\label{nfnbpos}
\end{center}
\end{figure}
\noindent The configurations are depicted in Figs~\ref{nfnbpos_a}-\ref{nfnbpos_e}. 
All position moduli along $T^4/\Z_2$ and $\tilde T^4/\Z_2$  are massive. Taking into account the Green--Schwarz mechanism, the gauge symmetry in six and four dimensions is  reduced to 
\be
\left[SU(5)^2\right]_{\rm DD}\times \left[SU(5)^2\right]_{\rm NN}\, ,
\ee
all twisted closed-string moduli are massive, and the positions along $\tilde T^2/I_{45}$ are either massive or massless, depending on the case at hand. 

The configuration in Fig.~\ref{nfnbpos_a} yields $\nF-\nB=40$. Notice that it may be considered in five dimensions by decompactifying the direction $X^4$.  In the case shown in Fig.~\ref{nfnbpos_b}, the direction $\tilde X^4$ is used to isolate one pair of D3-branes, which leads to  $\nF-\nB=24$. By  displacing a second pair of the same kind as shown in Fig.~\ref{nfnbpos_c}, one obtains $\nF-\nB=8$.  Starting from the distribution in Fig.~\ref{nfnbpos_c} and displacing a pair of D3-branes of the other kind as shown in Fig.~\ref{nfnbpos_d}, one obtains $\nF-\nB=10$. Finally, the configuration in Fig.~\ref{nfnbpos_e} yields $\nF-\nB=8$, but contains moduli fields in the ND sector whose masses need to be analysed at one loop in order to make a conclusion about its stability/instability.


\vspace{0.6cm}
\section{Conclusions}
\label{conclusion}

\noindent In this work, we have studied from various perspectives the generation at the quantum level of moduli masses in type~I string theory compactified on $T^2\times T^4/\Z_2$, when \mbox{$\N=2\to \N=0$} supersymmetry is spontaneously broken by the Scherk--Schwarz mechanism implemented along $T^2$. Our analysis is perturbative, restricted to the one-loop level, and our conclusions are valid when the supersymmetry breaking scale $M$ is the lowest mass scale of the background. We have considered  gauge-field backgrounds on the worldvolumes of the 32 D9-branes and 32 D5-branes, as well as positions of the D5-branes in $T^4/\Z_2$, that can be described from T-dual points of view as positions of 32+32 D3-branes distributed  on 64 O3-planes.  At such points in moduli space, the effective potential is extremal, except with respect to $M$  which runs away when $\nF\neq \nB$. 

We find that the D3-brane positions/moduli that are not already  heavy  thanks to a generalized Green--Schwarz mechanism in six dimensions can be stabilised at one loop. However, up to exponentially suppressed corrections, all degrees of freedom of the internal metric $G_{\I\J}$ (except  $M$ when $\nF\neq \nB$), of the two-form $C_{\I\J}$  and of  the dilaton remain flat directions. From heterotic/type~I duality, it is however possible to infer that some of the moduli $(G+C)_{\I\J}$ can be stabilised non-perturbatively at points where D1-branes  become massless~\cite{CoudarchetPartouche,APP}. When  moduli occur in the ND sector of the (non T-dualized) theory, their quantum masses can be derived by computing two-point functions. This will be presented elsewhere~\cite{wip}. Finally, the models contain blowing-up modes, which belong to quaternionic scalars arising in the twisted closed-string sector. While those involved in the Green--Schwarz mechanism are very heavy, we have not studied the masses generated for the remaining ones (if any). 

Among the plethora of allowed distributions of D3-branes on O3-planes, only two are tachyon free at one loop, with an  exponentially suppressed effective potential, \ie with \mbox{$\nF=\nB$}. Recall that such set-ups may be interesting candidates for generating, after stabilisation of $M$ and the dilaton, a cosmological constant which is orders of magnitude smaller than in generic models. Four more brane configurations lead to positive potentials, \ie $\nF>\nB$, where the only instabilities are associated with the run away of the supersymmetry-breaking no-scale modulus $M$. Finally, two brane distributions with similar properties contain moduli in the ND sector, whose one-loop masses remain to be analysed. It is worth mentioning that in a phenomenological setup, these moduli would naturally contain the Standard-Model Higgs field, so it is not \apriori obvious that one needs to banish tachyonic masses from these states entirely.  All of the above models are interesting in the sense that they describe non-Abelian gauge theories, with fermions that are massless at tree level  transforming in bifundamental representations. It would be interesting to derive the masses acquired at one loop by this fermionic matter.

To explore further possibilities, it would also be  interesting to relax some of the assumptions we have made. For instance, one may seek  type~I  vacua that include ``exotic'' orientifold planes, often referred to as O$_+$-planes, which can support  even or odd numbers of branes~\cite{triple}. O$_+$-planes have RR charges and tensions opposite to those of the O$_-$-planes we have used in the present work. Alternatively, when moduli in the ND sector are tachyonic and condense, branes recombine and the theory admits new vacua. Another possibility is to switch on discrete backgrounds for the internal components of the NSNS antisymmetric tensor (whose degrees of freedom are projected out by the orientifold action).
 Finally, one may analyze the theory when $T^4/\Z_2$ is deformed to a smooth K3 manifold. 


\vspace{0.6cm}
\section*{Acknowledgements} 

\noindent The authors would like to thank Carlo Angelantonj, Daniel Lewis and especially Emilian Dudas  for discussion and useful input during the realisation of this work. This work was funded by the Royal-Society/CNRS International Cost Share grant IE160590, and mutual hospitality from Durham University and the \'Ecole Polytechnique is acknowledged.  
 

\vspace{0.6cm}
\begin{appendices}
\makeatletter
\DeclareRobustCommand{\@seccntformat}[1]{%
  \def\temp@@a{#1}%
  \def\temp@@b{section}%
  \ifx\temp@@a\temp@@b
  \appendixname\ \thesection:\quad%
  \else
  \csname the#1\endcsname\quad%
  \fi
} 
\makeatother

\section{\bm One-loop effective potential}
\label{A0}
\renewcommand{\theequation}{A.\arabic{equation}}

\noindent  In this appendix, our goal is to present in some details the expression of the one-loop effective potential arising in a  four-dimensional orientifold model of type~IIB that realizes the $\N=2\to \N=0$ spontaneous breaking of supersymmetry. The background is
\begin{equation}
\label{back1}
\mathbb{R}^{1,3}\times T^{2}\times {T^4\over \Z_2}\, ,
\end{equation}
where a Scherk--Schwarz mechanism is implemented along one of the internal $T^2$ directions. 

In an orientifold theory (see Refs~\cite{orientifolds1,PradisiSagnotti,orientifolds3,orientifolds4,orientifolds5,orientifolds7,orientifolds8} for original papers and Refs~\cite{review-1,review-2,review-3} for reviews), 
the one-loop effective potential may be divided into the contributions arising from the torus, Klein bottle, annulus and M\"obius strip partition functions, 
\begin{align}
&\V=-{\Ms^4\over 2(2\pi)^4}\, (\T+\K+\A+\M)~,&&\where\espD\nonumber  \\
& \T=\half\int_\F{\dd\tau_1 \dd\tau_2\over \tau_2^{3}}\, \Str  \frac{1+g}{2} q^{L_0-\half}\bar q^{\tilde L_0-\half}~,  &&\K=\half\int_0^{+\infty}{\dd\tau_2\over \tau_2^{3}}\, \Str \Omega\frac{1+g}{2} q^{L_0-\half}\bar q^{\tilde L_0-\half}~,\label{Vdef}\\
&\A=\half\int_0^{+\infty}{\dd\tau_2\over \tau_2^{3}}\, \Str \frac{1+g}{2}q^{\half(L_0-1)}~,&&\!\!\M=\half\int_0^{+\infty}{\dd\tau_2\over \tau_2^{3}}\, \Str  \Omega\frac{1+g}{2} q^{\half(L_0-1)}~.\nonumber 
\end{align}
In the above formula, $\tau_1, \tau_2$ are the real and imaginary parts of the Teichm\"uller parameter~$\tau$, $q=e^{2i\pi\tau}$, $\F$ is the fundamental domain of $SL(2,\Z)$, $L_0,\tilde L_0$ are the zero frequency Virasoro operators, $\Omega$ is the orientifold generator and $g$ is the twist generator acting on the $T^{4}$ coordinates as $(X^{6},X^{7},X^{8},X^{9})\rightarrow(-X^{6},-X^{7},-X^{8},-X^{9})$. The factors $\half$ are due to the orientifold projection. In the following, we  first  introduce our notations and present the  amplitudes in the supersymmetric BSGP model compactified down to four dimensions. Then, we  implement discrete deformations as well as the spontaneous breaking of $\N=2$ supersymmetry, and display the associated amplitudes. 


\subsection{Summary of conventions and notations}
\label{conventions}

It is useful for reference to summarise the notation for the lattices of zero modes and for the characters that account for the oscillator excitations, that we use to write the one-loop amplitudes:


\paragraph{\em Indices:} The metric of $T^2\times T^4$ is defined as $G_{\I\J}$, $\I,\J=4,\dots 9$. However, due to the factorization of the internal space, it is convenient to introduce non-calligraphic indices that refer either to the $T^{2}$ or $T^4$ directions only. Hence, we will also use $G_{I'J'}$, $I',J'=4,5$ and $G_{IJ}$, $I,J=6,\dots,9$.

Internal momentum and winding numbers along $T^2\times T^4$ are organized in  six-vectors, $\vec{M}$ and $\vec{N}$, respectively. They can be split according to the tori factorization in the following way: $\vec{M}=(\vec m',\vec{m})$ and $\vec{N}=(\vec n',\vec{n})$, where primed vectors components are two-vectors and the not primed ones  are  four-vectors. 

\paragraph{\em Lattices: } 

For the genus-1 Riemann surface, the expression of the amplitude $\T$ involves
\be
\begin{aligned}
&\Lambda_{\vec M, \vec N}^{(6,6)}(\tau)=q^{{1\over 4}P^L_\I G^{\I\J}P^L_\J}\, \bar q^{{1\over 4}P^R_\I G^{\I\J}P^R_\J}~, \\
&P^L_\I=m_\I+G_{\I\J}n_\J ~,\quad P^R_\I\,\,=\,\,m_\I-G_{\I\J}n_\J~, \quad  \I=4,\dots, 9 ~,\esps
\end{aligned}
\ee
where $G^{\I\J}=G^{-1}_{\I\J}$. Due to the orientifold projection, the NS-NS antisymmetric tensor $B_{\I\J}$ present in the type~IIB string vanishes. The $(6,6)$ lattice can again be divided into $(2,2)$ and $(4,4)$ lattices of zero modes associated with $T^{2}$ and $T^{4}$, as follows:
\begin{align}
\begin{split}
\Lambda_{\vec{M},\vec{N}}^{(6,6)}(\tau)=\Lambda_{\vec{m}',\vec{n}'}^{(2,2)}(\tau)\Lambda_{\vec{m},\vec{n}}^{(4,4)}(\tau)=q^{{1\over 4}P^L_{I'} G^{I'J'}P^L_{J'}}\, \bar q^{{1\over 4}P^R_{I'} G^{I'J'}P^R_{J'}}\times q^{{1\over 4}P^L_I G^{IJ}P^L_J}\, \bar q^{{1\over 4}P^R_I G^{IJ}P^R_J}.
\end{split}
\end{align}

By contrast, the states that are running in the Klein bottle, annulus or M\"obius strip amplitudes have  a vanishing momentum or winding number along each internal direction, so the relevant lattices can be defined as 
\begin{align}
\begin{split}
P_{\vec M}^{(6)}(i\tau_2)&=\Lambda^{(6,6)}_{\vec M,\vec 0}(\tau)=e^{-\pi\tau_2m_\I G^{\I\J}m_\J}~,\\
W_{\vec n}^{(4)}(i\tau_2)&=\Lambda^{(4,4)}_{\vec 0,\vec n}(\tau)=e^{-\pi \tau_{2} n_I G_{IJ}n_J}~.\\
\end{split}
\label{pmlat}
\end{align}
As before, the momentum lattice can be factorized as 
\begin{equation}
P_{\vec M}^{(6)}(i\tau_2)=P_{\vec{m}'}^{(2)}(i\tau_{2})P_{\vec{m}}^{(4)}(i\tau_{2})=e^{-\pi\tau_2m_{I'}G^{I'J'}m_{J'}}\times e^{-\pi\tau_2m_{I}G^{IJ}m_{J}}~.
\end{equation}
Throughout this work, the implicit arguments of the lattices are as indicated in the above definitions.


\paragraph{\em Characters:} Our definitions of the Jacobi modular forms and Dedekind function are
\be
\label{th}
\vartheta\big[{}^\alpha_\beta\big](z| \tau)=\sum_m q^{{1\over 2}(m+\alpha)^2}e^{2i\pi(z+\beta)(m+\alpha)} ~, \;\;\quad \eta(\tau)= q^{1\over 24}\prod_{n=1}^{+\infty}(1-q^n)~.
\ee
It is standard to denote 
\be
\vartheta\big[{}^0_0\big](z| \tau)=\vartheta_3(z| \tau)~, ~~\vartheta\Big[{}^{{}^{\scriptstyle \phantom{.}\!0}}_{1\over 2}\Big](z| \tau)=\vartheta_4(z| \tau)~,~~ \vartheta\Big[{}^{1\over 2}_{\phantom{\!|}_{\scriptstyle0}}\Big](z| \tau)=\vartheta_2(z| \tau)~,~~ \vartheta\Big[{}^{1\over 2}_{1\over 2}\Big](z| \tau)=\vartheta_1(z| \tau)~,
\ee 
and to keep implicit both arguments  when $z=0$. In these notations, 
 the $SO(2n)$ affine characters can be written as
\begin{equation}
O_{2n}=\frac{\vartheta_{3}^{n}+\vartheta_{4}^{n}}{2\eta^{n}}~,\quad \;\;V_{2n}=\frac{\vartheta_{3}^{n}-\vartheta_{4}^{n}}{2\eta^{n}}~,\quad\;\; S_{2n}=\frac{\vartheta_{2}^{n}+i^{-n}\vartheta_{1}^{n}}{2\eta^{n}}~,\quad\;\; C_{2n}=\frac{\vartheta_{2}^{n}-i^{-n}\vartheta_{1}^{n}}{2\eta^{n}}~.\label{eq:CharacterDef}
\end{equation}
They satisfy the following modular properties:
\be
\begin{aligned}
&\begin{pmatrix}O_{2n}\\
V_{2n}\\
S_{2n}\\
C_{2n}
\end{pmatrix}\!\!(\tau+1)=e^{-in\pi/12}\diag\!\left(1,-1,e^{in\pi/4},e^{in\pi/4}\right)\begin{pmatrix}O_{2n}\\
V_{2n}\\
S_{2n}\\
C_{2n}
\end{pmatrix}\!\!\left(\tau\right)~,
\\
&\begin{pmatrix}O_{2n}\\
V_{2n}\\
S_{2n}\\
C_{2n}
\end{pmatrix}\!\!\Big(\!\!-\frac{1}{\tau}\Big)=\frac{1}{2}\begin{pmatrix}1 & 1 & 1 & 1\\
1 & 1 & -1 & -1\\
1 & -1 & i^{-n} & -i^{-n}\\
1 & -1 & -i^{-n} & i^{-n}
\end{pmatrix}\begin{pmatrix}O_{2n}\\
V_{2n}\\
S_{2n}\\
C_{2n}
\end{pmatrix}\!\!(\tau)~,
\end{aligned}
\label{eq:CharacterTransform}
\ee
which are relevant for the amplitudes $\T$, $\K$ and $\A$. 
For the M\"obius strip, it is convenient to switch from the characters $\chi$ to the real ``hatted'' characters $\hat \chi$ defined by \cite{review-1,review-2} 
\begin{equation}
\hat{\chi}\Big(\frac{1}{2}+i\tau_{2}\Big)=e^{-i\pi(h-{c\over 24})}\, \chi\Big(\frac{1}{2}+i\tau_{2}\Big)~,
\end{equation}
where $h$ is the weight of the associated primary state and
$c$ is the central charge. The so-called P-transformation then takes the form
\begin{equation}
\begin{pmatrix}\hat O_{2n}\\
\hat V_{2n}\\
\hat S_{2n}\\
\hat C_{2n}
\end{pmatrix}\!\!\Big(\frac{1}{2}+{i\over 2\tau_{2}}\Big)=\begin{pmatrix}c & s & 0 & 0\\
s & -c & 0 & 0\\
0 & 0 & \zeta c & i\zeta s\\
0 & 0 & i\zeta s & \zeta c
\end{pmatrix}\begin{pmatrix}\hat O_{2n}\\
\hat V_{2n}\\
\hat  S_{2n}\\
C_{2n}
\end{pmatrix}\!\!\Big(\frac{1}{2}+i\frac{\tau_{2}}{2}\Big)~, \quad \hat \eta\Big(\frac{1}{2}+{i\over 2\tau_{2}}\Big)=\sqrt{\tau_{2}}\, \hat \eta\Big(\frac{1}{2}+i\frac{\tau_{2}}{2}\Big)~,
\label{eq:hat chi}
\end{equation}
where $c=\cos(n\pi/4)$, $s=\sin(n\pi/4)$ and $\zeta=e^{-in\pi/4}$.
Throughout this work, the implicit arguments of the characters are $\tau$, $2i\tau_2$, $i\tau_2/2$ and $(1+i\tau_2)/2$ for the torus, Klein bottle, annulus and M\"obius strip amplitudes respectively. 


\subsection{Bianchi--Sagnotti--Gimon--Polchinski model}
\label{GPPS}

Let us first consider  the amplitudes arising in the simplest version of the BSGP model~\cite{BianchiSagnotti,GimonPolchinski,GimonPolchinski2} compactified on $T^2$. The background is as given in Eq.~(\ref{back1}), with at this stage no Wilson lines switched on in the worldvolumes of the D9- and D5-branes, all D5-branes coincident on a single O5-plane, and as yet no implementation of the Scherk--Schwarz mechanism. Of course, in the absence of any breaking of supersymmetry, ultimately the total effective potential vanishes. 

To write the one-loop vacuum amplitudes, we decompose the worldsheet fermion $SO(8)$ affine characters into characters of $SO(4)\times SO(4)$, where the first factor is the little group in six dimensions and the second  is associated with the internal directions $6,7,8,9$:
\begin{equation}
\begin{aligned}
&O_{8}=O_{4}O_{4}+V_{4}V_{4}~,\hspace{2cm}&&V_{8}=V_{4}O_{4}+O_{4}V_{4}~,\\
&S_{8}=S_{4}S_{4}+C_{4}C_{4}~,&&C_{8}=S_{4}C_{4}+C_{4}S_{4}~.
\end{aligned}
\end{equation}
It is convenient to define characters that mix NS and R sectors  but which diagonalize  the action of the $\Z_2$ orbifold generator $g$. The transformations of the 
$T_4/{\mathbb Z}_2$ characters  under~$g$~is 
\be
g~\cdot 
\begin{aligned}
\begin{pmatrix}
O_{4}\\
V_{4}\\
S_{4}\\
C_{4}
\end{pmatrix}
=
\begin{pmatrix}
O_{4}\\
-V_{4}\\
-S_{4}\\
C_{4}
\end{pmatrix}
\end{aligned}~,
\ee
so that defining
\begin{equation}
\begin{aligned}
&Q_{\text{O}}=V_{4}O_{4}-C_{4}C_{4}~,\hspace{2cm}&&Q_{\text{V}}=O_{4}V_{4}-S_{4}S_{4}~,\\
&Q_{\text{S}}=O_{4}C_{4}-S_{4}O_{4}~,&&Q_{\text{C}}=V_{4}S_{4}-C_{4}V_{4}~,
\end{aligned}
\end{equation}
the states belonging to the characters $Q_{\text{O}}, Q_{\text{S}}$ on the one hand, and  $Q_{\text{V}}, Q_{\text{C}}$ on the other,  have  $\Z_2$  eigenvalues $+1$ and $-1$ respectively.

With these definitions and the conventions of Appendix~\ref{conventions}, the torus and Klein bottle amplitudes read 
\begin{align}
&\T=\frac{1}{4}\int_{\F}\frac{\dd^{2}\tau}{\tau_{2}^{3}}\,\bigg\{\opv\lattice+\omv\tetad+16\spc\tetaq\nonumber\\
&\hspace{3.5cm}+16\smc\tetat\bigg\}\sum_{\vec{m}',\vec{n}'}\LAMBDA(\vec{m}',\vec{n}')~,\espDD\label{tknodef}\\
&\K=\frac{1}{4}\int_{0}^{+\infty}\frac{\dd\tau_{2}}{\tau_{2}^{3}}\, \bigg\{\opvo\latticek+32\spco\tetaqo\bigg\}\sum_{\vec{m}'}\frac{P^{(2)}_{\vec{m}'}}{\eta^{4}}~.\nonumber
\end{align} 
In the torus expression, the first term in the braces is the usual $|V_{8}-S_{8}|^2$ contribution occurring in  type~IIB. The second term is obtained by acting with the orbifold generator $g$, which imposes to be at the origin of the $T^{4}$ lattice. The last two terms correspond to the twisted sector and are also at the origin of the $T^{4}$ lattice.

The model contains D9-branes and D5-branes in order to cancel the RR charges of an O9-plane and 32 O5-planes that are respectively the fixed point loci of $\Omega$ and $\Omega g$. Denoting by $N$ and $D$ the numbers of D9-branes and D5-branes, and by $R_{N}$ and $R_{D}$ their counterparts under the action of $g$ on the associated Chan--Paton charges\cite{PradisiSagnotti,review-1,review-2}, the amplitudes are
\begin{align}
&\A=\frac{1}{4}\int_{0}^{+\infty}\frac{\dd\tau_{2}}{\tau_{2}^{3}}\, \bigg\{\opvo\latticeo+2ND\spco\tetaqo\nonumber\\
&\hspace{2.3cm}+(\RN^{2}+\RD^{2})\omvo\tetado+2\RN\RD\smco\tetato\bigg\}\sum_{\vec{m}'}\frac{P_{\vec{m}'}^{(2)}}{\eta^{4}}~,\nonumber\\
&\M=-\frac{1}{4}\int_{0}^{+\infty}\frac{\dd\tau_{2}}{\tau_{2}^{3}}\, \bigg\{\opvm\latticem\\
&\hspace{6.8cm}-(N+D)\omvm\tetadm\bigg\}\sum_{\vec{m}'}\frac{P_{\vec{m}'}^{(2)}}{\hat \eta^{4}}~.\nonumber
\end{align}
The first line in the amplitude $\A$ ($\M$) contains the contributions of the NN, DD and ND sectors (N and D sectors), while the second line arises by acting with the orbifold generator $g$ on these sectors.

The RR tadpole cancellation condition fixes the number of D9- and D5-branes to be $N=D=32$. Moreover, the structure of the open-string partition functions prevents orthogonal gauge groups. Unitary gauge group parameterisation of the Chan--Paton multiplicities is  the only possibility, with 
\begin{equation}
N=n+\bar{n}~,\qquad D=d+\bar{d}~,\qquad R_{\text{N}}=i(n-\bar{n})~,\qquad R_{\text{D}}=i(d-\bar{d})~,
\end{equation}
which gives $n=\bar n=d=\bar d=16$. In this undeformed model,  the open-string gauge group is $U(16)\times U(16)$.

\subsection{Deformations of the BSGP model}
\label{GPPSWL}

The previous model can be deformed in various ways. In particular, the D5-branes can be displaced in $T^{4}/\Z_2$, Wilson lines along $T^{2}$ can be turned on for the gauge group associated with  the D5-branes, and ``Wilson lines'' along all of the six internal directions can be switched on for the gauge group generated by the D9-branes. All these deformations spontaneously break the original gauge group. As described in Sect.~\ref{21} we are using a T-dual language in which 
all brane positions and WL's are understood as D3-brane positions, with the understanding that this is merely a convenience, and that there is no common physical prescription
where this is actually the case. 

We are mostly interested in the case where the deformations take discrete values corresponding to all 32+32 D3-branes (\mbox{T-dual} to the D9- and D5-branes) sitting on the corners of a six-dimensional box (T-dual to $T^2\times T^4/\Z_2$). The WL's are equal to the components of $\vec a_{ii'}\equiv (\vec a_{i'}, \vec a_{i})$ which are  0 or $\half$, where the corners of the box are labelled by a double index $ii'$, in the notation of  Sect.~\ref{21}.  The annulus amplitude in this case becomes 
\be
\begin{aligned}
\label{eq:ann}
\A=&~\;\frac{1}{4}\int_{0}^{+\infty}\frac{\dd\tau_{2}}{\tau_{2}^{3}}\sum_{i,i'\atop j,j'} \bigg\{\opvo\latticeowl\\
&~+2\ndwl\spco\tetaqo+\rnwl\omvo\tetado\\
&~+2\rnrdwl\smco\tetato\bigg\}\sum_{\vec{m}'}\frac{P^{(2)}_{\vec{m}'+\vec{a}_{i'}-\vec{a}_{j'}}}{\eta^{4}}~,
\end{aligned}
\ee
and the M\"obius amplitude reads
\begin{align}\begin{split}
\M=-\frac{1}{4}\int_{0}^{+\infty}\frac{\dd\tau_{2}}{\tau_{2}^{3}}\sum_{i,i'} \bigg\{\opvm&\latticemwl\\
&-(N_{ii'}+D_{ii'})\omvm\tetadm\bigg\}\sum_{\vec{m}'}\frac{P^{(2)}_{\vec{m}'}}{ \hat{\eta}^{4}}~.
\end{split}\end{align}
By contrast, the amplitudes $\T$ and $\K$ in the closed-string sector are independent of the deformations (discrete or otherwise) that we have introduced, and are the same as the  expressions given in Eq.~(\ref{tknodef}). 

There are two subtleties in the annulus amplitude of Eq.~\eqref{eq:ann}: first, in the term that corresponds to the action of the  generator $g$ on the NN  and DD sectors (the last term on the second line), the orbifold action enforces being at the origin of the $T^4$ or $\tilde T^4$ lattice. This explains the presence of a Kr\"onecker symbol $\delta_{ij}$. Second, the last contribution,  which arises from the action of $g$ on the ND sector, is dressed by signs $e^{4i\pi\vec{a}_{i}\cdot\vec{a}_{j}}$ which are necessary in the presence of discrete D9-brane WL's~\cite{GimonPolchinski2}. 

This leads to the following open-string gauge symmetry in the presence of discrete deformations: 
\be
\G_{\rm open}=   \prod_{ii' / n_{ii'}   \neq 0} U(n_{ii'})  \times \prod_{jj' / d_{jj'}   \neq 0} U(d_{jj'})~, \quad \where\quad  n_{ii'}={N_{ii'}\over 2}~,\quad d_{jj'}={D_{jj'}\over 2}~.
\ee


\subsection{Supersymmetry breaking}
\label{GPPSWLSS}

As anticipated in Sect.~\ref{sec22},
the $\N=2\to \N=0$ spontaneous breaking of supersymmetry is  induced by the Scherk--Schwarz mechanism~\cite{openSS1,openSS2,openSS3,openSS4,openSS5,openSS6,openSS7, openSS8}. Implementing the associated shifts in Eq.~\eqref{shifts}, the $T^2$ lattices of zero modes in presence of discrete WL's are modified as follows:
\be
\begin{aligned}
&\Lambda^{(2,2)}_{\vec m ',(n_{4},2n_{5}+h)}&&\longrightarrow\quad \Lambda^{(2,2)}_{\vec m '+F\vec{a}'_{\text{S}},(n_{4},2n_{5}+h)}~,\quad h=0,1~,\espD\\
&P^{(2)}_{\vec{m}'+\vec{a}_{i'}-\vec{a}_{j'}}  &&\longrightarrow\quad   P^{(2)}_{\vec{m}'+F\vec{a}'_{\text{S}}+\vec{a}_{i'}-\vec{a}_{j'}}~.
\end{aligned}
\ee
As a result, the mass of the gravitino, which we may take as defining the scale of spontaneous supersymmetry breaking, is $M=\Ms\sqrt{G^{55}}/2$. 

To write the amplitudes, we work in the so called ``Scherk--Schwarz basis''~\cite{review-1} and change $(G_{54},G_{55},G_{5I})\rightarrow(G_{54}/2,G_{55}/4,G_{5I}/2)$, $I=6,\dots,9$. Moreover, for the massless spectrum to be easily readable, we split the result into the contributions of the bosonic and fermionic degrees of freedom  running in the loops. The torus amplitude is lengthy, being given by
\begin{align}
\label{torus}
\T&=\frac{1}{4}\int_{\F}\frac{\dd^{2}\tau}{\tau_{2}^{3}} \,\bigg\{\bigg[\left(\carq(V,O,O,V,+)+\carq(S,S,C,C,+)\right)\lattice\nonumber\\
&+\left(\carq(V,O,O,V,-)+\carq(S,S,C,C,-)\right)\tetad\nonumber\\
&+16\left(\carq(O,C,V,S,+)+\carq(S,O,C,V,+)\right)\tetaq\nonumber\\
&+16\left(\carq(O,C,V,S,-)+\carq(S,O,C,V,-)\right)\tetat\bigg]\sum_{\vec{m}',\vec{n}'}\LAMBDA(\vec{m}',{(n_{4},2n_{5})})\nonumber\\
&-\bigg[\left(\cars(V,O,O,V,+)\carsbar(S,S,C,C,+)+\cars(S,S,C,C,+)\carsbar(V,O,O,V,+)\right)\lattice\nonumber\\
&+\left(\cars(V,O,O,V,-)\carsbar(S,S,C,C,-)+\cars(S,S,C,C,-)\carsbar(V,O,O,V,-)\right)\tetad\nonumber\\
&+16\left(\cars(O,C,V,S,+)\carsbar(S,O,C,V,+)+\cars(S,O,C,V,+)\carsbar(O,C,V,S,+)\right)\tetaq\\
&+16\left(\cars(O,C,V,S,-)\carsbar(S,O,C,V,-)+\cars(S,O,C,V,-)\carsbar(O,C,V,S,-)\right)\tetat\bigg]\sum_{\vec{m}',\vec{n}'}\LAMBDA(\vec{m}'+\vec{a}'_{\text{S}},{(n_{4},2n_{5})})\nonumber\\
&+\bigg[\left(\carq(O,O,V,V,+)+\carq(C,S,S,C,+)\right)\lattice+\left(\carq(O,O,V,V,-)+\carq(S,C,C,S,-)\right)\tetad\nonumber\\
&+16\left(\carq(O,S,V,C,+)+\carq(S,V,C,O,+)\right)\tetaq\nonumber\\
&+16\left(\carq(O,S,V,C,-)+\carq(S,V,C,O,-)\right)\tetat\bigg]\sum_{\vec{m}',\vec{n}'} \LAMBDA(\vec{m}',{(n_{4},2n_{5}+1)})\nonumber\\
&-\bigg[\left(\cars(O,O,V,V,+)\carsbar(C,S,S,C,+)+\cars(C,S,S,C,+)\carsbar(O,O,V,V,+)\right)\lattice\nonumber\\
&+\left(\cars(O,O,V,V,-)\carsbar(S,C,C,S,-)+\cars(S,C,C,S,-)\carsbar(O,O,V,V,-)\right)\tetad\nonumber\\
&+16\left(\cars(O,S,V,C,+)\carsbar(S,V,C,O,+)+\cars(S,V,C,O,+)\carsbar(O,S,V,C,+)\right)\tetaq\nonumber\\
&+16\left(\cars(O,S,V,C,-)\carsbar(S,V,C,O,-)+\cars(S,V,C,O,-)\carsbar(O,S,V,C,-)\right)\tetat\bigg]\sum_{\vec{m}',\vec{n}'}\LAMBDA(\vec{m}'+\vec a_{\rm S}',{(n_{4},2n_{5}+1)})\bigg\}.\nonumber
\end{align}
%
The proliferation of terms is due to the presence of an untwisted sector along with three twisted sectors, 
either twisted by $g$, the Scherk--Schwarz generator, or the combination of the two. 
The only states flowing in the Klein bottle are left/right-symmetric, leading to the simpler contribution
\begin{align}\begin{split}
\label{klein}
\K=&~\frac{1}{4}\int_{0}^{+\infty}\frac{\dd\tau_{2}}{\tau_{2}^{3}}\, \bigg\{\left(V_{4}O_{4}+O_{4}V_{4}\right)\latticek+32\left(O_{4}C_{4}+V_{4}S_{4}\right)\tetaqo\\
&-\left(S_{4}S_{4}+C_{4}C_{4}\right)\latticek-32\left(S_{4}O_{4}+C_{4}V_{4}\right)\tetaqo\bigg\}\sum_{\vec{m}'}\frac{P^{(2)}_{\vec{m}'}}{\eta^{4}}\,.
\end{split}\end{align}
%
Finally, the open-string amplitudes are
\begin{align}
\label{annulus}
\A=&~\frac{1}{4}\int_{0}^{+\infty}\frac{\dd\tau_{2}}{\tau_{2}^{3}} \sum_{\substack{i,i'\\ j,j'}}\bigg\{\bigg[\cars(V,O,O,V,+)\latticeowl\nonumber\\
&+\cars(V,O,O,V,-)\rnwl\tetado+2\ndwl\cars(O,C,V,S,+)\tetaqo\nonumber\\
&+2\rnrdwl\cars(O,C,V,S,-)\tetato\bigg]\sum_{\vec{m}'}\frac{P^{(2)}_{\vec{m}'+\vec{a}_{i'}-\vec{a}_{j'}}}{\eta^{4}} \\
&-\bigg[\cars(S,S,C,C,+)\latticeowl\nonumber\\
&+\cars(C,C,S,S,-)\rnwl\tetado+2\ndwl\cars(S,O,C,V,+)\tetaqo\nonumber\\
&+2\rnrdwl\cars(S,O,C,V,-)\tetato\bigg]\sum_{\vec{m}'}\frac{P^{(2)}_{\vec{m}'+\vec{a}'_{\text{S}}+\vec{a}_{i'}-\vec{a}_{j'}}}{\eta^{4}}\bigg\}~,\nonumber
\end{align}
\begin{align}\begin{split}
\label{mobius}
\hspace{-2.2cm}\M=-\frac{1}{4}\int_{0}^{+\infty}\frac{\dd\tau_{2}}{\tau_{2}^{3}}\sum_{i,i'} &\bigg\{\bigg[\carsm(V,O,O,V,+)\latticemwl\\
&-(N_{ii'}+D_{ii'})\carsm(V,O,O,V,-)\tetadm\bigg]\sum_{\vec{m}'}\frac{P^{(2)}_{\vec{m}'}}{\hat{\eta}^{4}}\\
&-\bigg[\carsm(C,C,S,S,+)\latticemwl\\
&-(N_{ii'}+D_{ii'})\carsm(C,C,S,S,-)\tetadm\bigg]\sum_{\vec{m}'}\frac{P^{(2)}_{\vec{m}'+\vec{a}'_{\text{S}}}}{\hat{\eta}^{4}}\bigg\}~.
\end{split}\end{align}
%


\vspace{0.6cm}
\section{Potential and continuous Wilson lines}
\renewcommand{\theequation}{B.\arabic{equation}}

\label{Apot}
\noindent In this appendix, we derive the effective potential of the model realizing the $\N=2\to \N=0$ spontaneous breaking of supersymmetry, when continuous open-string WL's are switched on. Our aim is to obtain expressions suitable for the derivation in Sect.~\ref{mT1}  of the WL mass terms by taking two derivatives with respect to these moduli at points in moduli space where all D3-branes are coincident with O3-planes.

When  generalizing the open-string amplitudes $\A$ and $\M$ given in Eqs~(\ref{annulus}) and~(\ref{mobius}) to arbitrary positions of the D3-branes, the lattice deformations cannot be defined anymore by the positions $2\pi \vec a_{ii'}\equiv (\vec a_{i'},\vec a_i)$ of the fixed points  $ii'$. Instead, the deformations must be parameterised by the locations $2\pi a_\alpha^\I$ and $2\pi b_\alpha^\I$, $\alpha=1,\dots,32$, of the D3-branes in their appropriate six-dimensional boxes. However, as described in Sect. \ref{21}, the moduli space of WL's admits disconnected components, themselves admitting various Higgs, Coulomb and mixed Higgs--Coulomb branches. The number of moduli fields at tree level is thus highly dependent on the branch under interest.  To capture the information needed to Taylor expand the potential at any point in moduli space where all D3-branes are stacked on O3-planes, we denote    
\be
\begin{aligned}
\vec a_\alpha'&\equiv (a_\alpha^4,a_\alpha^5)~,\qquad \vec a_\alpha\equiv (a_\alpha^6,a_\alpha^7,a_\alpha^8,a_\alpha^9)~, \\
\vec b_\alpha'&\equiv (b_\alpha^4,b_\alpha^5)~,\qquad\; \vec b_\alpha\equiv (b_\alpha^6,b_\alpha^7,b_\alpha^8,b_\alpha^9)~,
\end{aligned}
\ee
and  write the annulus amplitude as follows, 
\begin{align}
\label{annulus_stab}
\A=&~\frac{1}{4}\int_{0}^{+\infty}\frac{\dd\tau_{2}}{\tau_{2}^{3}}\sum_{\alpha,\beta}\sum_{\vec{m}'} \bigg\{\frac{\cars(V,O,O,V,+)}{\eta^{8}}\latticeowlab\nonumber\\
&+2\cars(O,C,V,S,+)\tetaqo\frac{P^{(2)}_{\vec{m}'+\vec{a}'_{\alpha}-\vec{b}'_{\beta}}}{\eta^{4}}\espD\\
&-\bigg[\frac{\cars(S,S,C,C,+)}{\eta^{8}}\latticeowlabshift\nonumber\\
&+2\cars(S,O,C,V,+)\tetaqo\frac{P^{(2)}_{\vec{m}'+\vec{a}'_{\text{S}}+\vec{a}'_{\alpha}-\vec{b}'_{\beta}}}{\eta^{4}}\bigg]\bigg\}\;.\nonumber 
\end{align}
Some remarks are in order:
\begin{itemize}
\item In this expression,  even if all components $a_\alpha^\I$, $b_\alpha^\I$ appear formally as independent variables, it is understood that they are  correlated 4 by 4 or 2 by 2, or identically equal to 0 or $\half$, according to the point in moduli space around which fluctuations are considered. 
\item All terms appearing in the braces are continuous  deformations of the contributions proportional to $N_{ii'}$ or $D_{ii'}$ coefficients in Eq.~(\ref{annulus}).
\item When continuous WL's are switched on only along $T^2$, the model sits in a Coulomb branch where the unitary nature of all gauge group factors persists. Hence, all terms proportional to coefficients $R_{ii'}^{\rm N}$ or $R_{ii'}^{\rm D}$ in Eq.~(\ref{annulus}) yield after deformation contributions vanishing identically.\footnote{\label{foo} This cancellation is only numerical, thanks to the pairing of degenerate modes of eigenvalues $\pm 1$ under the orbifold generator $g$.}
\item When continuous WL's are switched on only along $T^4/\Z_2$ or $\tilde T^4\Z_2$, the model sits in a Higgs branch where unitary and symplectic gauge group factors cohabit. In that case, the coefficients $R_{ii'}^{\rm N}$ and $R_{ii'}^{\rm D}$ need to be re-evaluated with the numbers of D3-branes that  remain localized on the O3-planes. Therefore, all terms proportional to coefficients $R_{ii'}^{\rm N}$ or $R_{ii'}^{\rm D}$ in Eq.~(\ref{annulus}) yield after deformation contributions vanishing identically.$^{\ref{foo}}$
\end{itemize}
Similarly, the M\"obius strip amplitude~(\ref{mobius})  reads in presence of continuous deformations
\begin{align}
\label{mobius_stab}
\M=&-\frac{1}{4}\int_{0}^{+\infty}\frac{\dd\tau_{2}}{\tau_{2}^{3}}\sum_{\alpha}\sum_{\vec{m}'} \bigg\{\frac{\carsm(V,O,O,V,+)}{\hat{\eta}^{8}}\latticemwlab\nonumber\\
&\,-\frac{\carsm(C,C,S,S,+)}{\hat{\eta}^{8}}\latticemwlabshift\bigg\}\;,
\end{align}
where all $a_\alpha^\I$, $b_\alpha^\I$ are again formally treated as free variables. In this expression, the terms proportional to the combinations of $SO(4)\times SO(4)$ characters $\hat V_4 \hat O_4-\hat O_4\hat V_4$ or $\hat C_4 \hat C_4-\hat S_4\hat S_4$ are omitted, since they vanish identically.$^{\ref{foo}}$

Next, we may expand the characters as follows,
\begin{align}
\begin{split}
&\frac{V_{4}O_{4}+O_{4}V_{4}}{\eta^{8}}=\frac{C_{4}C_{4}+S_{4}S_{4}}{\eta^{8}}=8\sum_{k\geq 0}c_{k}e^{-\pi k\tau_{2}}~,\\
&\frac{\hat{V}_{4}\hat{O}_{4}+\hat{O}_{4}\hat{V}_{4}}{\hat{\eta}^{8}}=\frac{\hat{C}_{4}\hat{C}_{4}+\hat{S}_{4}\hat{S}_{4}}{\hat{\eta}^{8}}=8\sum_{k\geq 0}(-1)^{k}c_{k}e^{-\pi k\tau_{2}}~,\\
&2(O_{4}C_{4}+V_{4}S_{4})\left(\frac{\eta}{\vartheta_{4}}\right)^{2}\frac{1}{\eta^{4}}=2(S_{4}O_{4}+C_{4}V_{4})\left(\frac{\eta}{\vartheta_{4}}\right)^{2}\frac{1}{\eta^{4}}=4\sum_{k\geq 0}d_{k}e^{-\frac{\pi}{2}k\tau_{2}}~,
\end{split}
\end{align}
where $c_{0}=d_{0}=1$, to obtain 
\begin{align}
\begin{split}
\label{annulus_B}
\A=2\int_{0}^{+\infty}\frac{\dd\tau_{2}}{\tau_{2}^{3}}\sum_{k\geq 0}\sum_{\alpha,\beta}\sum_{\vec{m}'} \bigg\{c_{k}e^{-\pi k\tau_{2}}&\bigg[\sum_{\vec{m}}P^{(4)}_{\vec{m}+\vec{a}_{\alpha}-\vec{a}_{\beta}}\left(P^{(2)}_{\vec{m}'+\vec{a}'_{\alpha}-\vec{a}'_{\beta}}-P^{(2)}_{\vec{m}'+\vec{a}'_{\text{S}}+\vec{a}'_{\alpha}-\vec{a}'_{\beta}}\right)\\
&+\sum_{\vec{n}}W^{(4)}_{\vec{n}+\vec{b}_{\alpha}-\vec{b}_{\beta}}\left(P^{(2)}_{\vec{m}'+\vec{b}'_{\alpha}-\vec{b}'_{\beta}}-P^{(2)}_{\vec{m}'+\vec{a}'_{\text{S}}+\vec{b}'_{\alpha}-\vec{b}'_{\beta}}\right)\bigg]\\
&+d_{k}e^{-\frac{\pi}{2}k\tau_{2}}\left(P^{(2)}_{\vec{m}'+\vec{a}'_{\alpha}-\vec{b}'_{\beta}}-P^{(2)}_{\vec{m}'+\vec{a}'_{\text{S}}+\vec{a}'_{\alpha}-\vec{b}'_{\beta}}\right)\bigg\}\;,
\end{split}
\end{align}
and
\begin{align}
\begin{split}
\label{mobius_B}
\M=-2\int_{0}^{+\infty}\frac{\dd\tau_{2}}{\tau_{2}^{3}}\sum_{k\geq 0}\sum_{\alpha}\sum_{\vec{m}'}\bigg\{(-1)^{k}c_{k}&\bigg[\sum_{\vec{m}}P^{(4)}_{\vec{m}+2\vec{a}_{\alpha}}\left(P^{(2)}_{\vec{m}'+2\vec{a}'_{\alpha}}-P^{(2)}_{\vec{m}'+\vec{a}'_{\text{S}}+2\vec{a}'_{\alpha}}\right)\\
&+\sum_{\vec{n}}W^{(4)}_{\vec{n}+2\vec{b}_{\alpha}}\left(P^{(2)}_{\vec{m}'+2\vec{b}'_{\alpha}}-P^{(2)}_{\vec{m}'+\vec{a}'_{\text{S}}+2\vec{b}'_{\alpha}}\right)\bigg]\bigg\}\;.
\end{split}
\end{align}


The moduli space region in which we are interested to find the WL masses is where the lightest non-vanishing scale of the model is the supersymmetry breaking scale \mbox{$M=\Ms\sqrt{G^{55}}/2$}. In terms of internal metric components, this means that  
\begin{equation}
G^{55}\ll G_{44},|G_{IJ}|\ll G_{55}~,~\quad |G_{45}|, |G_{5J}|\ll\sqrt{G_{55}}~,~\quad  I,J\in\{6,\dots,9\}~,~\quad G_{55}\gg 1~.
\end{equation}
The Scherk--Schwarz compact direction $X^5$ being large, it is convenient to Poisson sum over the momentum $m_{5}$ (the new sum index is denoted $l_{5}$). The annulus amplitude becomes 
\begin{align}
\begin{split}
\A&=\left(G^{55}\right)^{2}\frac{\Gamma\big(\frac{5}{2}\big)}{\pi^{\frac{5}{2}}}\,4\sum_{k\geq 0}\sum_{\alpha,\beta}\sum_{m_{4}}\sum_{l_{5}}\frac{1}{|2l_{5}+1|^{5}}\\
&\bigg\{\sum_{\vec{m}}c_{k}\cos\left[2\pi |2l_{5}+1|\left(a_{\alpha}^{5}-a_{\beta}^{5}+\frac{G^{54}}{G^{55}}(m_{4}+a_{\alpha}^{4}-a_{\beta}^{4})\right)\right]\cH_{\frac{5}{2}}\left(\pi |2l_{5}+1|\frac{\M_{\A_{1}}}{\sqrt{G^{55}}}\right)\\
&+\sum_{\vec{n}}c_{k}\cos\left[2\pi |2l_{5}+1|\left(b_{\alpha}^{5}-b_{\beta}^{5}+\frac{G^{54}}{G^{55}}(m_{4}+b_{\alpha}^{4}-b_{\beta}^{4})\right)\right]\cH_{\frac{5}{2}}\left(\pi |2l_{5}+1|\frac{\M_{\A_{2}}}{\sqrt{G^{55}}}\right)\\
&+\frac{d_{k}}{2}\cos\left[2\pi |2l_{5}+1|\left(a_{\alpha}^{5}-b_{\beta}^{5}+\frac{G^{54}}{G^{55}}(m_{4}+a_{\alpha}^{4}-b_{\beta}^{4})\right)\right]\cH_{\frac{5}{2}}\left(\pi |2l_{5}+1|\frac{\M_{\A_{3}}}{\sqrt{G^{55}}}\right)\bigg\}\;,
\end{split}
\label{aamp}
\end{align}
where the function $\cH_{\nu}$ can be expressed in terms of $K_\nu$, a modified Bessel function of the second kind,
\be 
\label{cH}
\cH_\nu(z)= {1\over \Gamma(\nu)}\int_0^{+\infty}{dx\over x^{1+\nu}}\, e^{-{1\over x}-z^2x}={2\over \Gamma(\nu)}\, z^\nu K_\nu(2z)~.
\ee
In Eq.~(\ref{aamp}), $\M_{\A_1}$, $\M_{\A_2}$ and $\M_{\A_3}$ define three characteristic mass scales (in string units) satisfying
\begin{align}
\begin{split}
\M_{\A_{1}}^{2}&\;=(m_{I}+a_{\alpha}^{I}-a_{\beta}^{I})G^{IJ}(m_{J}+a_{\alpha}^{J}-a_{\beta}^{J})+(m_{4}+a_{\alpha}^{4}-a_{\beta}^{4})^{2}\hat{G}^{44}+k~,\\
\M_{\A_{2}}^{2}&\;=(n_{I}+b_{\alpha}^{I}-b_{\beta}^{I})G_{IJ}(n_{J}+b_{\alpha}^{J}-b_{\beta}^{J})+(m_{4}+b_{\alpha}^{4}-b_{\beta}^{4})^{2}\hat{G}^{44}+k~,\\
\M_{\A_{3}}^{2}&\;=(m_{4}+a_{\alpha}^{4}-b_{\beta}^{4})^{2}\hat{G}^{44}+\frac{k}{2}~,
\end{split}
\label{m123}
\end{align}
where 
\be
\hat{G}^{44}=G^{44}-\frac{G^{45}}{G^{55}}\,G^{55}\,\frac{G^{54}}{G^{55}}~.
\label{ghat}
\ee
Because we are interested in motions of D3-brane around O3-planes, we split the WL moduli into background values and fluctuations,
\begin{align}
\begin{split}
\label{WL}
a_{\alpha}^{\I}&=\langle a_{\alpha}^{\I}\rangle+\epsilon_{\alpha}^{\I}~,\qquad \langle a_{\alpha}^{\I}\rangle\in\left\{0,\frac{1}{2}\right\}\,,\\
b_{\alpha}^{\I}&=\langle b_{\alpha}^{\I}\rangle+\xi_{\alpha}^{\I}~,\,\qquad\langle b_{\alpha}^{\I}\rangle\in\left\{0,\frac{1}{2}\right\}\,,
\end{split}
\end{align}
which allow us to determine when the masses~(\ref{m123}) are large or small compared to $M$. This is relevant since $\cH_\nu$ is finite for small argument and exponentially suppressed for large argument:
\be 
\label{expsup}
 \cH_\nu(z)=1-{z^2\over \nu-1}+\O(z^4)\quad \mbox{as}\quad |z|\ll 1~,\quad\,\cH_\nu(z)~\sim~ {\sqrt{\pi}\over \Gamma(\nu)}\, z^{\nu-{1\over 2}}\, e^{-2z}\quad \mbox{as} \quad z\gg 1~.
\ee
For $\M_{\A_{1}}/\sqrt{G^{55}}$ not to yield exponentially suppressed contributions to $\A$, we need $k=0$, $m_{I}+\langle a_{\alpha}^{I}\rangle-\langle a_{\beta}^{I}\rangle=0$ and $m_{4}+\langle a_{\alpha}^{4}\rangle-\langle a_{\beta}^{4}\rangle=0$. This amounts to having $\vec{m}=\vec{0}$, $m_{4}=0$ and $(\alpha,\beta)$ in the set $L_{\rm NN}$ such that the D3-branes $\alpha,\beta$ T-dual to D9-branes
\renewcommand{\labelitemi}{$\bullet$}
\begin{itemize}
\setlength{\itemsep}{0pt}
\item belong to the same stack of $N_{ii'}$ branes, $i=1,\dots,16$, $i'=1,\dots,4$,
\item or belong respectively to stacks of $N_{i,2i''-1}$ and $N_{i,2i''}$ branes, $i=1,\dots,16$, $i''=1,2$,
\item or belong respectively to stacks of  $N_{i,2i''}$ and $N_{i,2i''-1}$ branes, $i=1,\dots,16$, $i''=1,2$.
\end{itemize}
Similarly, for $\M_{\A_{2}}/\sqrt{G^{55}}$ not to yield exponentially suppressed terms in $\A$, we need $k=0$, $\vec n=\vec 0$, $m_4=0$ and 
$(\alpha,\beta)$ in the set $L_{\rm DD}$ such that the D3-branes $\alpha,\beta$ T-dual to D5-branes
\renewcommand{\labelitemi}{$\bullet$}
\begin{itemize}
\setlength{\itemsep}{0pt}
\item belong to the same stack of $D_{ii'}$ branes, $i=1,\dots,16$, $i'=1,\dots,4$,
\item or belong respectively to stacks of $D_{i,2i''-1}$ and $D_{i,2i''}$ branes, $i=1,\dots,16$, $i''=1,2$,
\item or belong respectively to stacks of  $D_{i,2i''}$ and $D_{i,2i''-1}$ branes, $i=1,\dots,16$, $i''=1,2$.
\end{itemize}
Finally, terms involving $\M_{\A_{3}}/\sqrt{G^{55}}$ are relevant when 
$k=0$ and $m_{4}+\langle a_{\alpha}^{4}\rangle-\langle b_{\beta}^{4}\rangle=0$. This is achieved if $m_{4}=0$ and $(\alpha,\beta)$ is  in the set $L_{\rm ND}$ such that the D3-branes $\alpha,\beta$ T-dual to a D9-brane and a D5-brane 
\begin{itemize}
\setlength{\itemsep}{0pt}
\item belong respectively to stacks of $N_{ii'}$ and $D_{ji'}$ branes, $i,j=1,\dots,16$, $i'=1,\dots,4$,
\item or belong respectively to stacks of $N_{i,2i''-1}$ and $D_{j,2i''}$ branes, $i,j=1,\dots,16$, $i''=1,2$,
\item or belong respectively to stacks of $N_{j,2i''}$ and $D_{i,2i''-1}$ branes, $i,j=1,\dots,16$, $i''=1,2$.
\end{itemize}
Up to exponentially suppressed terms, we thus obtain
\begin{align}
\A=&\,\left(G^{55}\right)^{2}\frac{\Gamma\big(\frac{5}{2}\big)}{\pi^{\frac{5}{2}}}\sum_{l_{5}}\frac{4}{|2l_{5}+1|^{5}}\Bigg\{\hspace{0.3cm}\sum_{\mathclap{(\alpha,\beta)\in L_{\rm NN}}}(-)^{2(\langle a_{\alpha}^{5}\rangle-\langle a_{\beta}^{5}\rangle)}\cos\left[2\pi |2l_{5}+1|\left(\epsilon_{\alpha}^{5}-\epsilon_{\beta}^{5}+\frac{G^{54}}{G^{55}}(\epsilon_{\alpha}^{4}-\epsilon_{\beta}^{4})\right)\right]\nonumber\\
&\qquad\qquad\qquad\quad\times\cH_{\frac{5}{2}}\Bigg(\pi |2l_{5}+1|\frac{\left[(\epsilon_{\alpha}^{I}-\epsilon_{\beta}^{I})G^{IJ}(\epsilon_{\alpha}^{J}-\epsilon_{\beta}^{J})+(\epsilon_{\alpha}^{4}-\epsilon_{\beta}^{4})^{2}\hat{G}^{44}\right]^{\frac{1}{2}}}{\sqrt{G^{55}}}\Bigg)\nonumber\\
&\;+\sum_{\mathclap{(\alpha,\beta)\in L_{\rm DD}}}(-)^{2(\langle b_{\alpha}^{5}\rangle-\langle b_{\beta}^{5}\rangle)}\cos\left[2\pi |2l_{5}+1|\left(\xi_{\alpha}^{5}-\xi_{\beta}^{5}+\frac{G^{54}}{G^{55}}(\xi_{\alpha}^{4}-\xi_{\beta}^{4})\right)\right]\nonumber\\
&\qquad\qquad\qquad\quad\times\cH_{\frac{5}{2}}\Bigg(\pi |2l_{5}+1|\frac{\left[(\xi_{\alpha}^{I}-\xi_{\beta}^{I})G_{IJ}(\xi_{\alpha}^{J}-\xi_{\beta}^{J})+(\xi_{\alpha}^{4}-\xi_{\beta}^{4})^{2}\hat{G}^{44}\right]^{\frac{1}{2}}}{\sqrt{G^{55}}}\Bigg)\label{annulus_epsilon}\\
&\;+\half\sum_{\mathclap{(\alpha,\beta)\in L_{\rm ND}}}(-)^{2(\langle a_{\alpha}^{5}\rangle-\langle b_{\beta}^{5}\rangle)}\cos\left[2\pi |2l_{5}+1|\left(\epsilon_{\alpha}^{5}-\xi_{\beta}^{5}+{G^{54}\over G^{55}}(\epsilon_{\alpha}^{4}-\xi_{\beta}^{4})\right)\right]\nonumber\\
&\qquad\qquad\qquad\quad\times\cH_{\frac{5}{2}}\Bigg(\pi |2l_{5}+1|\frac{\left[(\epsilon_{\alpha}^{4}-\xi_{\beta}^{4})^{2}\hat{G}^{44}\right]^{\frac{1}{2}}}{\sqrt{G^{55}}}\Bigg)\Bigg\}+\O\left(G^{55}e^{-\frac{2\pi c}{\sqrt{G^{55}}}}\right)\,,\nonumber
\end{align}
where $c$ is positive  of order one.

Proceeding in a similar way with the M\"obius amplitude, we may write
\begin{align}
\begin{split}
\M=&-\left(G^{55}\right)^{2}\frac{\Gamma\big(\frac{5}{2}\big)}{\pi^{\frac{5}{2}}}\,4\sum_{k\geq 0}(-1)^{k}c_{k}\sum_{\alpha}\sum_{m_{4}}\sum_{l_{5}}\frac{1}{|2l_{5}+1|^{5}}\\
&\Bigg\{\sum_{\vec{m}}\cos\left[2\pi |2l_{5}+1|\left(2a_{\alpha}^{5}+\frac{G^{54}}{G^{55}}(m_{4}+2a_{\alpha}^{4})\right)\right]\cH_{\frac{5}{2}}\left(\pi |2l_{5}+1|\frac{\M_{\M_{1}}}{\sqrt{G^{55}}}\right)\\
&+\sum_{\vec{n}}\cos\left[2\pi |2l_{5}+1|\left(2b_{\alpha}^{5}+\frac{G^{54}}{G^{55}}(m_{4}+2b_{\alpha}^{4})\right)\right]\cH_{\frac{5}{2}}\left(\pi |2l_{5}+1|\frac{\M_{\M_{2}}}{\sqrt{G^{55}}}\right)\Bigg\}\;,
\end{split}
\end{align}
which involves characteristic mass scales 
\begin{align}
\begin{split}
\M_{\M_{1}}^{2}&=(m_{I}+2a_{\alpha}^{I})G^{IJ}(m_{J}+2a_{\alpha}^{J})+(m_{4}+2a_{\alpha}^{4})^{2}\hat{G}^{44}+k~,\\
\M_{\M_{2}}^{2}&=(n_{I}+2b_{\alpha}^{I})G_{IJ}(n_{J}+2b_{\alpha}^{J})+(m_{4}+2b_{\alpha}^{4})^{2}\hat{G}^{44}+k~.
\end{split}
\end{align}
The functions $\cH_{5\over 2}$ are exponentially suppressed unless their arguments satisfy $k=0$ and $m_{I}=-2\langle a_{\alpha}^{I}\rangle$, $m_{4}=-2\langle a_{\alpha}^{4}\rangle$, or $n_{I}=-2\langle b_{\alpha}^{I}\rangle$, $m_{4}=-2\langle b_{\alpha}^{4}\rangle$. As a result, the amplitude takes the following form
\begin{align}
\M=&-\left(G^{55}\right)^{2}\frac{\Gamma\big(\frac{5}{2}\big)}{\pi^{\frac{5}{2}}}\sum_{\alpha}\sum_{l_{5}}\frac{4}{|2l_{5}+1|^{5}}\vast\{\cos\left[4\pi |2l_{5}+1|\left(\epsilon_{\alpha}^{5}+\frac{G^{54}}{G^{55}}\epsilon_{\alpha}^{4}\right)\right]\nonumber \\
&\qquad\qquad\qquad\quad\times\cH_{\frac{5}{2}}\Bigg(2\pi |2l_{5}+1|\frac{\left[\epsilon_{\alpha}^{I}G^{IJ}\epsilon_{\alpha}^{J}+\left(\epsilon_{\alpha}^{4}\right)^{2}\hat{G}^{44}\right]^{\frac{1}{2}}}{\sqrt{G^{55}}}\Bigg)\nonumber \\
&+\cos\left[4\pi |2l_{5}+1|\left(\xi_{\alpha}^{5}+\frac{G^{54}}{G^{55}}\xi_{\alpha}^{4}\right)\right]\label{mobius_epsilon}\\
&\qquad\qquad\qquad\quad\times\cH_{\frac{5}{2}}\Bigg(2\pi |2l_{5}+1|\frac{\left[\xi_{\alpha}^{I}G_{IJ}\xi_{\alpha}^{J}+\left(\xi_{\alpha}^{4}\right)^{2}\hat{G}^{44}\right]^{\half}}{\sqrt{G^{55}}}\Bigg)\Bigg\}+\O\left(G^{55}e^{-\frac{2\pi c}{\sqrt{G^{55}}}}\right)\,.\nonumber 
\end{align}

Adding the annulus and M\"obius strip amplitudes, the contribution of the open-string sector to the one-loop effective potential can be written as 
\begin{equation}
\label{potential_open}
-{\Ms^4\over 2(2\pi)^4}\, (\A+\M)=\frac{\Gamma\big(\frac{5}{2}\big)}{\pi^{\frac{13}{2}}}M^{4}\sum_{l_{5}}\frac{\N^{\rm open}_{2l_{5}+1}(\epsilon,\xi,G)}{|2l_{5}+1|^{5}}+\O\left((\Ms M)^{2}e^{-2\pi c\frac{\Ms }{M}}\right)\,,
\end{equation}
where $\N^{\rm open}_{2l_{5}+1}(\epsilon,\xi,G)$ is given by 
\begin{align}
\N^{\rm open}_{2l_{5}+1}(\epsilon,\xi,G&)=2\, \Bigg\{-\sum_{\mathclap{(\alpha,\beta)\in L_{\rm NN}}}(-)^{F}\cos\left[2\pi |2l_{5}+1|\frac{G^{5I'}}{G^{55}}\left(\epsilon_{\alpha}^{I'}-\epsilon_{\beta}^{I'}\right)\right]\nonumber\\
&\qquad \qquad\times\cH_{\frac{5}{2}}\Bigg(\pi |2l_{5}+1|\frac{\left[(\epsilon_{\alpha}^{I}-\epsilon_{\beta}^{I})G^{IJ}(\epsilon_{\alpha}^{J}-\epsilon_{\beta}^{J})+(\epsilon_{\alpha}^{4}-\epsilon_{\beta}^{4})^{2}\hat{G}^{44}\right]^{\frac{1}{2}}}{\sqrt{G^{55}}}\Bigg)\nonumber\\
&-\sum_{\mathclap{(\alpha,\beta)\in L_{\rm DD}}}(-)^{F}\cos\left[2\pi |2l_{5}+1|\frac{G^{5I'}}{G^{55}}\left(\xi_{\alpha}^{I'}-\xi_{\beta}^{I'}\right)\right]\nonumber\\
&\qquad\qquad\times\cH_{\frac{5}{2}}\Bigg(\pi |2l_{5}+1|\frac{\left[(\xi_{\alpha}^{I}-\xi_{\beta}^{I})G_{IJ}(\xi_{\alpha}^{J}-\xi_{\beta}^{J})+(\xi_{\alpha}^{4}-\xi_{\beta}^{4})^{2}\hat{G}^{44}\right]^{\frac{1}{2}}}{\sqrt{G^{55}}}\Bigg)\label{N}\\
&-\half\sum_{\mathclap{(\alpha,\beta)\in L_{\rm ND}}}(-)^{F}\cos\left[2\pi |2l_{5}+1|\frac{G^{5I'}}{G^{55}}\left(\epsilon_{\alpha}^{I'}-\xi_{\beta}^{I'}\right)\right]\cH_{\frac{5}{2}}\Bigg(\pi |2l_{5}+1|\frac{\left[(\epsilon_{\alpha}^{4}-\xi_{\beta}^{4})^{2}\hat{G}^{44}\right]^{\frac{1}{2}}}{\sqrt{G^{55}}}\Bigg)\nonumber\\
&+\sum_{\alpha}\cos\left[4\pi |2l_{5}+1|\frac{G^{5I'}}{G^{55}}\epsilon_{\alpha}^{I'}\right]\cH_{\frac{5}{2}}\Bigg(2\pi |2l_{5}+1|\frac{\left[\epsilon_{\alpha}^{I}G^{IJ}\epsilon_{\alpha}^{J}+\left(\epsilon_{\alpha}^{4}\right)^{2}\hat{G}^{44}\right]^{\frac{1}{2}}}{\sqrt{G^{55}}}\Bigg)\nonumber\\
&+\sum_{\alpha}\cos\left[4\pi |2l_{5}+1|\frac{G^{5I'}}{G^{55}}\xi_{\alpha}^{I'}\right]\cH_{\frac{5}{2}}\Bigg(2\pi |2l_{5}+1|\frac{\left[\xi_{\alpha}^{I}G_{IJ}\xi_{\alpha}^{J}+\left(\xi_{\alpha}^{4}\right)^{2}\hat{G}^{44}\right]^{\frac{1}{2}}}{\sqrt{G^{55}}}\Bigg)\Bigg\}\;.\nonumber
\end{align}
In this expression, $F$ is the fermionic number of the string $(\alpha,\beta)\in L_{\rm NN}\cup L_{\rm DD}\cup L_{\rm ND}$.

\end{appendices}



\end{document}